\documentclass[12pt,prd,preprint,nofootinbib]{revtex4}

\usepackage[pdftex]{graphicx}
\usepackage{latexsym,amsmath,amssymb,lmodern,float,url}
\usepackage{natbib}
\usepackage[pdftex,bookmarks,linktocpage,pdfpagelabels,plainpages=false,hyperfigures,linkcolor=blue,citecolor=blue]{hyperref} 
\hypersetup{colorlinks=true}

\newcommand\ee{\end{equation}}
\newcommand\be{\begin{equation}}
\newcommand\eea{\end{eqnarray}}
\newcommand\bea{\begin{eqnarray}}

\begin{document}
MI-TH-169

\title{No $\nu$ floors: Effective field theory treatment of the neutrino background in direct dark matter detection experiments}
\author{James B.~Dent$^{\bf a}$}
\author{Bhaskar Dutta$^{\bf b}$}
\author{Jayden L.~Newstead$^{\bf c}$}
\author{Louis E. Strigari$^{\bf b}$}

\affiliation{$^{\bf a}$ Department of Physics, University of Louisiana at Lafayette, Lafayette, LA 70504, USA}
\affiliation{$^{\bf b}$ Mitchell Institute for Fundamental Physics and Astronomy,  Department of Physics and Astronomy, Texas A\&M University, College Station, TX 77845, USA}
\affiliation{$^{\bf c}$ Department of Physics, Arizona State University, Tempe, AZ 85287, USA}
 
\begin{abstract}
Distinguishing a dark matter interaction from an astrophysical neutrino-induced interaction will be major challenge for future direct dark matter searches. In this paper, we consider this issue within non-relativistic Effective Field Theory (EFT), which provides a well-motivated theoretical framework for determining nuclear responses to dark matter scattering events. We analyze the nuclear energy recoil spectra from the different dark matter-nucleon EFT operators, and compare to the nuclear recoil energy spectra that is predicted to be induced by astrophysical neutrino sources. We determine that for 11 of the 14 possible operators, the dark matter-induced recoil spectra can be cleanly distinguished from the corresponding neutrino-induced recoil spectra with moderate size detector technologies that are now being pursued, e.g., these operators would require 0.5 tonne years to be distinguished from the neutrino background for low mass dark matter. Our results imply that in most models detectors with good energy resolution will be able to distinguish a dark matter signal from a neutrino signal, without the need for much larger detectors that must rely on additional information from timing or direction. 
\end{abstract}

\maketitle

\section{Introduction}
\par For over two decades, direct dark matter detection experiments have made great progress in searching for dark matter in the form of Weakly Interacting Massive Particles (WIMPs). The most stringent bounds now constrain the spin-independent and spin dependent  WIMP-nucleon direct detection cross sections to be less than $\sim 10^{-46}$ cm$^2$~\cite{Akerib:2015rjg} and $\sim 10^{-39}$ cm$^2$~\cite{Amole:2015lsj} respectively. Larger scale detectors in development are expected to further improve the cross section bounds by 2-3 orders of magnitude~\cite{Aprile:2015uzo,Akerib:2015cja}. From a theoretical perspective, experiments are now probing dark matter that interacts with nucleons through tree-level Higgs exchange. 

\par A theoretical interpretation of the experimental limits depends on a detailed modeling of the WIMP-nucleus interaction. The WIMP-nucleus interaction is  traditionally approximated by multiplying the cross section at zero-moment transfer, i.e. the point nucleus model, by the form factor, which represents the extended structure of the nucleus and encodes the momentum dependence of the interaction~\cite{Goodman:1984dc,Lewin:1995rx}. The WIMP-nucleon interaction is approximated as a sum of a spin-independent (SI) and spin-dependent (SD) cross section. The SI interaction is coherent on the nucleus, leading to an enhanced sensitivity, so that current experimental limits on the SI interaction are much stronger than on the SD interaction. 

\par In a series of recent papers, this standard theoretical formalism has been generalized within a non-relativistic effective field theory (EFT) model for the nucleus, in which the WIMP interacts with a nucleon via a larger sample of operators~\cite{Fitzpatrick:2012ix,Anand:2013yka}. Additional nuclear responses were identified that augment the standard SI and SD responses. Though these operators induce nuclear recoil energy spectra that differ from the traditional SI/SD models, the upper limits on the WIMP-nucleon cross section are not strongly affected, except for experiments with relatively high recoil energy thresholds, greater than tens of keV~\cite{Gresham:2014vja}. The non-relativistic EFT formalism also provides unique signatures in direct detection experiments with directional sensitivity~\cite{Catena:2015vpa,Kavanagh:2015jma}. 

\par Larger volume, next generation direct dark matter searches that detect WIMPs only via the energy deposition from the WIMP to the nucleus will be affected by a background from neutrinos produced in the Sun, atmosphere, and diffuse supernovae~\cite{Monroe:2007xp,Strigari:2009bq,Billard:2013qya}. Neutrinos from these sources will interact primarily with nuclei through the coherent scattering process, which is induced by neutrinos with energies of tens of MeV. Considering the nuclear recoil energy spectrum alone, within the traditional SI/SD formalism Solar neutrinos mimic a WIMP with mass $\sim 6$ GeV, while atmospheric neutrinos mimic a WIMP with mass $\sim 100$ GeV~\cite{Billard:2013qya,Ruppin:2014bra}. 

\par Identifying and reducing the neutrino backgrounds presents a significant challenge for direct detection experiments, in particular those which strive to reach the ton scale and beyond. Several recent studies have discussed methods to distinguish WIMPs from neutrinos in next generation detectors. Ruppin et al. discussed the prospects for exploiting the complementarity between detectors that use different nuclear targets to detect energy deposition, considering both SI and SD interactions~\cite{Ruppin:2014bra}. Davis~\cite{Davis:2014ama} considered the difference between time variation of the WIMP signal, due to the well-known annual modulation~\cite{Freese:2012xd} from the rotation of the Earth around the Sun, and the Solar neutrino signal, which is due to the small but non-zero eccentricity of the Earth's orbit around the Sun. These time variations generate a phase difference between the Solar neutrino and the WIMP signal. Grothaus~\cite{Grothaus:2014hja} and O'Hare et al.~\cite{O'Hare:2015mda} discussed the prospects for exploiting the difference in the direction of the nuclear recoil energy induced by the WIMPs and Solar neutrinos. For these timing and directional-based techniques, an exposure on the scale of 100 tonne years is required to distinguish the Solar neutrino background from a WIMP signal. 

\par In this paper, we calculate the expected WIMP signal in future detectors using non-relativistic EFT, and compare to the predictions of the astrophysical neutrino backgrounds. For each of the operators that describe the WIMP coupling to the nucleons within EFT, and for a wide range of WIMP masses, we compare the nuclear recoil energy spectrum to the neutrino backgrounds. Using the nuclear recoil energy spectrum, we categorize the operators that both can and cannot be distinguished from the neutrino backgrounds. We find that the majority of the operators can in fact be distinguished from the neutrino backgrounds over the entire WIMP mass range. For the few operators that cannot be distinguished, we identify the specific WIMP mass that best matches the neutrino background, and highlight the scatter in this best matching mass between the operators. Our results imply, for detectors with good nuclear recoil energy resolution, that the neutrino background is less significant than it is when using the traditional SI/SD formalism. 

\par This paper is organized as follows. In Section~\ref{sec:scattering} we briefly review both the physics of non-relativistic EFT and of neutrino coherent scattering. In Section~\ref{sec:spectra} we calculate the nuclear recoil spectra for EFT operators, and identify the operators that induce nuclear recoils that mimic the neutrino backgrounds. In Section~\ref{sec:discovery} we calculate the discovery limit for each operator in light of the neutrino background, and show that many of the operators can in fact be distinguished from the neutrino background over a wide range of masses. In Section~\ref{sec:conclusions}, we present our summary and conclusions. 

\section{WIMP and neutrino scattering with nuclei} 
\label{sec:scattering}
\par In this section we review the WIMP-nucleus and the neutrino-nucleus scattering formalism that is required for our analysis. For  WIMP-nucleus scattering, we describe the necessary ingredients of non-relativistic EFT, while for neutrino-nucleus scattering we describe the cross section that is predicted in the Standard Model. 

\subsection{Non-relativistic EFT WIMP-nucleus scattering} 

Dark matter-nucleus scattering is expected to occur due to the presence of a dark matter distribution in our galaxy, with the interaction rate being sensitive to both the local dark matter density (for reviews of observations and theoretical models of the local density see for example \cite{Read:2014qva,Strigari:2013iaa}) and the velocity distribution of the dark matter, as well as the nuclear properties of the target material.  The precise form of the velocity distribution is unknown, but can be modeled using N-body simulations~\cite{Mao:2012hf,Mao:2013nda}. The speed of the dark matter is predicted to be in the $\mathcal{O}$(few 100km/s) region, with an upper limit corresponding to the galactic escape velocity (the RAVE survey gave a value of $533^{+54}_{-41}$ km/s at 90\% confidence \cite{Piffl:2013mla}).  Standard momentum exchanged in such collisions is in the MeV range, which lends direct detection interactions to a non-relativistic effective field theory treatment for mediator particles with masses above this value, which is the case for a large variety of dark matter models.

Traditionally WIMP-nucleus scattering has been formulated as an incident dark matter particle scattering off a nucleus through either SI or SD interactions. However, as theoretical investigations into the particle nature of dark matter have broadened in scope to include a more general set of interactions, including a variety of velocity and momentum dependence, it has been recognized that the SI/SD interaction categorization insufficiently captures the range of the possible relevant interaction properties. Importantly, not including the full array of interactions and nuclear responses could lead to a misinterpretation of any future direct detection observations if carried out within the conventional framework \cite{Catena:2014hla}. 

Following standard semi-leptonic electroweak treatments of the nuclear physics involved in the WIMP-nucleus scattering, it has been shown~\cite{Fitzpatrick:2012ix,Anand:2013yka} that the SI and SD interactions are only a portion of a larger set of nuclear responses which must be considered for a proper consideration of direct detection studies.  In addition to responses giving rise to SI (the vector charge nuclear operator) and SD interactions (which is a sum of two responses: the axial and longitudinal spin-dependent responses), there are also nuclear responses sensitive to orbital angular momentum and spin-orbit coupling.  Different WIMP-nucleus scattering models will correspond to different nuclear responses, which often include a sum of responses contributing, and can also lead to interference terms between the responses \cite{Catena:2015uua,Schneck:2015eqa}.  A study of a variety of spin-1/2 dark matter UV complete models whose responses are described by these non-standard responses was given in \cite{Gresham:2014vja}, and a general survey of simplified models of spin-0, spin-1/2, and spin-1 dark matter models was carried out in \cite{Dent:2015zpa}. There exist additional nuclear responses beyond these five, but are typically not considered due to P and CP properties of the nuclear ground state and an assumption of CP conservation of the interaction.

Within this general EFT framework, the WIMP-nucleus interaction is written as a sum over the individual WIMP-nucleon interactions, whose Lagrangian is of the form
\bea
\mathcal{L} = \sum_{\tau=0,1}\sum_{i=1}^{15}c_i^{\tau}\mathcal{O}_it^{\tau}
\eea
where $t^0$ is the identity matrix, thus giving the isoscalar interaction, and $t^1$ is the third Pauli matrix giving the isovector interaction.  It can be seen in general treatments that interference effects can arise not only between operators giving rise to different nuclear responses but also between the same operators characterized by different $c_i^{\tau}$ and $c_i^{\tau'}$ \cite{Catena:2015uua,Schneck:2015eqa}. The coefficients $c_i^{\tau}$ can be related to the familiar neutron and proton couplings by
\bea
c_i^n = \frac{c_i^0-c_i^1}{2} \;\;\;;\;\;\; c_i^p = \frac{c_i^0+c_i^1}{2}
\label{eq:cnp}
\eea

The nucleon-level interactions arise from WIMP-quark interactions (either at the Lagrangian level including mediator particles, which are subsequently integrated out, or by directly writing down bi-linear terms suppressed by some high mass scale) where quarks are then embedded into the nucleons through standard techniques \cite{Agrawal:2010fh,Dienes:2013xya,Hill:2014yxa,Hoferichter:2015ipa}, and all operators are treated non-relativistically. 

We consider the operators in Table~\ref{tabHaxtonOp}, with the exception of $\mathcal{O}_2$ which cannot be generated at leading order from Lorentz invariant relativistic operators~\cite{Anand:2013yka}. There are two additional operators which need to be included if the WIMP under consideration has spin-1~\cite{Dent:2015zpa}, but for this work we are assuming a spin-1/2 dark matter particle. Assuming Galilean invariance (for a treatment which includes operators which are constrained by Lorentz invariance rather than Galilean invariance see Ref.~\cite{Hill:2014yxa}), time-reversal symmetry, and Hermiticity, these operators only depend on four quantities: the exchanged momentum, $\vec{q}$, in the dimension-less, Hermitian form $i\vec{q}/m_N$, the velocity $\vec{v}^{\perp} = \vec{v} + \vec{q}/2\mu_N$, where $\vec{v}$ is the WIMP velocity in the target nucleon rest frame and $\mu_N$ is the WIMP-nucleon reduced mass, the WIMP spin $S_{\chi}$, and the nuclear spin $S_{N}$.

Although we retain the remaining 14 operators, it should be kept in mind that not all of these non-relativistic operators arise at leading order from simple UV models~\cite{Dent:2015zpa}, and therefore may not be relevant when a complete Lagrangian picture of dark matter is formulated.  Additionally the recoil response of the operators can vary by many orders of magnitude on a given target material, and operator responses can vary greatly between various detector materials, which demonstrates the premium placed on target complementarity~\cite{Ruppin:2014bra,Newstead:2013pea}.

\begin{table}[ht]
\caption{List of NR effective operators described in~\cite{Fitzpatrick:2012ix}}
\begin{tabular}{rc}
$\mathcal{O}_1$ & $1_\chi 1_N$ \\
$\mathcal{O}_2$ & $(\vec{v}^{\perp})^2$ \\
$\mathcal{O}_3$ & $i\vec{S}_N\cdot(\frac{\vec{q}}{m_N}\times \vec{v}^{\perp})$ \\
$\mathcal{O}_4$ & $\vec{S}_\chi \cdot \vec{S}_N$ \\
$\mathcal{O}_5$ & $i\vec{S}_\chi\cdot(\frac{\vec{q}}{m_N}\times \vec{v}^{\perp})$ \\
$\mathcal{O}_6$ & $(\frac{\vec{q}}{m_N}\cdot\vec{S}_N)(\frac{\vec{q}}{m_N}\cdot\vec{S}_\chi)$ \\
$\mathcal{O}_7$ & $\vec{S}_N \cdot \vec{v}^{\perp}$ \\
$\mathcal{O}_8$ & $\vec{S}_\chi \cdot \vec{v}^{\perp}$ \\
$\mathcal{O}_9$ & $i\vec{S}_\chi \cdot (\vec{S}_N \times \frac{\vec{q}}{m_N}) $ \\
$\mathcal{O}_{10}$ & $i\frac{\vec{q}}{m_N} \cdot \vec{S}_N$ \\
$\mathcal{O}_{11}$ & $i\frac{\vec{q}}{m_N} \cdot \vec{S}_\chi$ \\
$\mathcal{O}_{12}$ & $\vec{S}_\chi \cdot(\vec{S}_N \times \vec{v}^{\perp})$ \\
$\mathcal{O}_{13}$ & $i(\vec{S}_\chi \cdot \vec{v}^{\perp})(\frac{\vec{q}}{m_N} \cdot \vec{S}_N)$ \\
$\mathcal{O}_{14}$ & $i(\vec{S}_N \cdot \vec{v}^{\perp})(\frac{\vec{q}}{m_N} \cdot \vec{S}_\chi)$ \\
$\mathcal{O}_{15}$ & $-(\vec{S}_\chi \cdot \frac{\vec{q}}{m_N})\left( (\vec{S}_N\times \vec{v}^{\perp})\cdot\frac{\vec{q}}{m_N}\right)$ \\
\end{tabular}
\label{tabHaxtonOp}
\end{table}

\subsection{Neutrino-nucleus scattering} 
The theoretical prediction for neutrino interaction with the nucleus is much more simple than the WIMP interaction described above. Neutrino-nucleus coherent scattering is a straightforward prediction of the Standard Model, and has been theoretically studied for many years~\cite{Freedman:1973yd,Cabrera:1984rr}. The coherent cross section is
\begin{equation}
  \frac{\textrm{d} \sigma}{\textrm{d}E_r}(E_r,E_\nu) = \frac{G_F^2}{4 \pi} Q_W m_N \left(1-\frac{m_N E_r}{2 E_\nu^2} \right) F^2(E_r) \,,
  \label{eq:cns}
\end{equation}
where $Q_W = \mathcal{N} - (1-4\sin^2\theta_W) \mathcal{Z}$ is the weak nuclear hypercharge of a nucleus with $\mathcal{N}$ neutrons and $\mathcal{Z}$ protons, $G_F$ is the Fermi coupling constant, $\theta_W$ is the weak mixing angle and $m_N$ is the target nucleus mass. There are few percent corrections to Equation~\ref{eq:cns} for non isoscalar nuclei ($\mathcal{N} \ne \mathcal{Z}$) arising from axial couplings~\cite{Dutta:2015vwa}. In addition there is an angular dependence in the recoil direction of the nucleus which we do not consider in this paper.

\section{Matching the WIMP and neutrino recoil spectra} 
\label{sec:spectra}
In this section we analyze the nuclear recoil spectrum that is induced by WIMPs within non-relativistic EFT, and by neutrinos through coherent scattering. We identify operators which admit recoil spectra that are degenerate with the neutrino backgrounds, and for these operators we find the corresponding WIMP masses that provide the ``best-fit" which is defined below. We classify operators into groups based on their induced recoil spectra, and compare to the neutrino-induced spectra. 

\subsection{Best fit rates} 
We begin by matching the nuclear recoil spectra from the various WIMP-nucleon operators described above to the predicted Solar and atmospheric neutrino-induced recoil energy spectrum. For the Solar neutrinos, we consider the $^8$B component. The predicted recoil energy spectra in dark matter detectors due to these neutrinos are taken from Refs.~\cite{Strigari:2009bq,Billard:2013qya}. To find the ``best-fit" WIMP masses for a given operator we maximize the Poisson likelihood,
\be
\mathcal{L}_{Poisson} = \prod_{i=1}^b \frac{\nu_i^{n_i} e^{\nu_i}}{n_i!}
\label{eq:Lpoisson} 
\ee
where $b$ is the number of nuclear recoil energy bins, $n_i$ is the expected number of WIMP events and $\nu_i$ is the expected number of neutrino events in the bin. We consider several detector targets, which are indicated in Table~\ref{tab:targets} along with the corresponding nuclear energy recoil range. For our likelihood analysis we choose an exposure such that we obtain 200 neutrino events for each target~\cite{Billard:2013qya}, binned into 16 energy bins.

\begin{table}[ht]
\caption{List of detector targets considered in this work}
\begin{tabular}{l|r|r}
&   low region (keV) &  high region (keV) \\
\hline
 xenon     &    0.003 - 3 & 4.0 - 100 \\
 germanium &    0.0053 - 7 & 7.9 - 120 \\
 silicon   &    0.014 - 18 &  20 - 300 \\
 flourine  &    0.033 - 25 &  28 - 500 \\
\end{tabular}
\label{tab:targets}
\end{table}

\par
\begin{figure}[ht]
\begin{tabular}{cc}
\includegraphics[height=5cm]{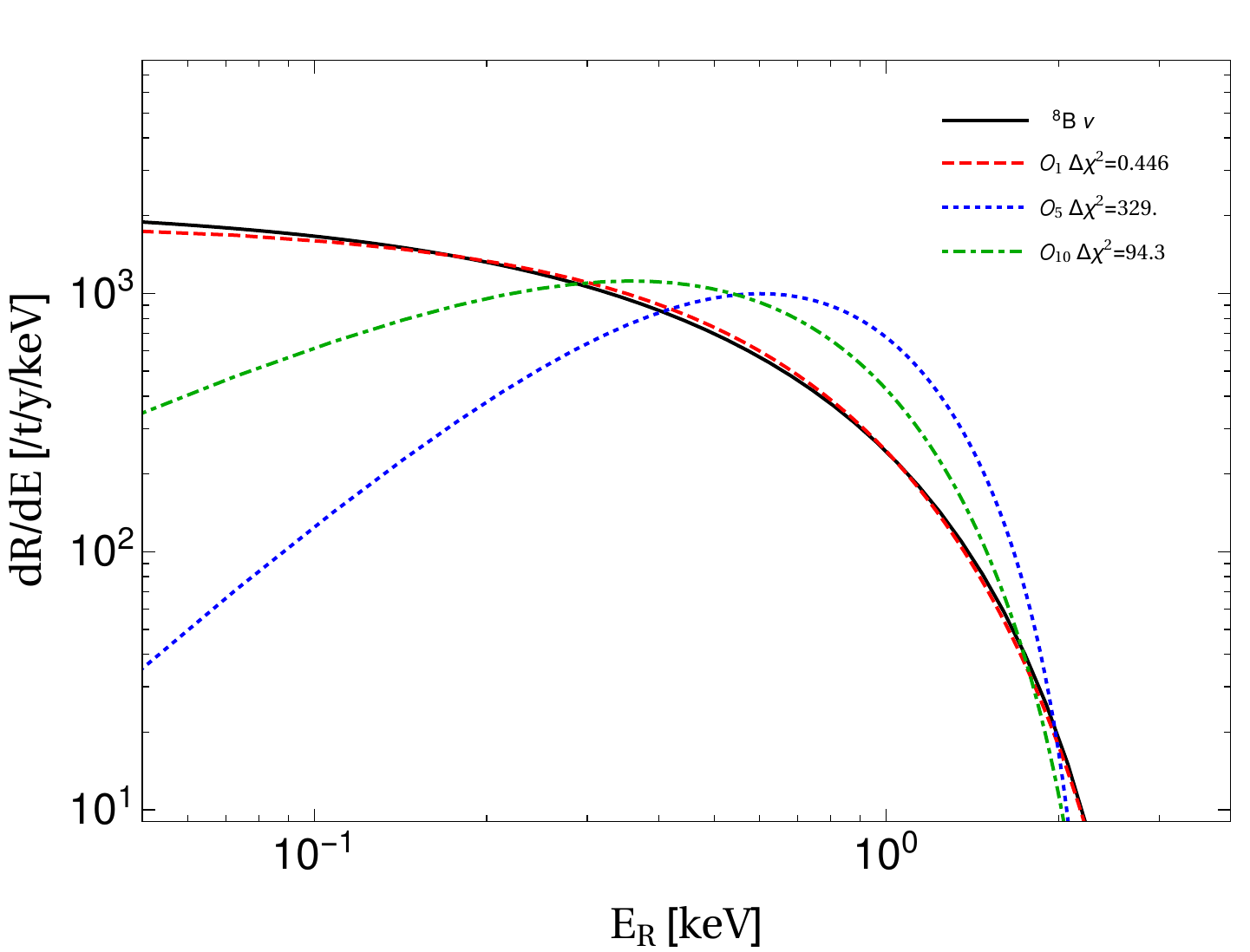} &
\includegraphics[height=5cm]{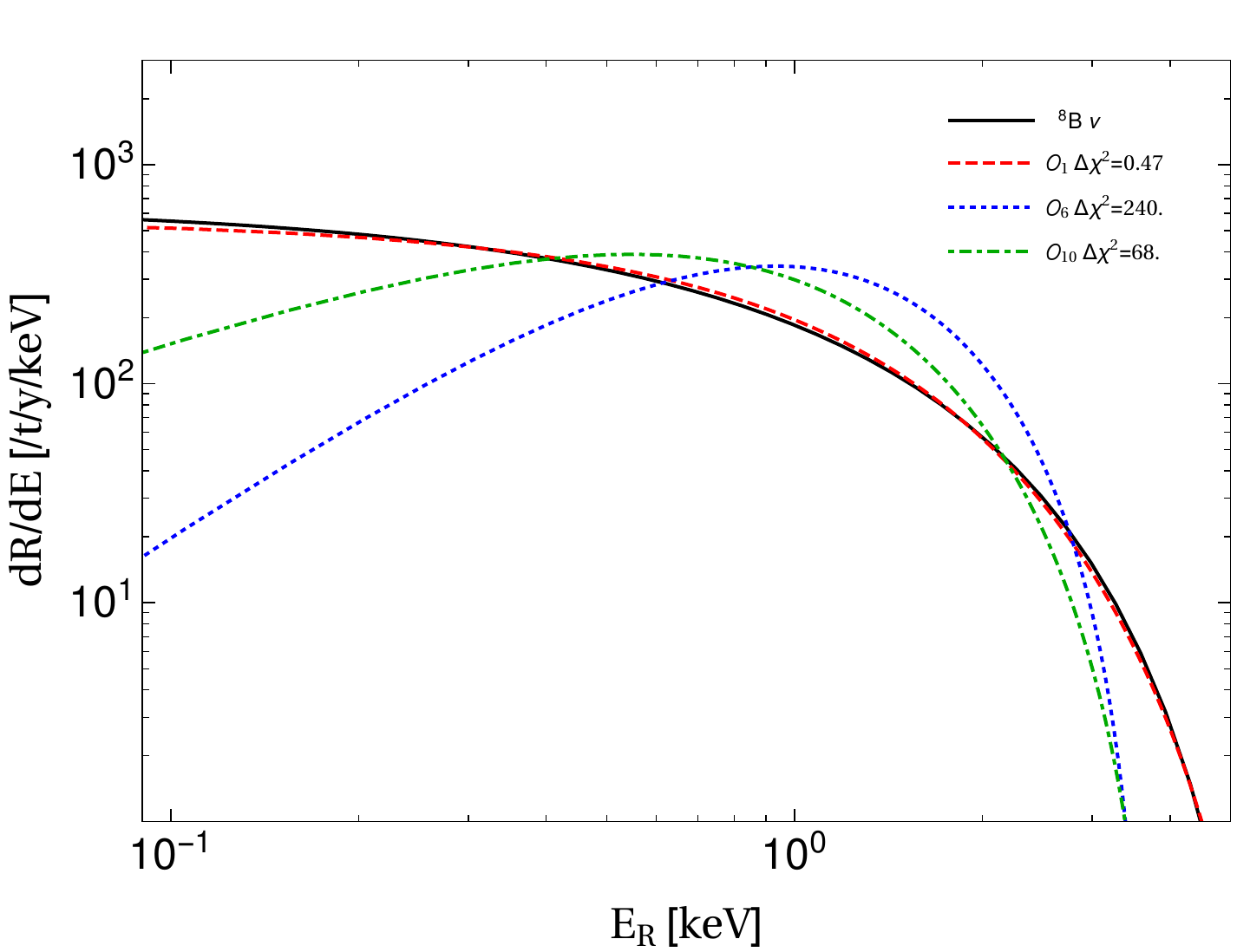} \\
\end{tabular}
\caption{Sample maximum likelihood fits to the $^8$B Solar neutrino-induced nuclear recoil event rate spectrum in Xenon (left) and Germanium (right). Three different operators are shown, one operator from each of the groupings in Table~\ref{tabMassExp}.}
\label{fig:8BbestFitRates}
\end{figure}

\begin{figure}[ht]
\begin{tabular}{cc}
\includegraphics[height=5cm]{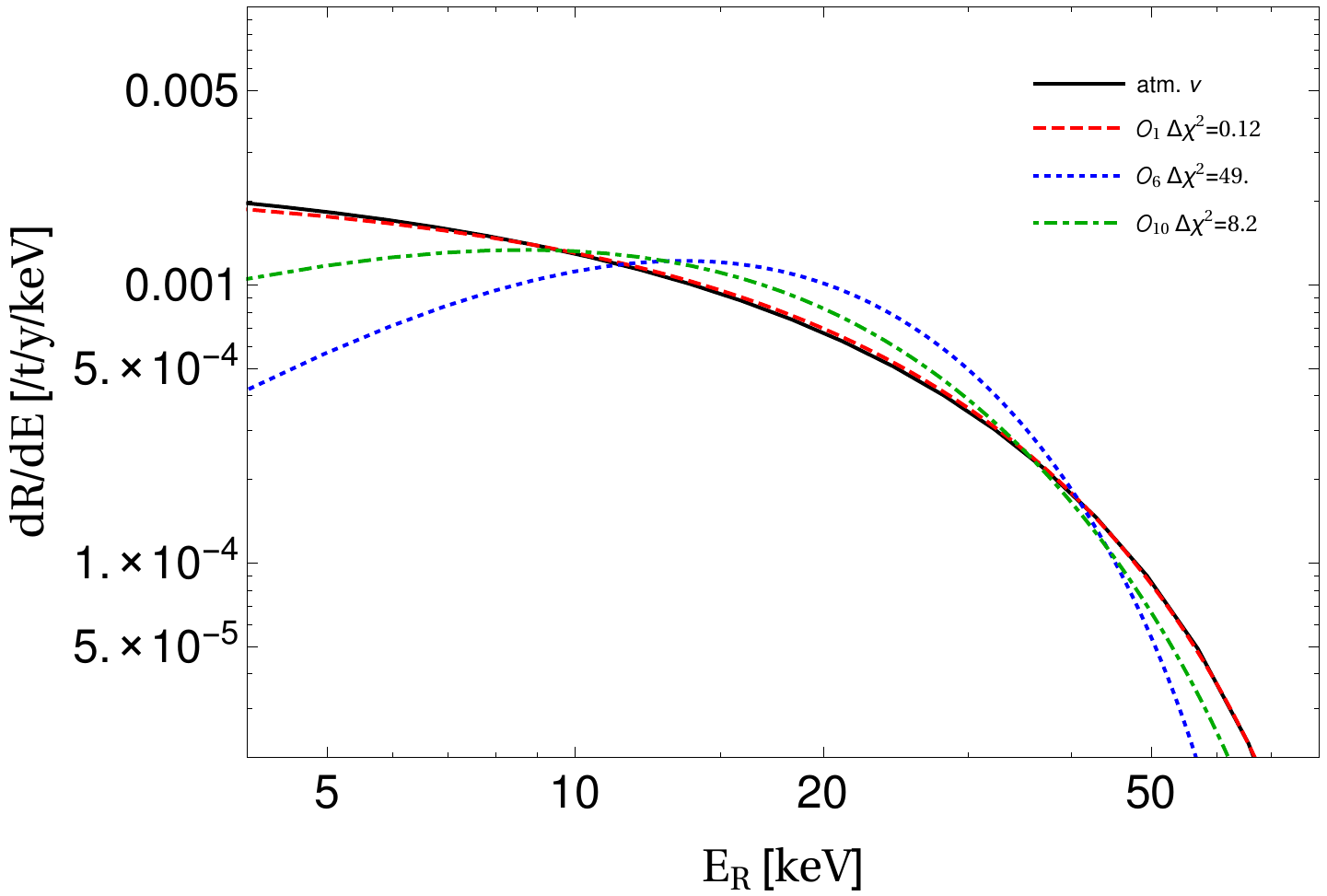} &
\includegraphics[height=5cm]{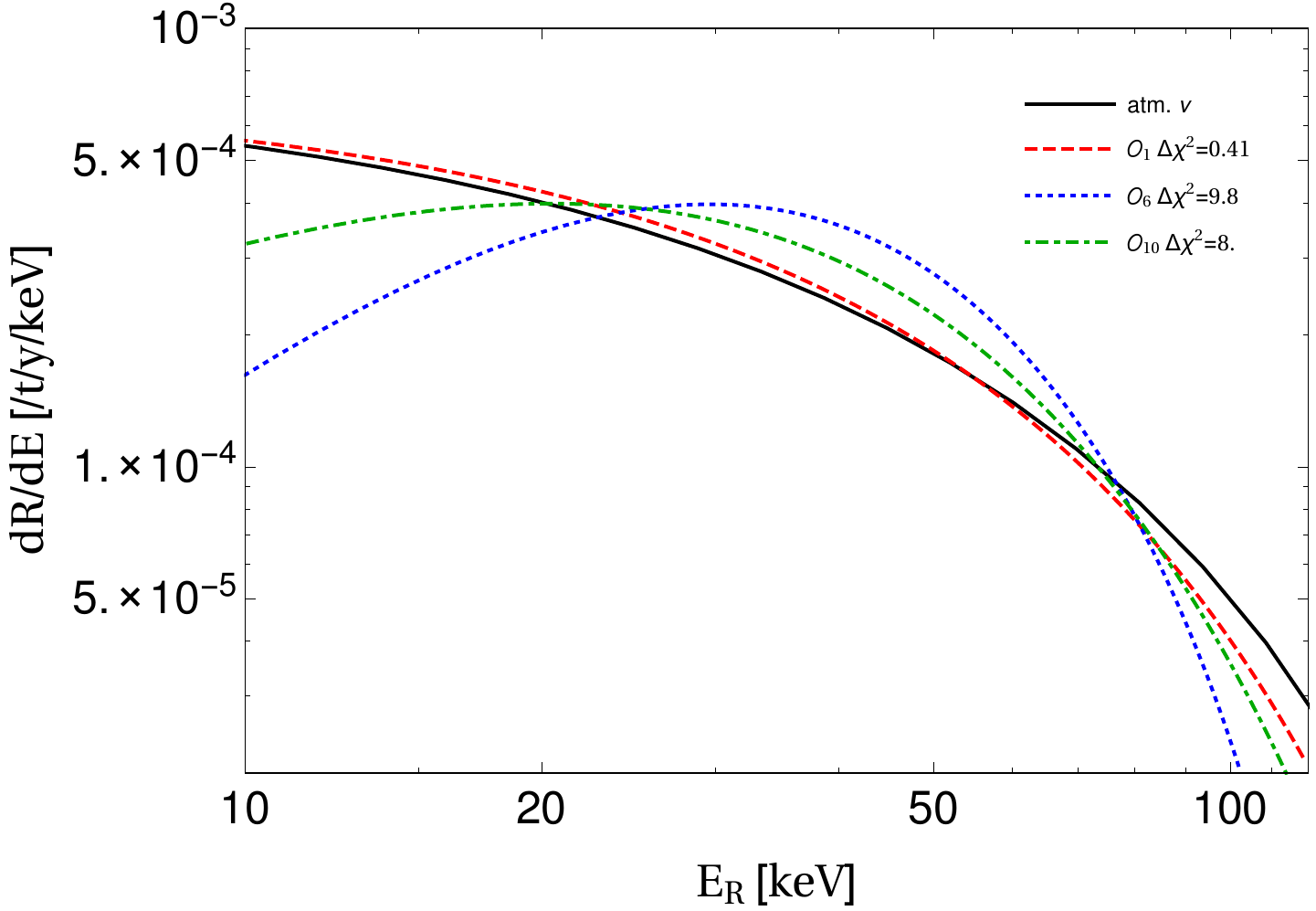} \\
\end{tabular}
\caption{Sample maximum likelihood fits to the atmospheric neutrino-induced nuclear recoil event rate spectrum in xenon (left) and germanium (right). The same operators are used here as in Figure~\ref{fig:8BbestFitRates}.}
\label{fig:ATMbestFitRates}
\end{figure}

 Figure~\ref{fig:8BbestFitRates} shows a sample of the best fitting WIMP-induced recoil energy spectra when comparing to the predicted $^8$B spectrum, and Figure~\ref{fig:ATMbestFitRates} shows a sample when comparing to the predicted atmospheric-induced recoil spectrum. In both figures we have used one operator representative from each group where the groups are  defined in Table~\ref{tabMassExp}. As is shown for several operators, in particular $\mathcal{O}_1$, we find a good match to both the $^8$B Solar and atmospheric spectra. This is quantified by the $\Delta \chi^2$ indicated in Figure~\ref{fig:8BbestFitRates} and~\ref{fig:ATMbestFitRates}, which is calculated as the negative log likelihood in Equation~\ref{eq:Lpoisson}. Note that $\mathcal{O}_1$ ($\mathcal{O}_4$) from group 1 is the SI (SD) response used in the standard analyses, and our result agrees with previous results~\cite{Billard:2013qya,Ruppin:2014bra}. 

\par On the other hand, the nuclear recoil spectra from many WIMP-nucleon operators are clearly distinct from the $^8$B and atmospheric-induced neutrino spectra, even when taken at the best-fit WIMP masses. For example, as is indicated in Figures~\ref{fig:8BbestFitRates} and~\ref{fig:ATMbestFitRates}, the $\mathcal{O}_6$ (belongs to group 3) and $\mathcal{O}_{10}$ (belongs to group 2) best fit WIMP mass gives a poor $\Delta \chi^2$ relative to the neutrino backgrounds. This indicates that for essentially all WIMP masses and cross sections, $\mathcal{O}_6$ and $\mathcal{O}_{10}$ can be distinguished from the neutrino backgrounds. We return to this point below when we discuss the evolution of the discovery limit. 

\par
\begin{figure}[ht]
\centering
\begin{tabular}{lll}
\hspace{-1.5cm}
\includegraphics[height=4cm]{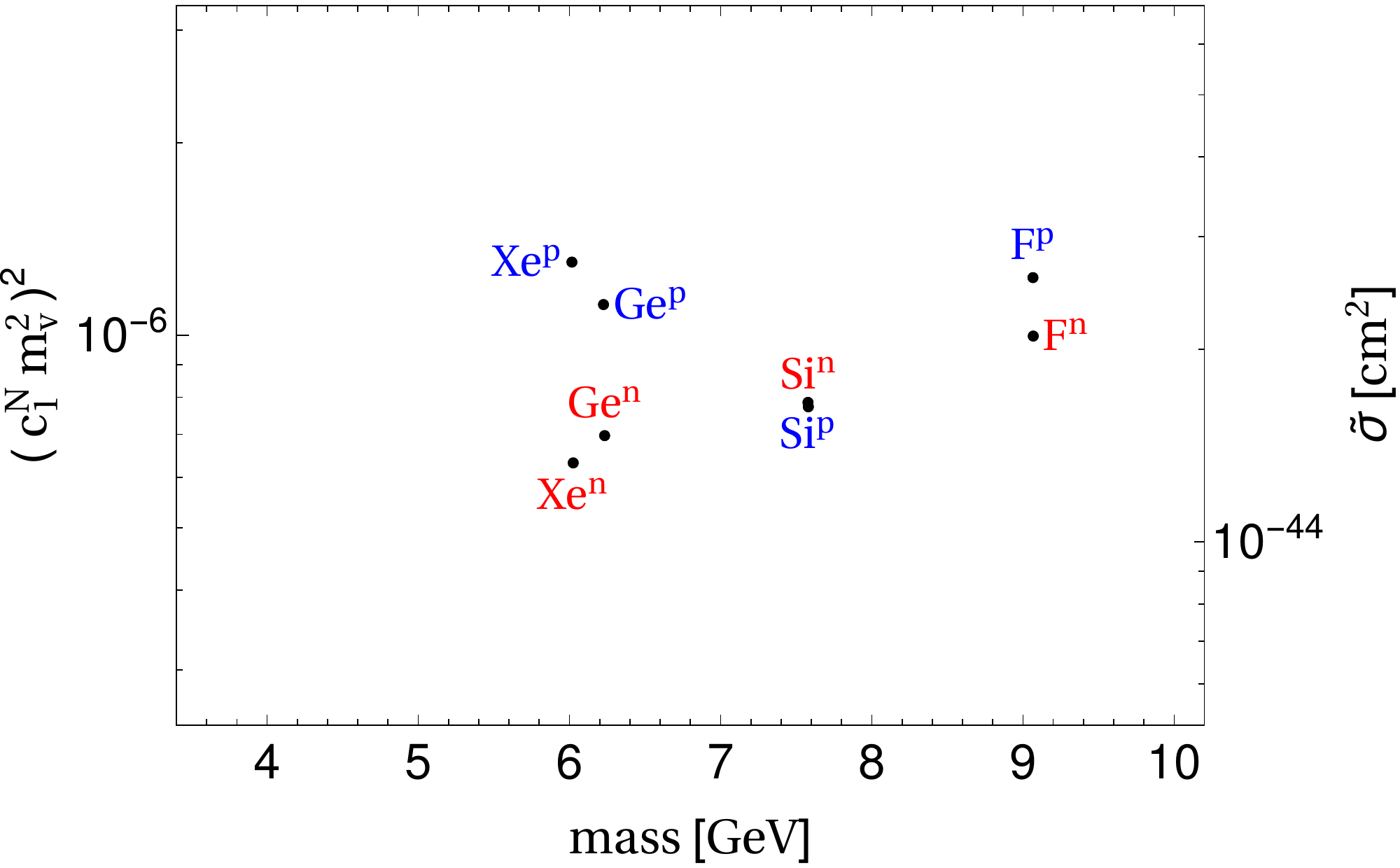} &
\includegraphics[height=4cm]{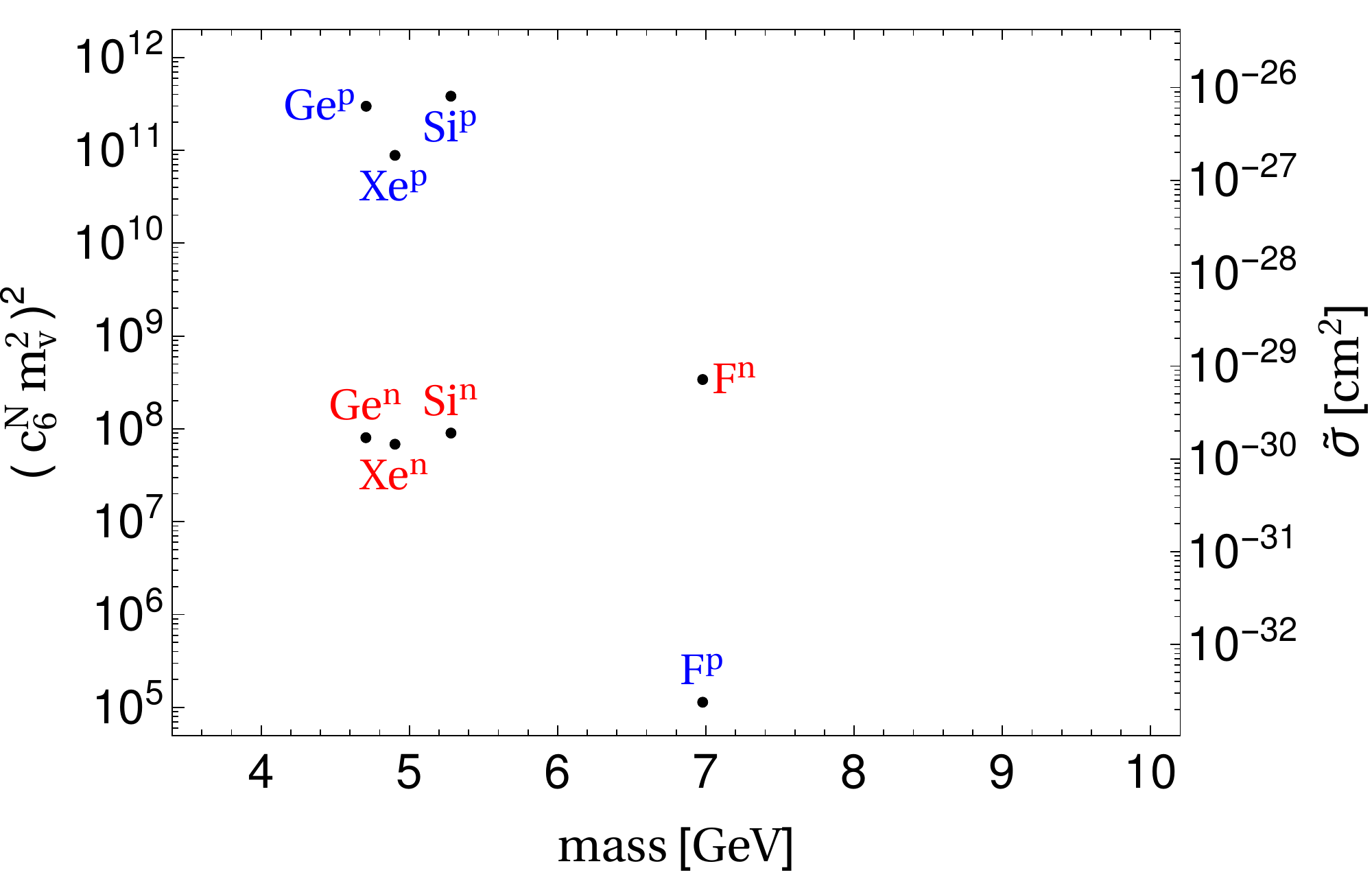} &
\includegraphics[height=4cm]{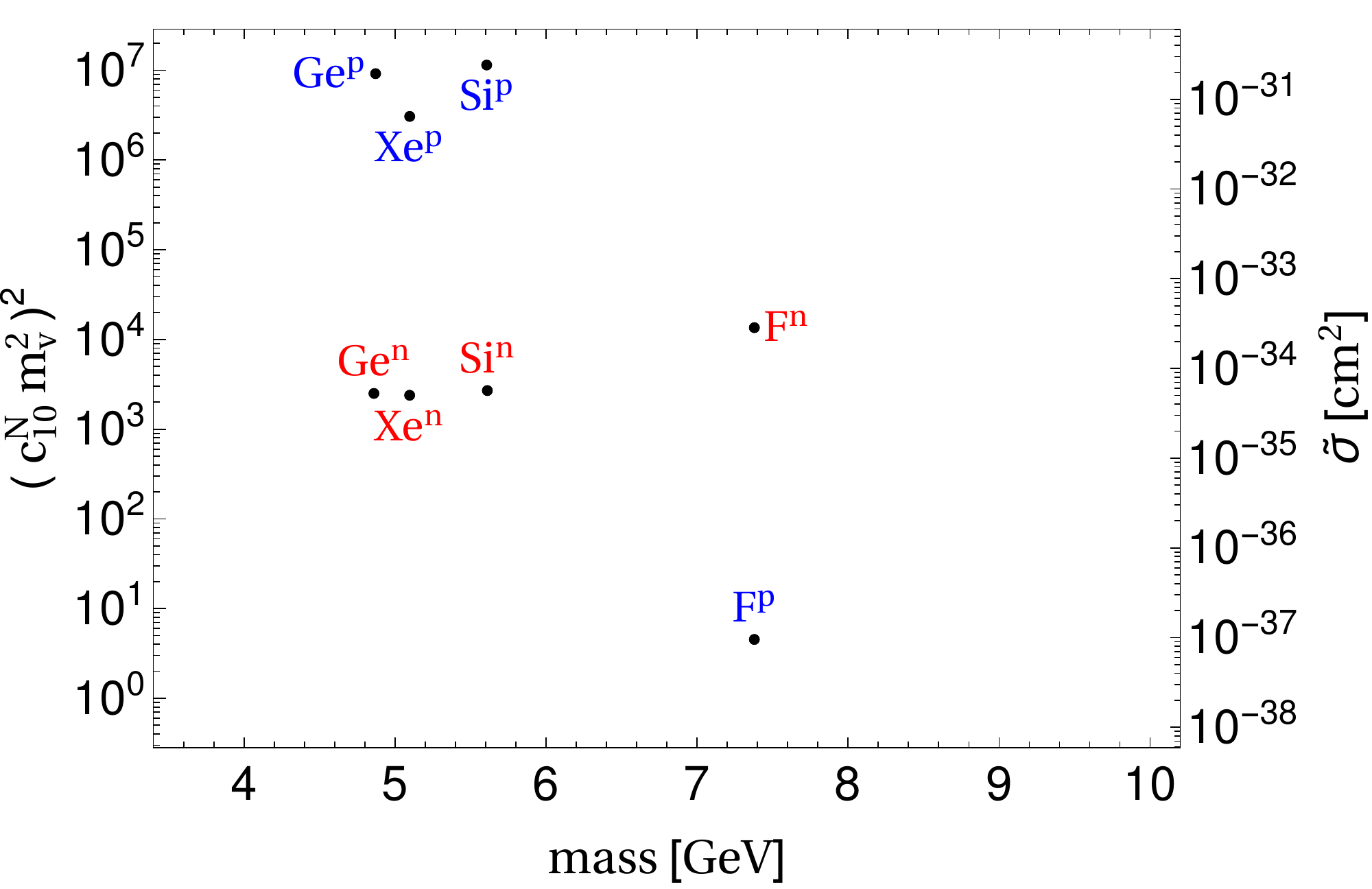} \\
\end{tabular}
\caption{Best fit WIMP mass to the $^8B$ Solar neutrino induced nuclear recoil spectrum for Xe, Ge, Si, and F targets. For each target and each operator, we show the best fitting WIMP mass for neutron and proton couplings, defined as in Equation~\ref{eq:cnp}. The quantity on the vertical axis of the left-hand side in each figure is dimensionless, since the $c_\imath$'s as defined in Ref.~\cite{Anand:2013yka} have dimensions of inverse mass-squared. An exposure is assumed to produce 200 neutrino events for each target.  
}
\label{fig:8BbestFitMasses}
\end{figure}

 The WIMP masses that provide the best fit to the $^8$B recoil spectrum for the operators $\mathcal{O}_1$, $\mathcal{O}_6$, $\mathcal{O}_{10}$ are shown in Figure~\ref{fig:8BbestFitMasses}. As discussed above we assume an exposure to produce 200 neutrino events for each target. Each point in Figure~\ref{fig:8BbestFitMasses} represents either the proton or neutron coupling as defined in Equation~\ref{eq:cnp}. In Figure~\ref{fig:8BbestFitMasses} we have scaled the coupling by a factor $m_v = 246$ GeV, so that the resulting quantity $c_\imath m_v^2$ is dimensionless (the $c_\imath$'s as defined in Ref.~\cite{Anand:2013yka} have dimensions of inverse mass-squared). Also shown are the corresponding WIMP-nucleon cross sections calculated as $\sigma_i = \frac{c_i^2 \mu^2}{m_v^4}$. For Si, Ge, and Xe, the excess spin in the nucleus is carried by the neutron, so that for a fixed number of neutrino events the neutron coupling corresponds to a lower cross section. For flourine the excess spin is carried by the proton, so in this case for a fixed number of neutrino events the proton coupling corresponds to a lower cross section. Note here that the $\mathcal{O}_1$ operator corresponds to the standard SI interaction and is in agreement with previous studies~\cite{Billard:2013qya,Ruppin:2014bra}.

\subsection{Grouping of operators} 
\par Although we consider 15 operators, each of which coupling to protons and neutrons, the nuclear recoil energy spectra that is induced by many of these operators are similar. This is evident from their best fitting WIMP masses shown Figure~\ref{fig:8BbestFitMasses} and in Figure~\ref{fig:8BbestFitMasses_all} in Appendix A, which shows the best fitting masses for the operators that are not shown in Figure~\ref{fig:8BbestFitMasses}. These figures motivate a grouping of operators based on their best fit WIMP mass, which are shown in Table~\ref{tabMassExp}. Operators $\mathcal{O}_{1,\,4,\,7,\,8}$, $\mathcal{O}_{5,\,9,\,10,\,11,\,12,\,14}$ and $\mathcal{O}_{3,\,6,\,13,\,15}$ form group 1, 2 and 3 respectively (this is a similar grouping to that found in~\cite{Catena:2015vpa}, although $\mathcal{O}_{13}$ is in our third group along with $\mathcal{O}_{15}$, rather than in a fourth). For the entries in this table we have assumed a Xe target, though we have checked that these results do not strongly depend on the nature of the target. We again emphasize that for many operators, the $\chi^2$ is large when comparing the neutrino-induced spectra to the WIMP spectra, so that even these ``worst case" scenarios should be easily distinguishable from the neutrino backgrounds, provided an experiment can obtain a robust measurement of the recoil energy spectrum. 

\begin{table}
\caption{List of NR effective operators categorized by the best fit mass to $^8$B Solar neutrinos in Xenon (the other targets follow suit). The third column gives the 
exposure to reach saturation due to the neutrino background, as defined in Section~\ref{sec:discovery}. 
}
\begin{tabular}{lccc}
\hline
& Operator         &  Mass (GeV) & Exp. (t.y)\\
\hline
&$\mathcal{O}_1$ & 6     &   2.9 \\
&$\mathcal{O}_4$ & 6     &   3.5 \\
\rotatebox{90}{\rlap{Group 1}}
&$\mathcal{O}_7$ & 6.2   &   4.3 \\
&$\mathcal{O}_8$ & 6.3   &   3.6 \\
\hline
&$q^2$ and $q^2 v_T^2$  &     \\
\hline
&$\mathcal{O}_5$    & 4.8    &  0.43 \\
&$\mathcal{O}_9$    & 4.6    &  0.34 \\
&$\mathcal{O}_{10}$ & 4.6    &  0.36 \\
\rotatebox{90}{\rlap{Group 2}}
&$\mathcal{O}_{11}$ & 4.6    &  0.40 \\
&$\mathcal{O}_{12}$ & 4.6    &  0.44 \\
&$\mathcal{O}_{14}$ & 4.8    &  0.43 \\
\hline
&$q^2 v_T^2$, $q^4$ and $q^4 v_T^2$ &   \\
\hline
&$\mathcal{O}_3$    & 4.2    &  0.27 \\
&$\mathcal{O}_6$    & 4.2    &  0.29 \\
\rotatebox{90}{\rlap{Group 3}}
&$\mathcal{O}_{13}$ & 4.2    &  0.27 \\
&$\mathcal{O}_{15}$ & 4.1    &  0.21 \\
\end{tabular}
\label{tabMassExp}
\end{table}

\clearpage
\section{Discovery Bounds}
\label{sec:discovery}
\par With the nuclear recoil spectrum in non-relativistic EFT understood, we now move on to determine the bounds on the discovery of WIMPs in the presence of the neutrino background. We determine the exposure at which each operator is maximally affected by the neutrino background. As above we distinguish between those operators that are most  and least affected by the neutrino background.

\subsection{Formalism} 
\par The statistical formalism that we employ follows that of Ref.~\cite{Billard:2013qya}. Here we review the relevant aspects for our analysis. The discovery potential of an experiment is defined as the smallest WIMP-nucleon cross section which produces a 3$\sigma$ fluctuation above the background 90\% of the time. To calculate this limit we use the following test statistic for the null hypothesis and try to reject it,
\be
q_0 =
\begin{cases}
   -2 \mathrm{log}  \frac{\mathcal{L}(\sigma=0,\hat\theta)}{\mathcal{L}(\hat\sigma,\hat{\hat\theta})}  & \sigma  \geq \hat\sigma \\
   0	& \sigma < \hat\sigma \\
\end{cases}
\ee
where $\sigma$ is the WIMP-nucleon cross section, $\theta$ represents the nuisance parameters (neutrino fluxes), and the hatted parameters are maximized. By Wilks' theorem, under background only experiments, $q_0$ is chi-square distributed and the equivalent gaussian significance is simply $\sqrt{q_0}$~\cite{Cowan:2010js}.  To include the uncertainty of the neutrino flux normalization the likelihood function is modified to include a gaussian term~\cite{Billard:2013qya}:
\be
\mathcal{L} = \mathcal{L}_{Poisson} \prod_{j} e^{-\frac{1}{2}(1-N_j)^2\left(\frac{\phi_j}{\sigma_j}\right)^2}
\ee
where $N_j$ is the flux normalization and $\phi_j$ and $\sigma_j$ are the fluxes and their uncertainties given in Table~\ref{tab:nuFlux}. The Poisson likelihood $\mathcal{L}_{Poisson}$ is defined as in Equation~\ref{eq:Lpoisson}. 

\begin{table}
\caption{Neutrino flux components and their respective uncertainties in the flux normalizations. For the Solar components we utilize the high metallicity Solar model as outlined in Ref.~\cite{Robertson:2012ib}.}
\begin{tabular}{c|c}
 component  & $\nu$ flux (cm$^{-2}$s$^{-1}$) \\
\hline
PP          &  $5.98(1\pm0.006) \times 10^{10}$ \\
$^7$Be      & $5.00(1\pm 0.07) \times 10^9$ \\ 
$^8$B       & $5.58(1 \pm 0.14) \times 10^6$ \\
pep         & $1.44(1 \pm 0.012) \times 10^8$ \\  
DSNB        & $85.5 \pm 42.7$ \\  
Atmospheric & $10.5 \pm 2.1$ \\  
\hline
\end{tabular}
\label{tab:nuFlux}
\end{table}

\par We calculate the evolution of the discovery potential for all operators using a Xe based experiment, in the low and high recoil energy regions as defined above. The WIMP mass considered for each operator was taken from Table~\ref{tabHaxtonOp} as this is the worst case scenario where the WIMP spectrum most closely resembles the neutrino background. Note that while in the low region the best fit WIMP mass is very similar for the neutron and proton scattering rates, this is not the case in the high region. Thus in the low region the discovery potential curves remain parallel, but this is not necessarily the case for the high region. The discovery evolution for three of the operators from three groups is shown in Figure~\ref{figDiscEvo} and the remaining operators can be found in the Appendix. For operators which are sufficiently neutrino like (group 1), the evolution exhibits saturation when the systematic uncertainty in the neutrino flux becomes relevant. Note that this saturation is achieved at a smaller cross section than in previous studies~\cite{Ruppin:2014bra}, because the analysis in this paper separates proton and neutron couplings, thereby reducing the coherence factor and providing a less stringent limit. This saturation is then broken when the exposure becomes large enough that small differences in the WIMP and neutrino-induced recoil spectra become distinguishable~\cite{Ruppin:2014bra}. \\

\par For the other operators with recoil spectra that are sufficiently different than the neutrino-induced recoil spectra (group 2 and 3), no significant saturation is observed. For these cases a weak inflection point defines the exposure at which the saturation is a maximum. The corresponding saturations are listed for each operator in Table~\ref{tabMassExp}. From this table we see that operators in the same category reach the inflection point at very similar exposures. The operators that reach the inflection point at the lowest exposures are those that are most easily distinguishable from the neutrino backgrounds. These operators then return quickly to a $1/\sqrt{MT}$ evolution as the exposure is increased. \\

\begin{figure}[ht]
\begin{tabular}{ccc}
\includegraphics[height=4cm]{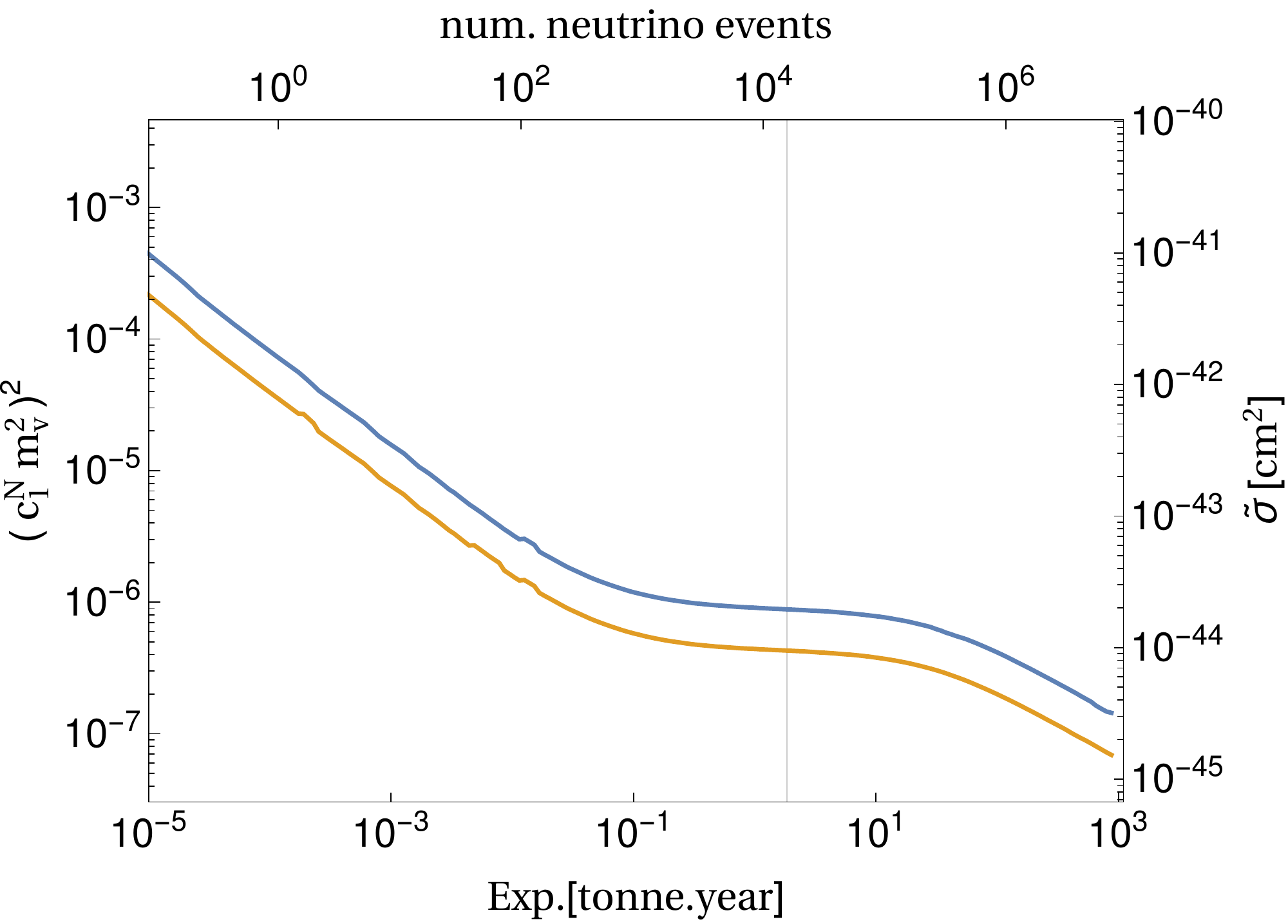}  &
\includegraphics[height=4cm]{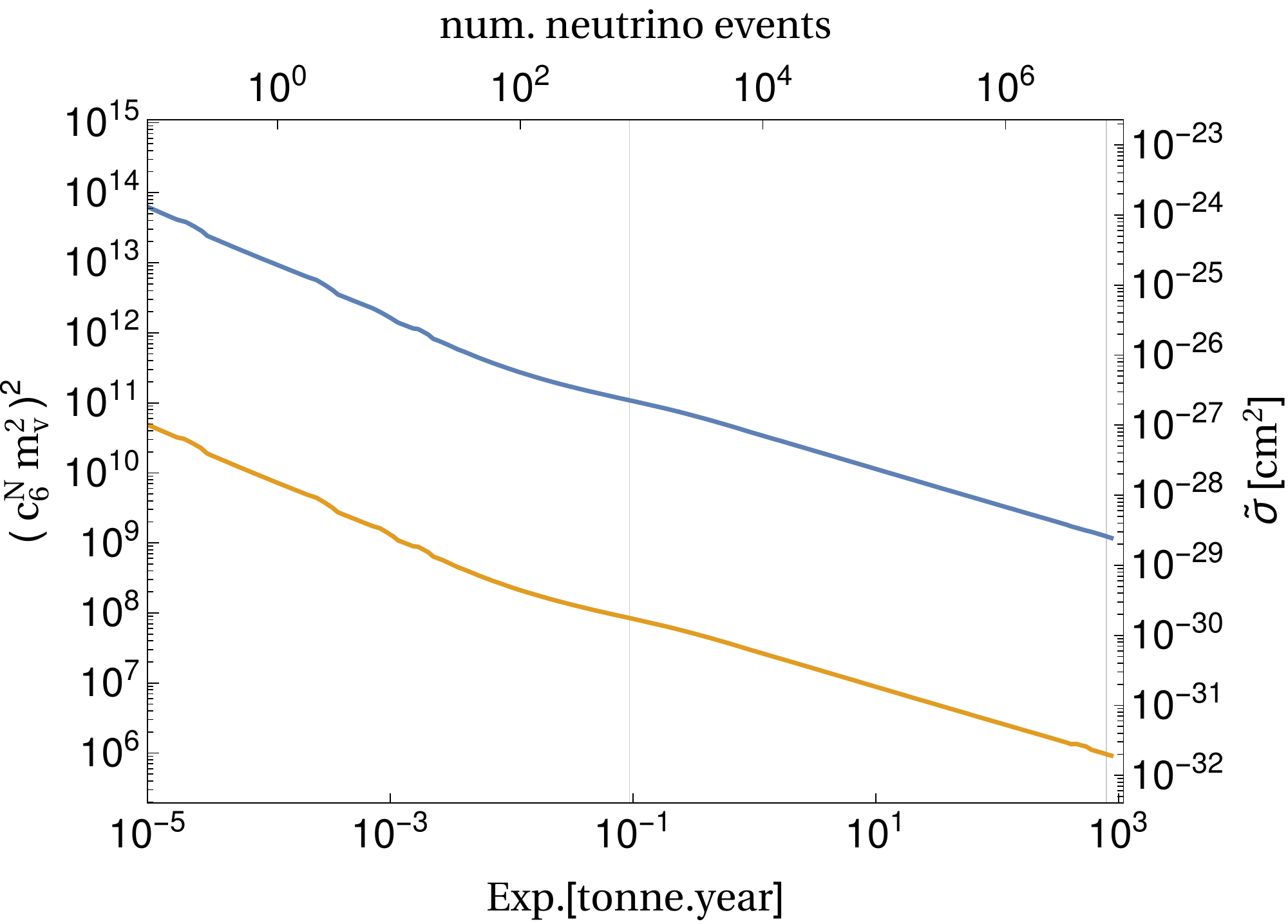}  &
\includegraphics[height=4cm]{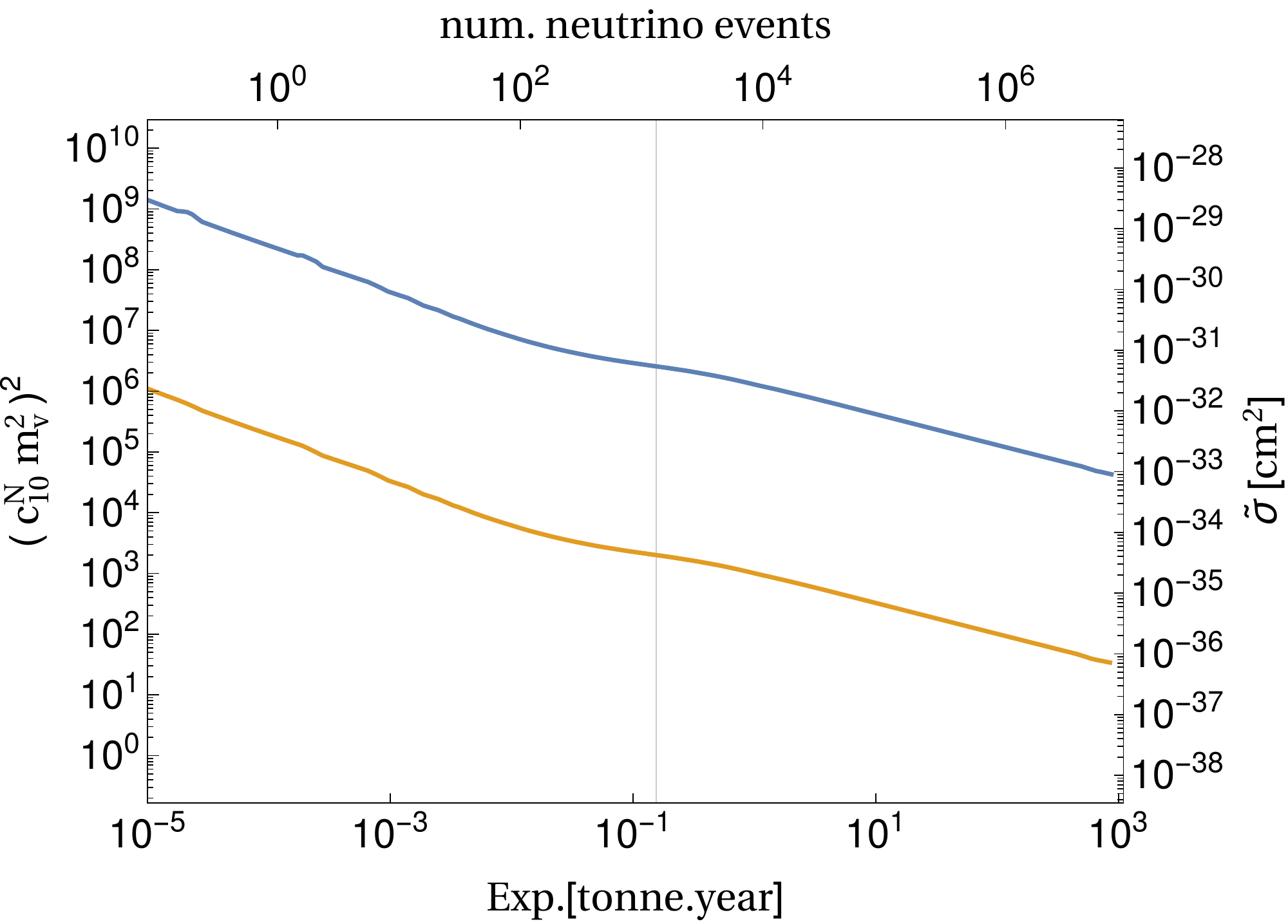} \\
\includegraphics[height=4cm]{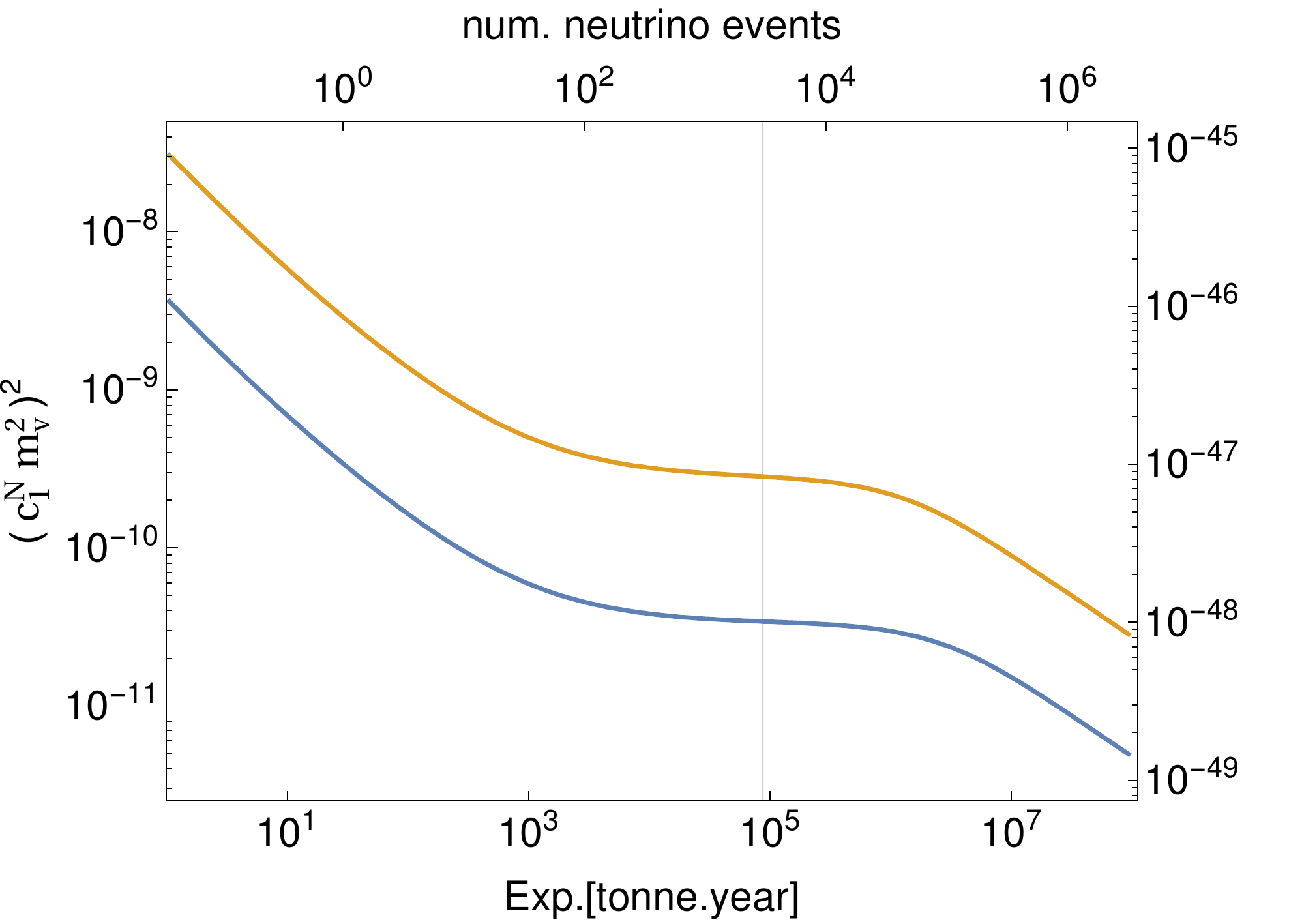} &
\includegraphics[height=4cm]{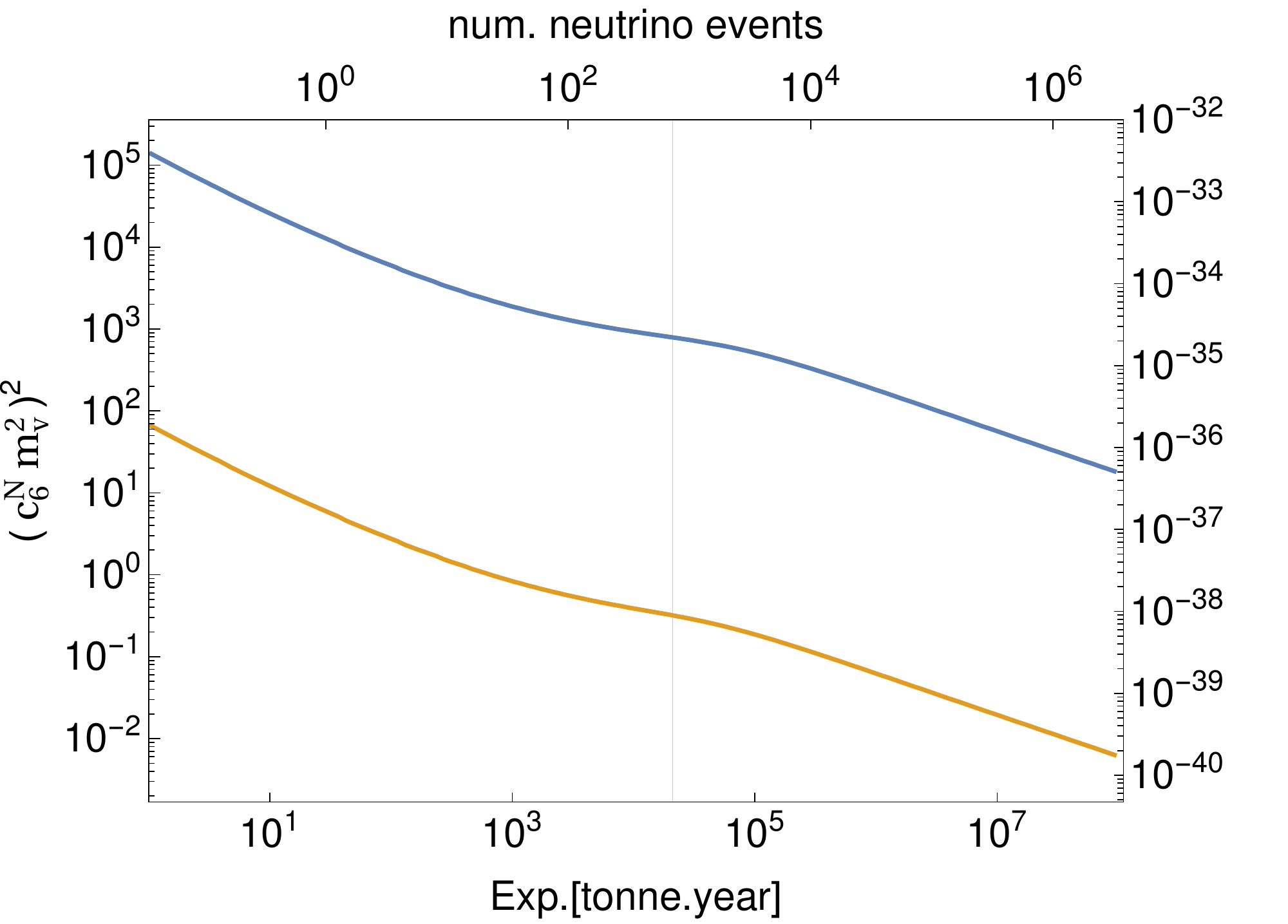} &
\includegraphics[height=4cm]{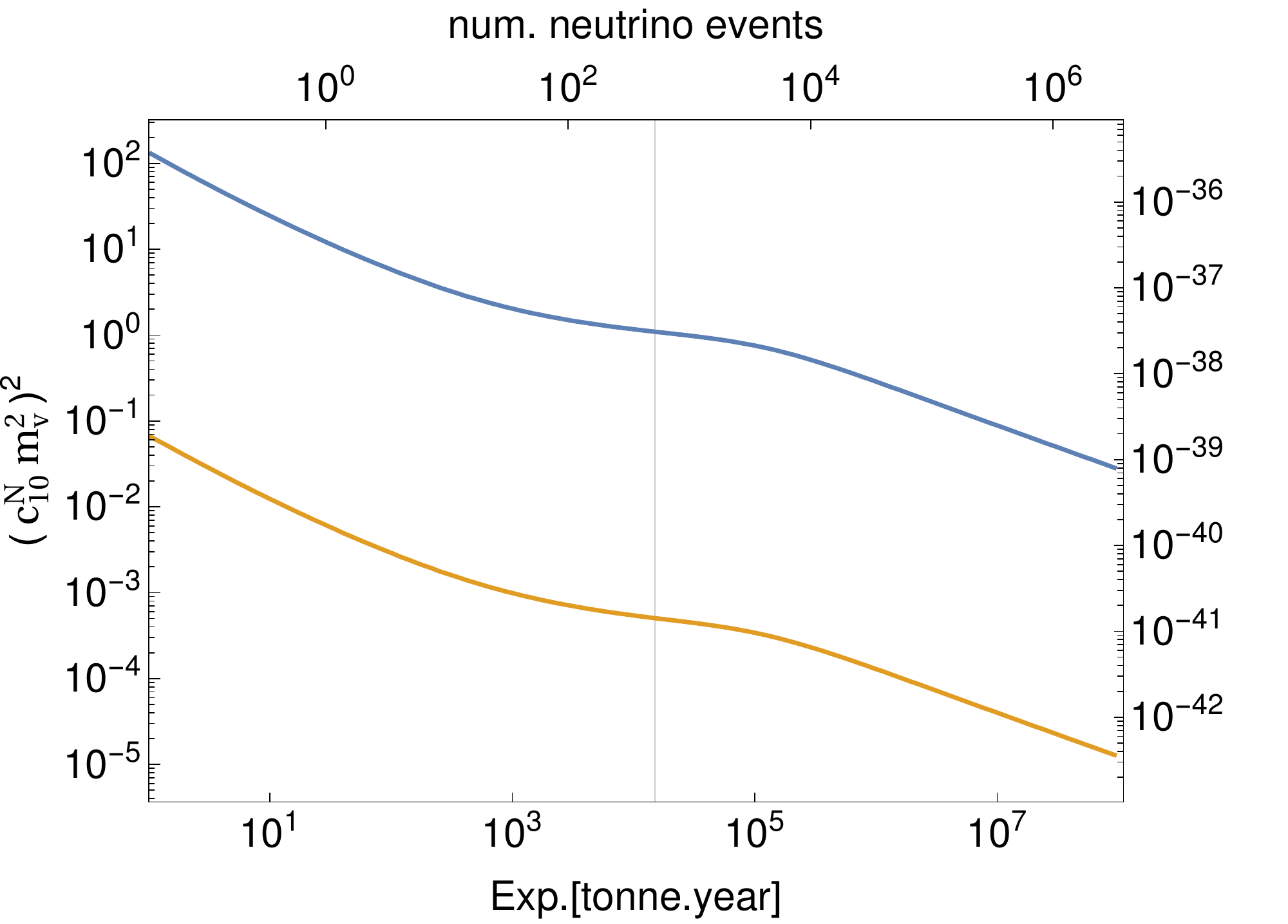} \\
\end{tabular}
\caption{Discovery evolution of $\mathcal{O}_1$ (left), $\mathcal{O}_6$ (middle), $\mathcal{O}_{10}$ (right) operators, for the low region (top) and high region (bottom). The blue and yellow curves show the limits for proton and neutron scattering respectively}
\label{figDiscEvo}
\end{figure}

\clearpage
\subsection{Discovery limits and exclusion regions}
\par We now move on to find the 3$\sigma$ discovery limit across the entire WIMP mass range. We calculate the discovery limits for all operators using a single exposure which saturates $\mathcal{O}_1$ $\sim10^4$( $\sim10^3$) neutrino events in the low (high) region. The motivation for this choice is primarily a simplification of the analysis, noting that the discovery evolution of group 2 and 3 operators do not experience saturation as strongly as group 1 operators.  The exposures for the different targets are given in Table~\ref{tabDetectors}. 

\begin{table}[ht]
\caption{List of exposures used to calculate the neutrino floor}
\begin{tabular}{l|r|r}
Target   &   exposure$^{low}$ (t.y) &  exposure$^{high}$ (kt.y) \\
\hline
 xenon     &    1.76  & 58 \\
 germanium &    3.26  & 87  \\
 silicon   &    10.4  & 206 \\
 flourine  &    16.3  & 278 \\
\end{tabular}
\label{tabDetectors}
\end{table}

In addition to the discovery limits we also determine the $90\%$ exclusion regions from the most recent LUX results~\cite{Akerib:2015rjg}. To calculate exclusion limits we use the profile likelihood method with test statistic,
$$
q_\sigma =
\begin{cases}
   -2 \mathrm{log} \frac{ \mathcal{L}(\sigma,\hat\theta)}{\mathcal{L}(\hat\sigma,\hat{\hat{\theta}})}  & \sigma  \geq \hat\sigma \\
   0	& \sigma < \hat\sigma \\
\end{cases}
$$
where we now use a likelihood which includes gaussian terms for the astrophysical errors: $\rho_\chi = 0.3\pm0.1$GeV/cm$^3$, $v_0=220\pm20$km/s and $v_{esc}=544\pm40$km/s. Under repeated background-only experiments $q_\sigma$ is half-chi-square distributed and the significance is $\sqrt{q_\sigma}$. 

\par
\begin{figure}[ht]
\includegraphics[height=6cm]{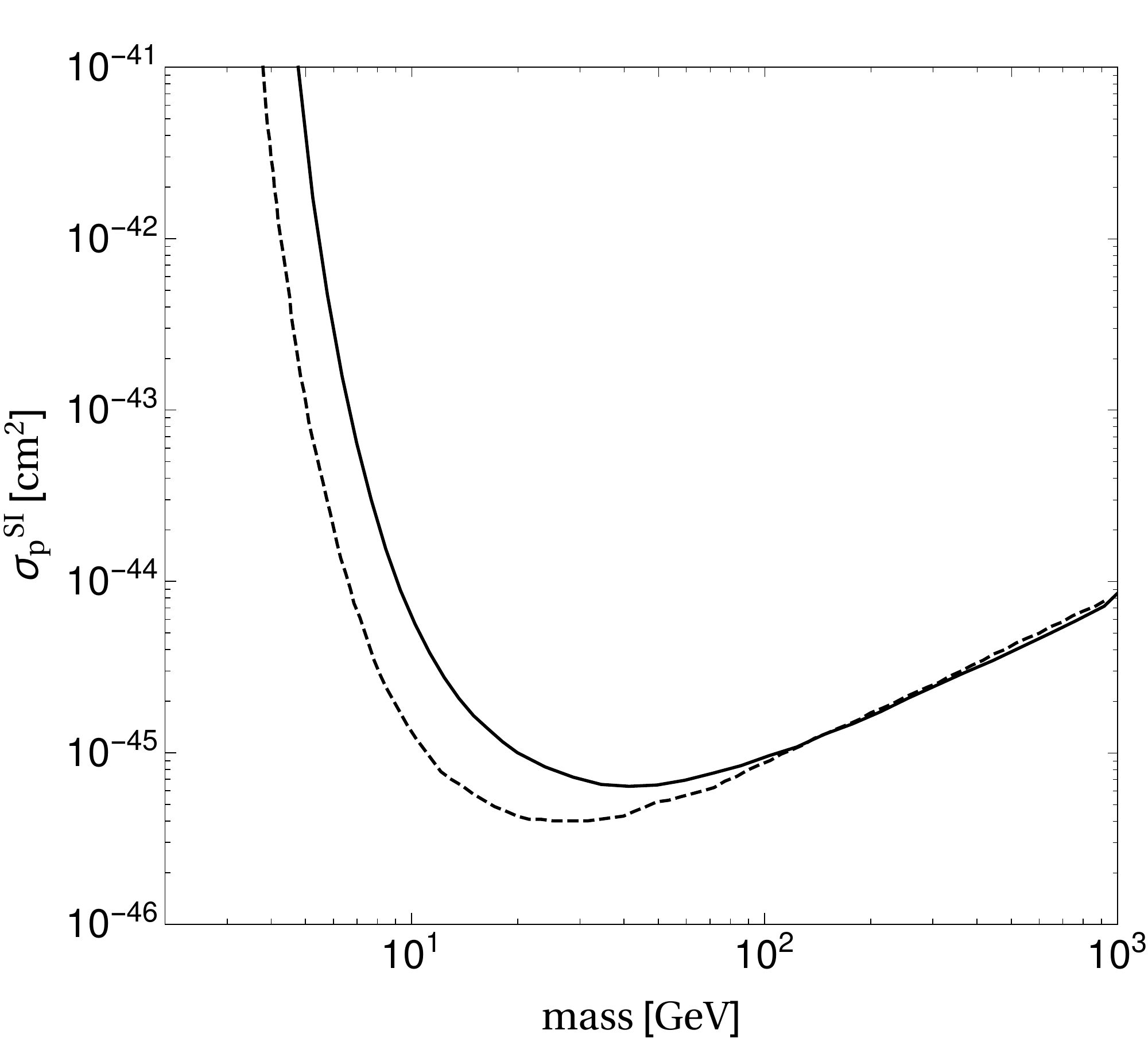} 
\caption{Comparison of our exclusion limits (solid) with the official LUX result (dashed)}
\label{figLUXcompare}
\end{figure}

 For each of the operators, we calculate 90\% confidence limits for the inner 18cm fiducial volume (117kg) over the 95 day LUX run, which resulted in a 30.5 kg day exposure. For simplicity we will assume that the background prediction is uniform throughout the fiducial volume. While this is actually likely not the case, it is a conservative estimate given the background is lower within the inner fiducial volume. After the 99.6$\%$ electronic recoil discrimination efficiency, 1.9 events were expected in the nuclear recoil region, and 2 were actually observed. The energy dependent detector efficiency was taken from LUXcalc~\cite{Savage:2015xta}, which takes into account detector resolution and threshold effects. While we have reduced the threshold to 1.1 keV, this efficiency curve is based on the 2013 LUX analysis which causes us to undercover the confidence limit at low WIMP mass, as illustrated in Fig.~\ref{figLUXcompare}. A summary of the experimental specifications are given in Table~\ref{tabExperiments}.

\begin{table}[ht]
\caption{Experiments used to generate exclusion curves}
\begin{tabular}{r|r|r|r|r|r|r|r|}
Name      & Target &  Exp. (kg.y) &     ROI & efficiency &  background & observed    \\
\hline
LUX       &     Xe &     30.5   &   1.1-41 keV &      0.5 &       1.9  &        2    \\
\hline
\end{tabular}
\label{tabExperiments}
\end{table}

\begin{figure}[ht]
\begin{tabular}{ccc}
\includegraphics[height=5.2cm]{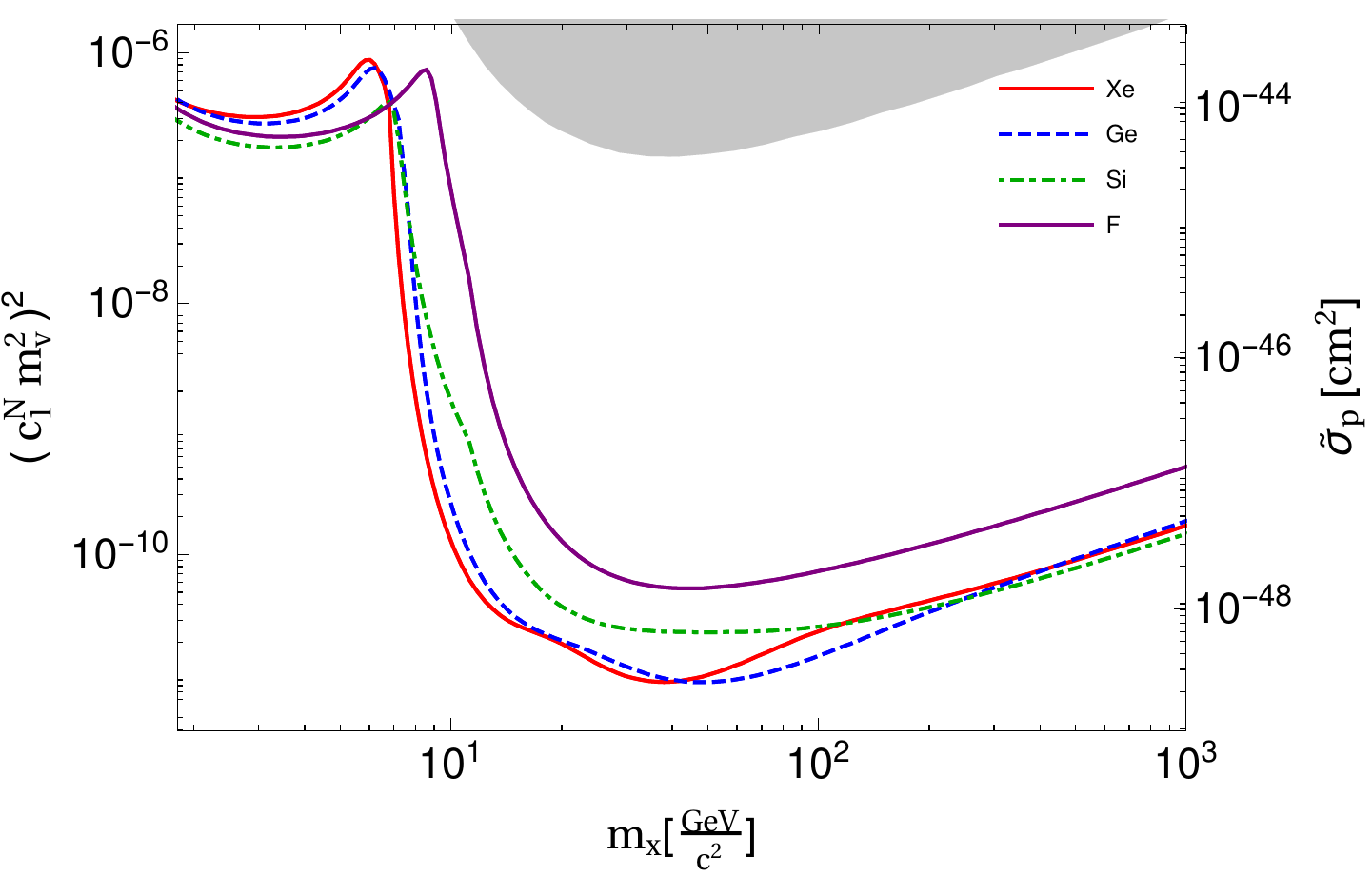} &
\includegraphics[height=5.2cm]{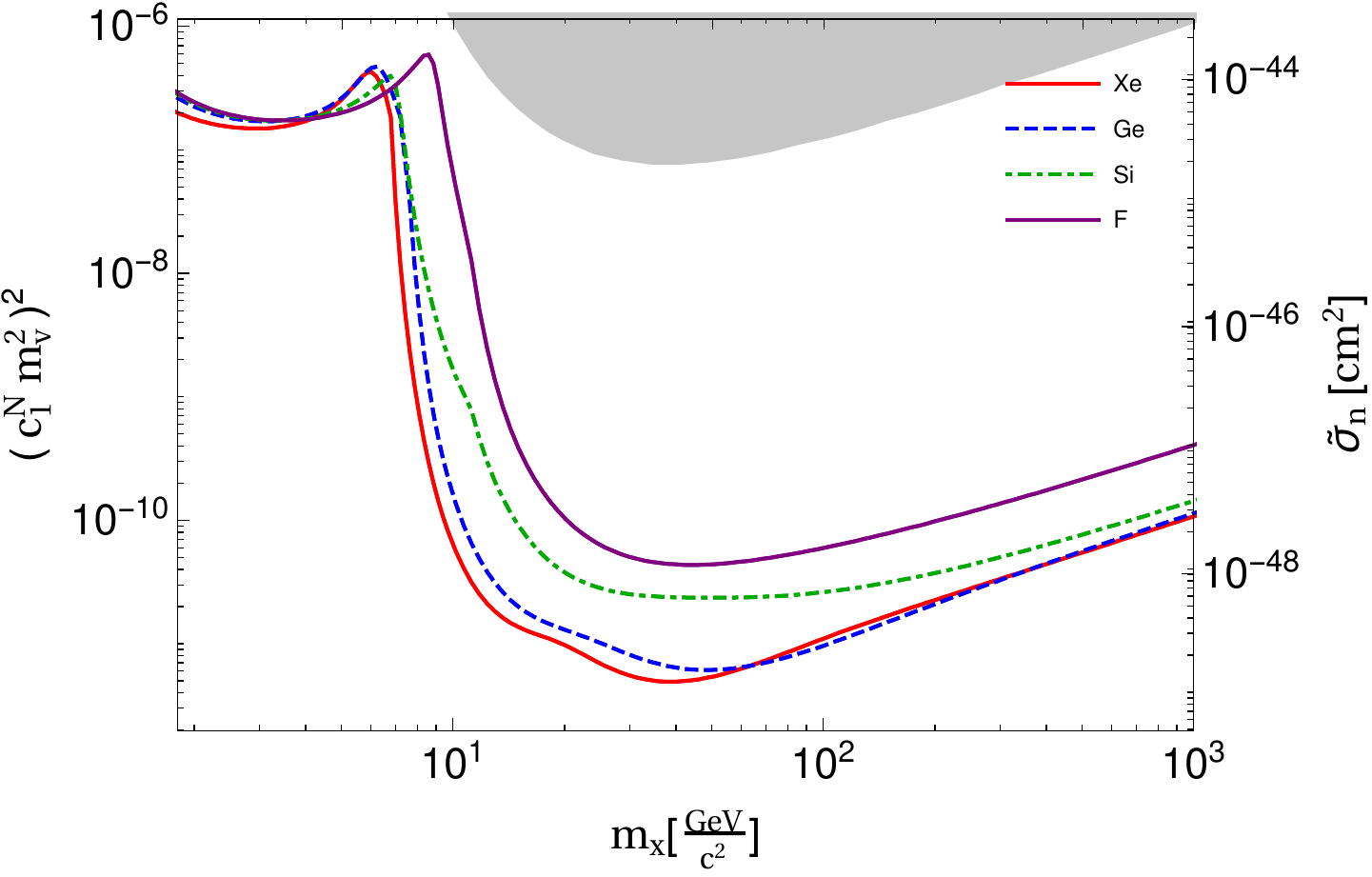} \\
\includegraphics[height=5.2cm]{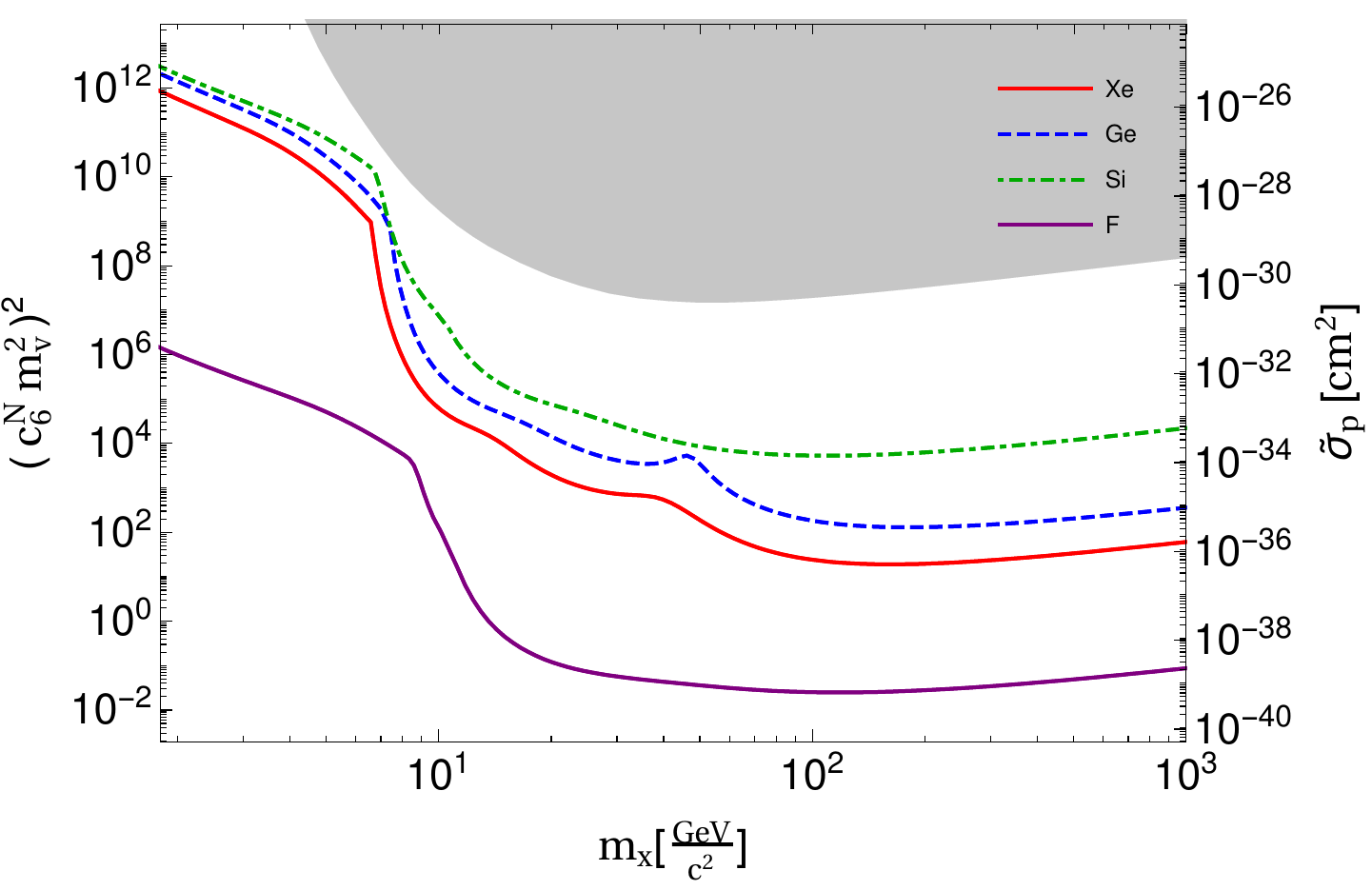} &
\includegraphics[height=5.2cm]{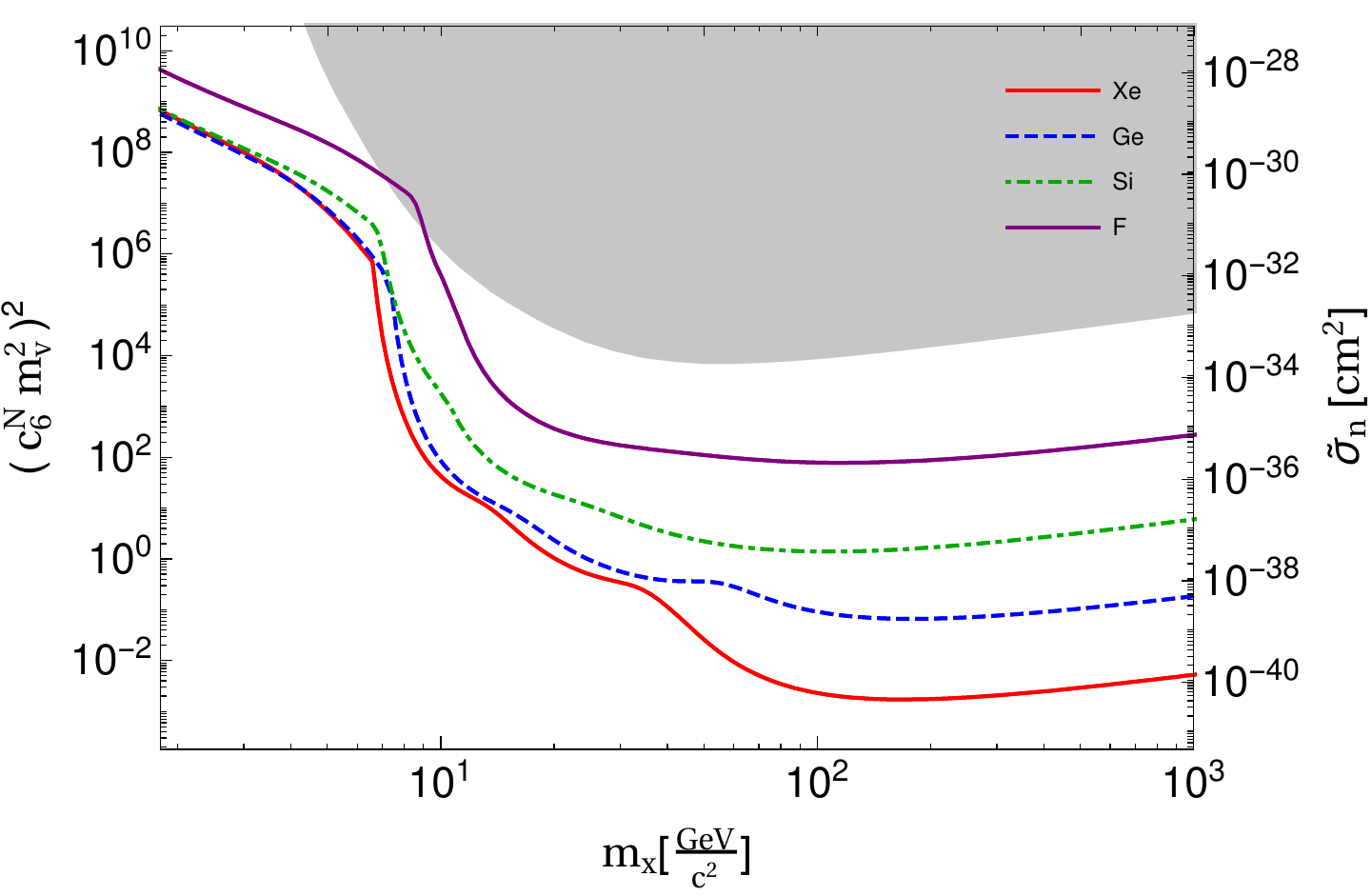} \\
\includegraphics[height=5.2cm]{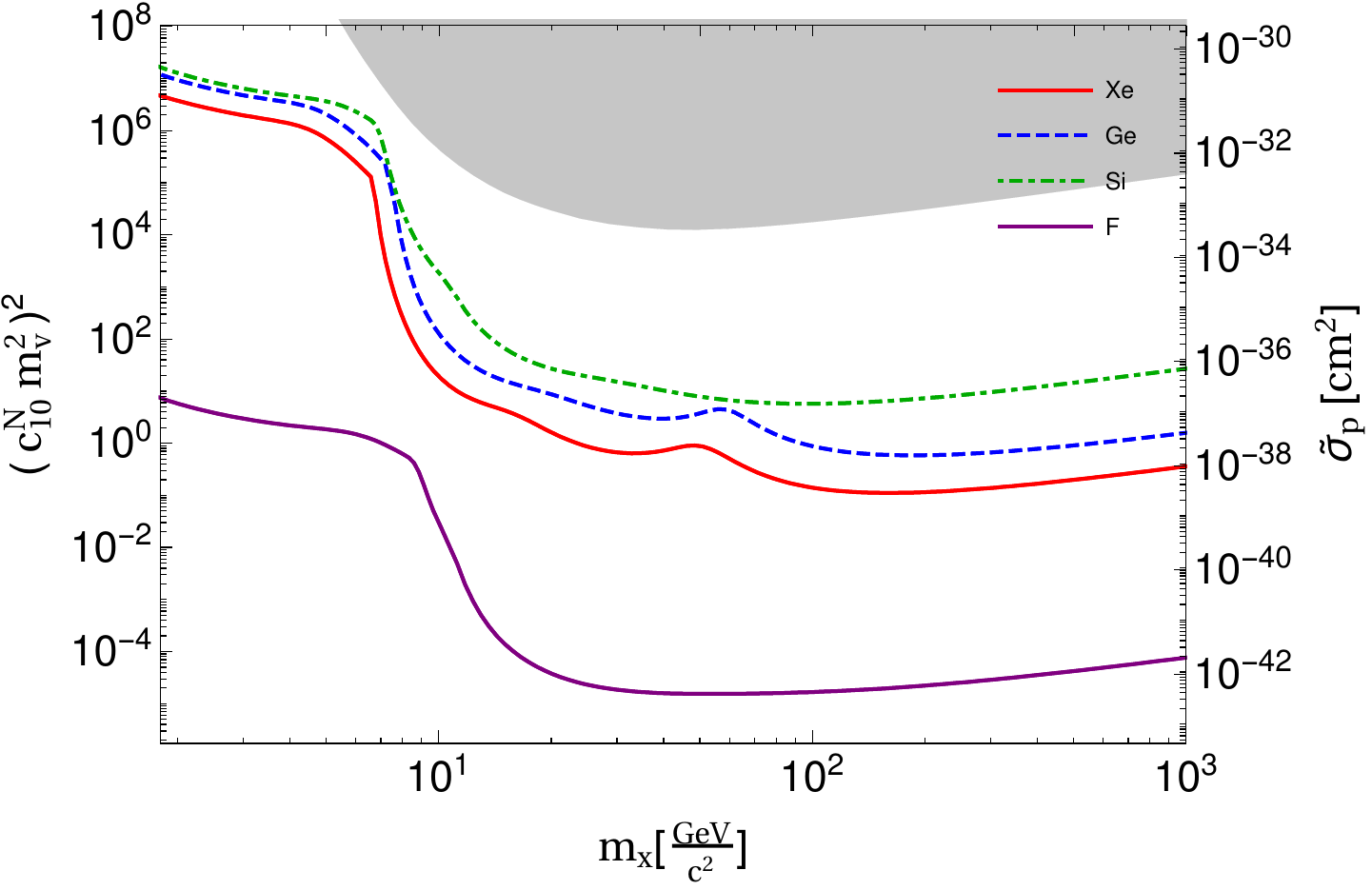} &
\includegraphics[height=5.2cm]{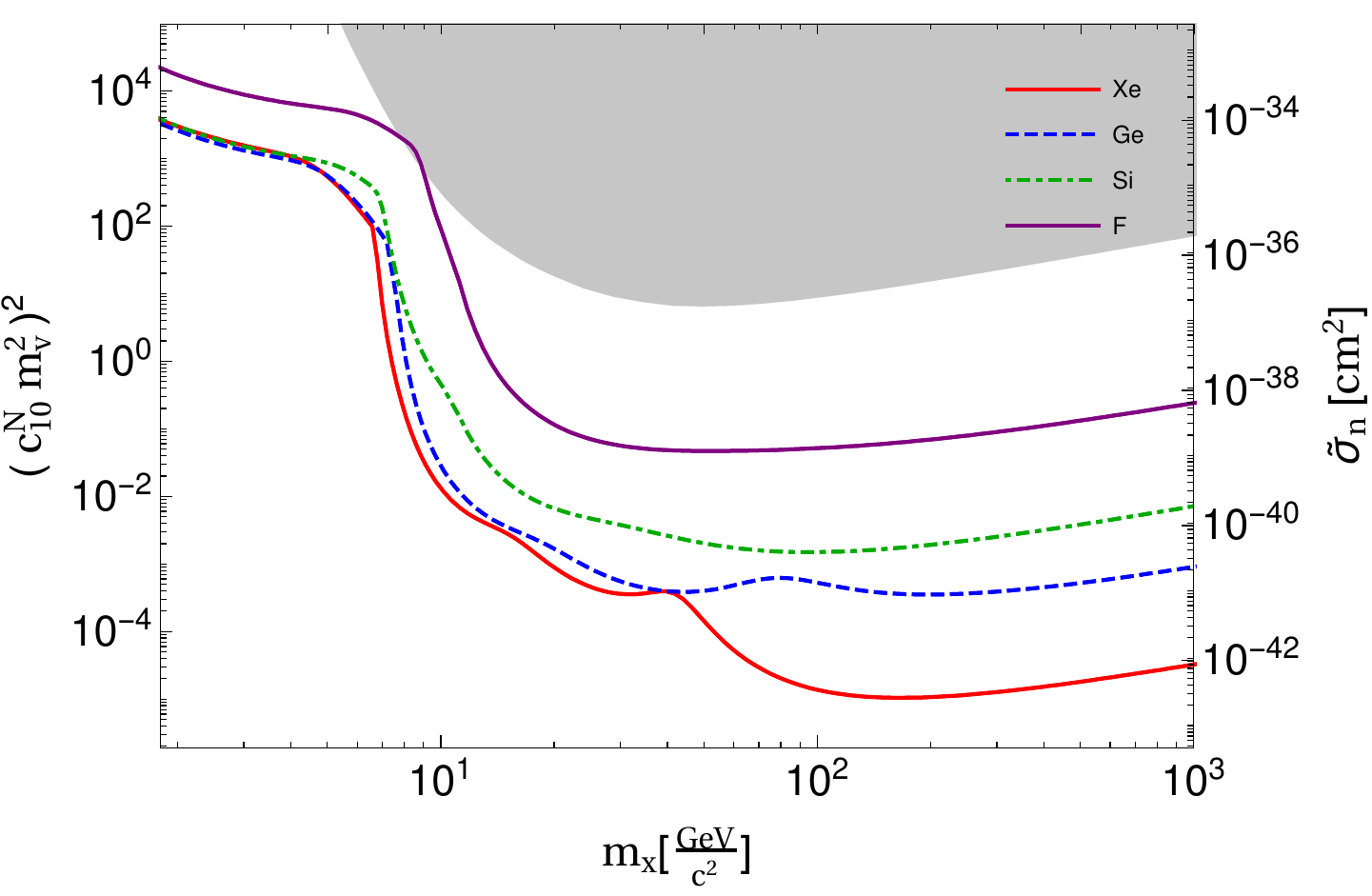} \\
\end{tabular}
\caption{Discovery limits for $\mathcal{O}_1$ (top), $\mathcal{O}_6$ (middle), $\mathcal{O}_{10}$ (bottom), for protons (left) and neutrons (right). The shaded region shows the 90\% confidence limits for a 30.5 kg day exposure at LUX.}
\label{figFloors}
\end{figure}

Figure~\ref{figFloors} shows the discovery limits and exclusion curves for $\mathcal{O}_1$ (top), $\mathcal{O}_6$ (middle), $\mathcal{O}_{10}$. The corresponding discovery limits and exclusion curves for the remaining operators are shown in the Appendix. For several operators, for example $\mathcal{O}_6$ coupling to neutrons, we find that the calculated limits (grey shaded regions) are overlapping with the discovery limits curves for low mass where the discovery limit is dominated by Solar neutrinos. This does not imply a contradiction, as the exclusion curves, which only apply to xenon targets, do not overlap the xenon discovery limits. The proximity of the exclusion curves to the discovery limits (which have vastly larger exposures) is a reflection of the different statistical procedures used to generate the two sets of curves. In particular, the calculated discovery limits are a more statistically demanding criteria than an exclusion limit at $90\%$ confidence, so for a given WIMP mass and cross section a larger exposure would be required to claim a 3$\sigma$ fluctuation. In future larger scale detectors for which the neutrino signal will be non-negligible, it will be necessary to include neutrinos into the statistical model that determines exclusion regions. 

\section{Discussion and Conclusion}
\label{sec:conclusions}
\par In order to continue to improve bounds on the WIMP-nucleon cross section, future larger scale detectors must become effective at distinguishing a WIMP interaction from a neutrino interaction. In this paper we considered this issue within the well-motivated EFT framework. Within this framework, the standard SI and SD interactions represent only a portion of a larger set of nuclear responses which must be considered in direct dark matter detection. We specifically focus on the set of EFT operators that respect Galilean invariance, time-reversal symmetry, and Hermiticity. 

\par We have shown that for 10 of the 14 operators, the energy spectrum induced by WIMPs is distinct from that induced by neutrinos. For these operators, we show that a clean statistical separation between WIMPs and neutrinos will be possible. For only 4 of the 14 WIMP-nucleon operators that we consider do we find that the WIMP and neutrino spectrum can be highly degenerate. For these 4 operators (which belong to group 1) we specifically calculate the ``worst-case scenario" WIMP mass which most closely matches the neutrino spectra. Our results show that an experiment with good spectral energy resolution and exposure near the ton scale should have little trouble distinguishing certain WIMP interactions from neutrino-induced nuclear recoil events. The group 2 and 3 operators would require an exposure of about 0.5 tonne years to be distinguished from the neutrino background for a low mass WIMP (as can be surmised from the linear region of Figure~\ref{figDiscEvo} beyond the saturation region/inflection point). The group 1 operators can be distinguished from the neutrino backgrounds for a sufficiently large exposure, $\sim 10^3$ tonne years. 

\par  Relative to previous results that considered energy deposition, our theoretical framework is more complete and encompasses a wider range of possible nuclear responses.  In its most general form, the WIMP nucleon cross section is a superposition of all of the operators that we have discussed, with the observable being a superposition of the corresponding nuclear recoil spectrum for each operator. The limiting case that we have studied here in which a single operator dominates the cross section will provide guidance and intuition for future more detailed studies that consider more complicated superpositions of operators. In order to extract information about the particle properties of dark matter from a detection of events, the challenge that future detectors will face not only lies in reducing the neutrino backgrounds, but also in understanding the degeneracies that are incurred when attempting to map the detected energy spectrum onto a particular superposition of operators~\cite{Gluscevic:2015sqa}. 

\section{Acknowledgements} 
BD acknowledges support from DOE Grant DE-FG02-13ER42020. LES acknowledges support from NSF grant PHY-1522717. J.B.D. thanks Dr. and Mrs. Sammie W. Cosper at the University of Louisiana at Lafayette, and the Louisiana Board of Regents for support. J.L.N acknowledges support from the DOE for this work under grant No.\,DE-SC0008016.

\appendix
\section{Analysis for all operators}
\par
\begin{figure}[ht]
\centering
\begin{tabular}{lll}
\hspace{-1.5cm}\includegraphics[height=4cm]{figures/8BmaxLike_O1.pdf} &
\includegraphics[height=4cm]{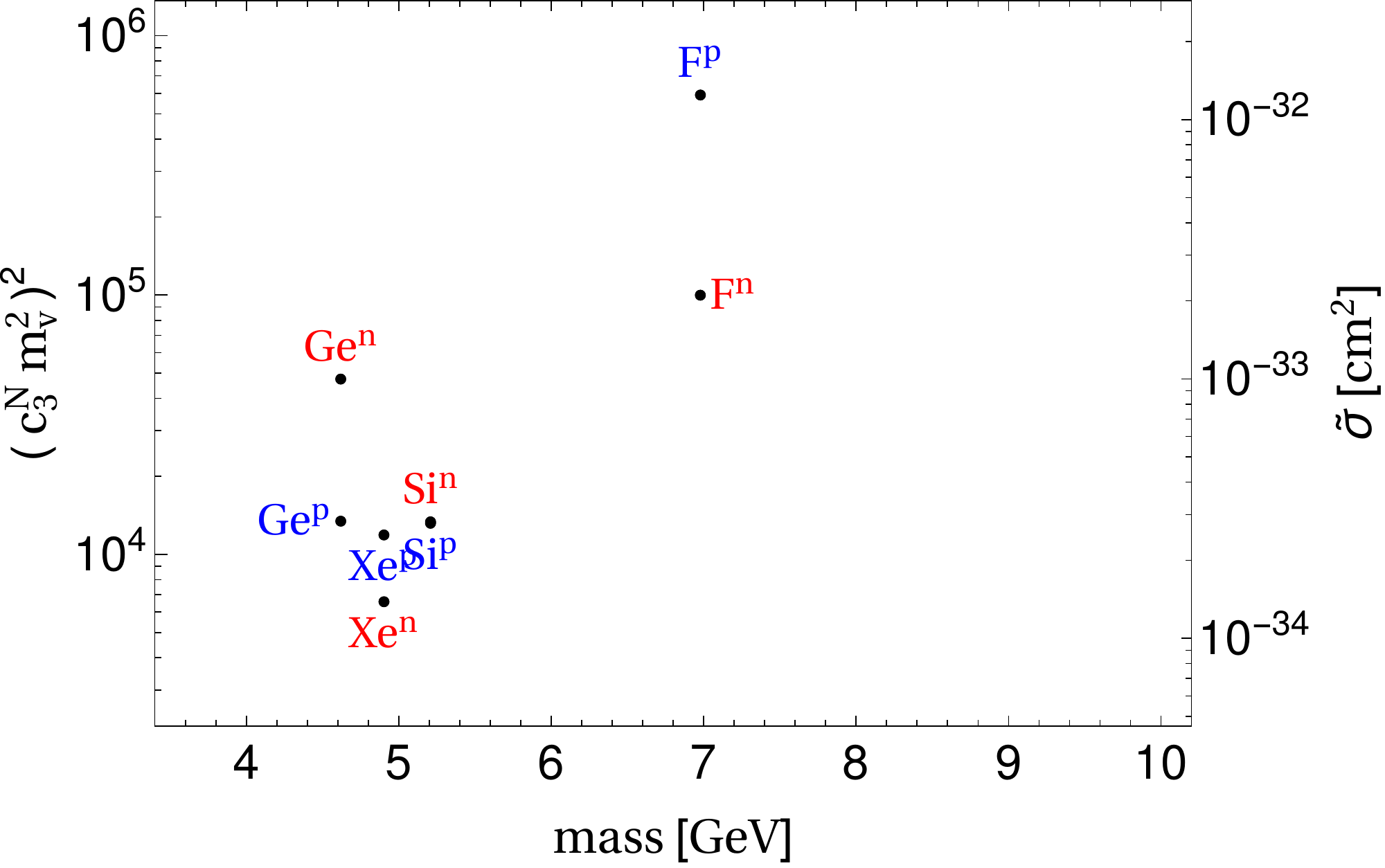} &
\includegraphics[height=4cm]{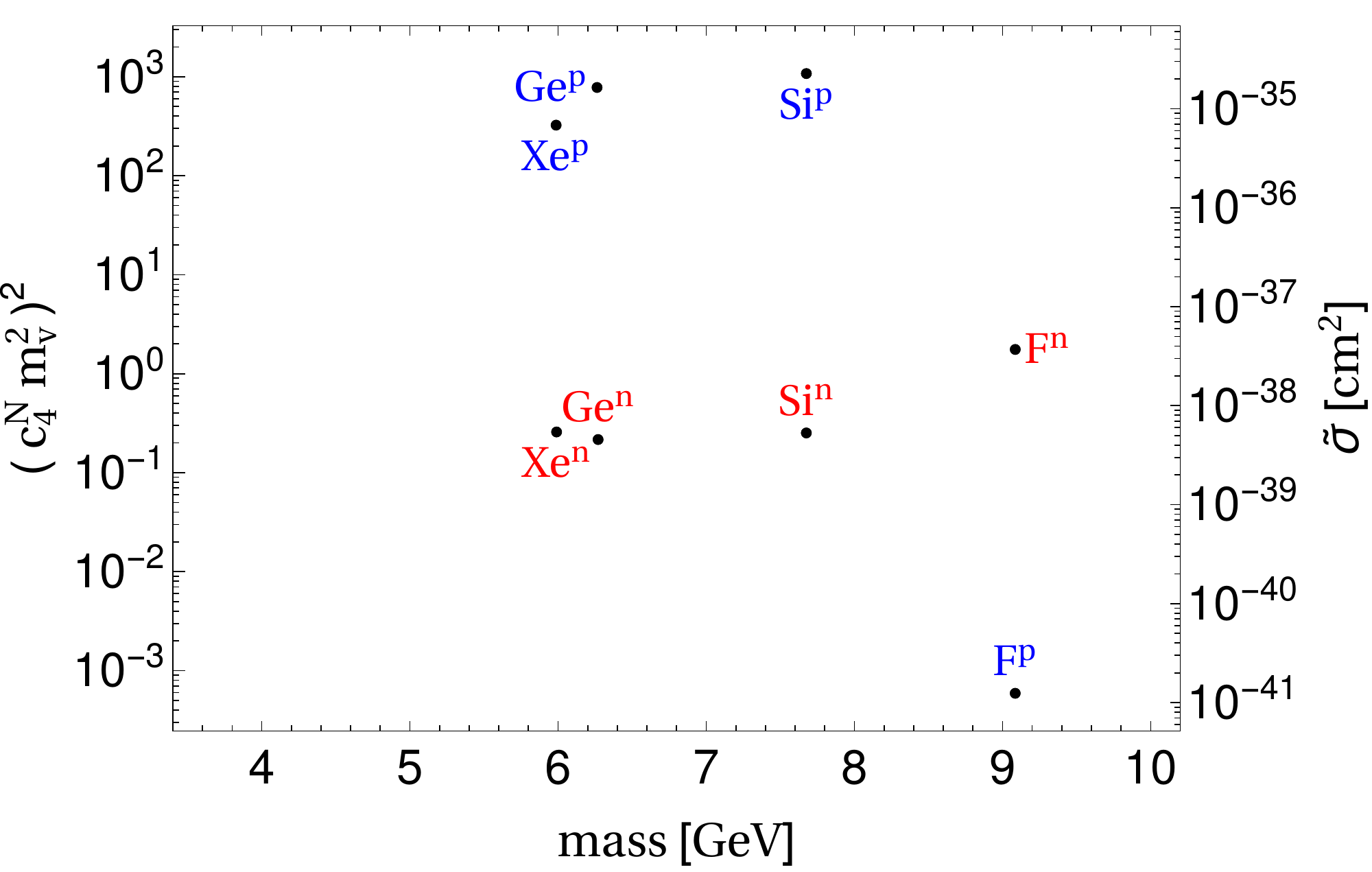} \\
\hspace{-1.5cm}\includegraphics[height=4cm]{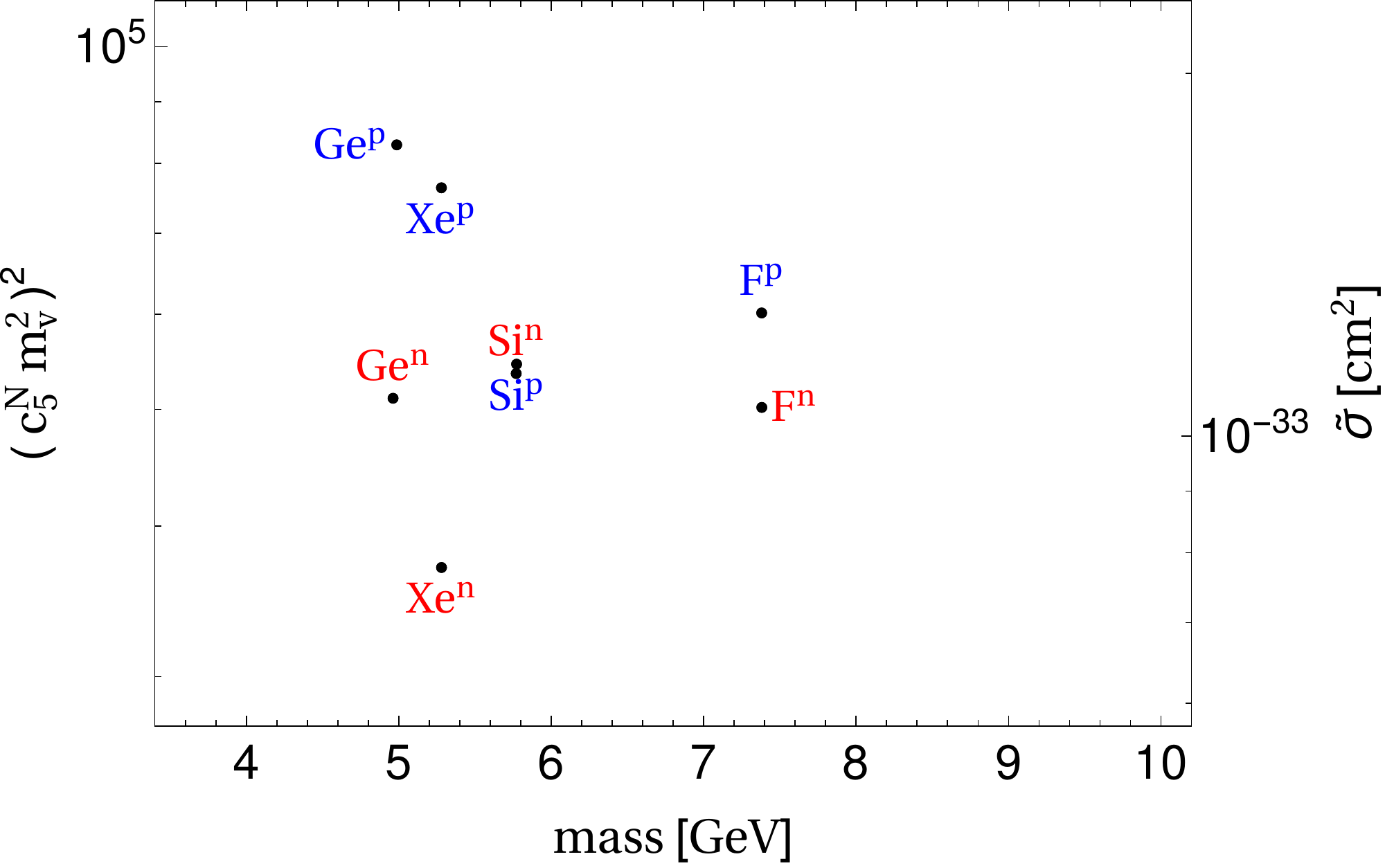} &
\includegraphics[height=4cm]{figures/8BmaxLike_O6.pdf} &
\includegraphics[height=4cm]{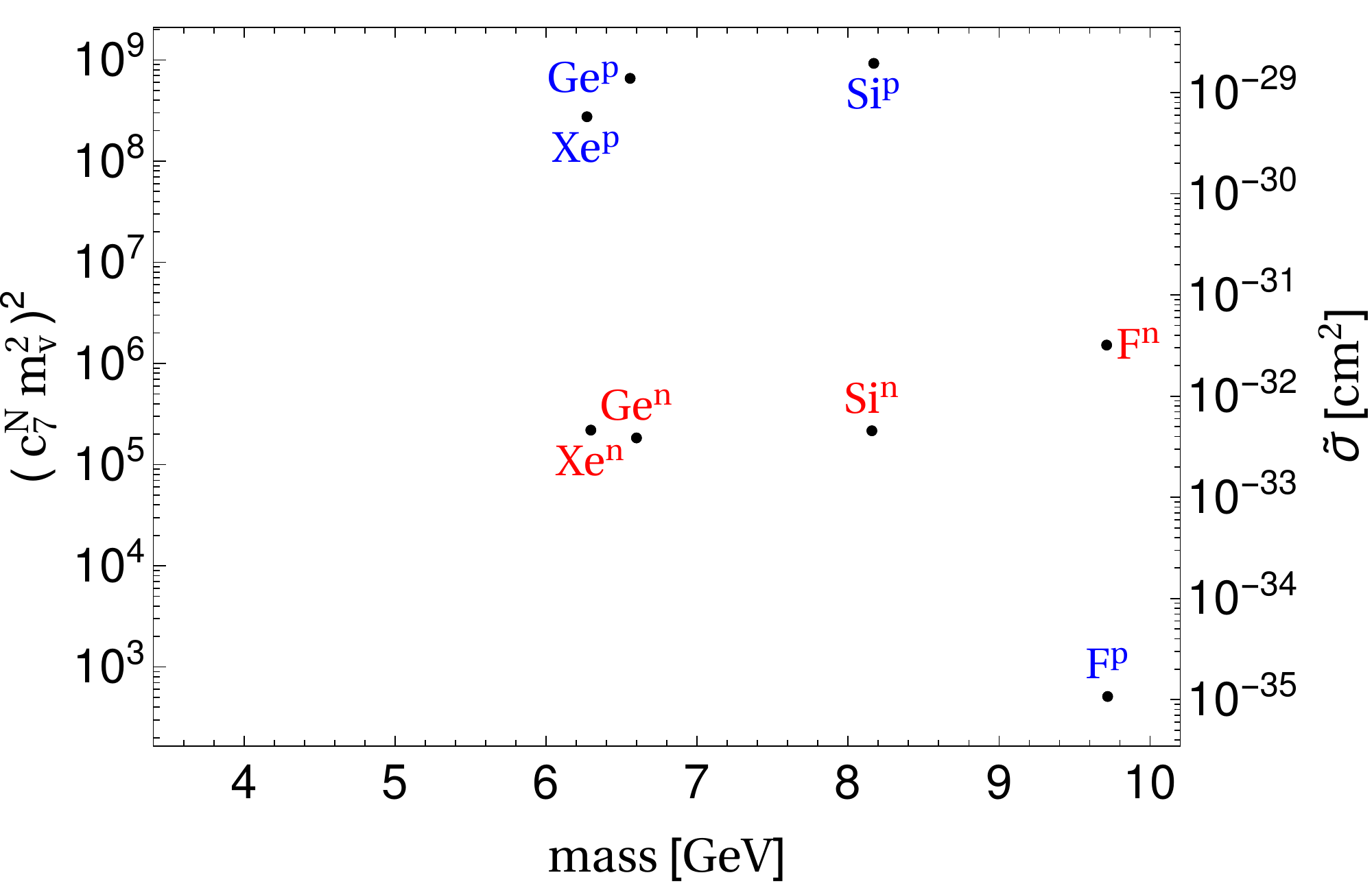} \\
\hspace{-1.5cm}\includegraphics[height=4cm]{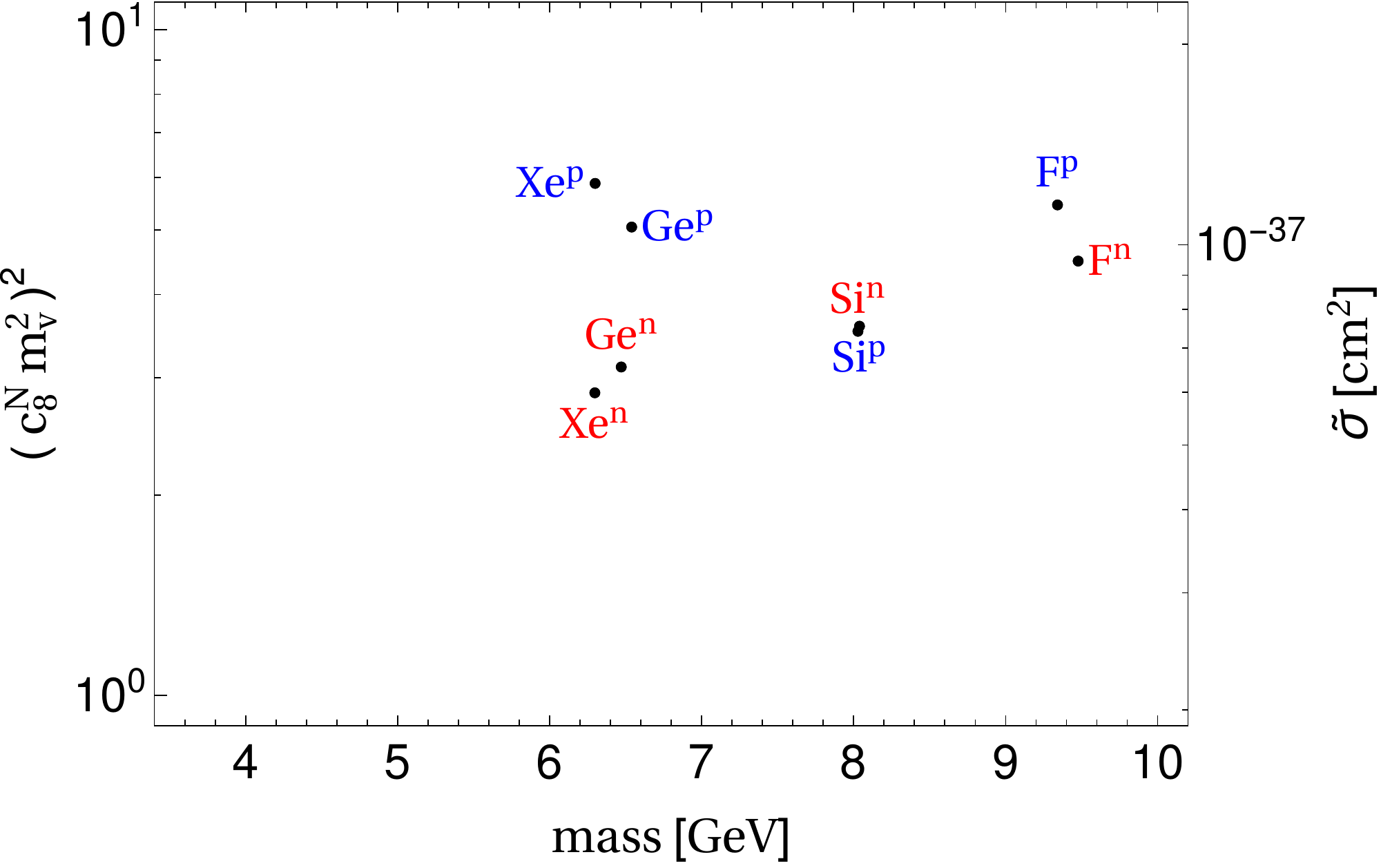} &
\includegraphics[height=4cm]{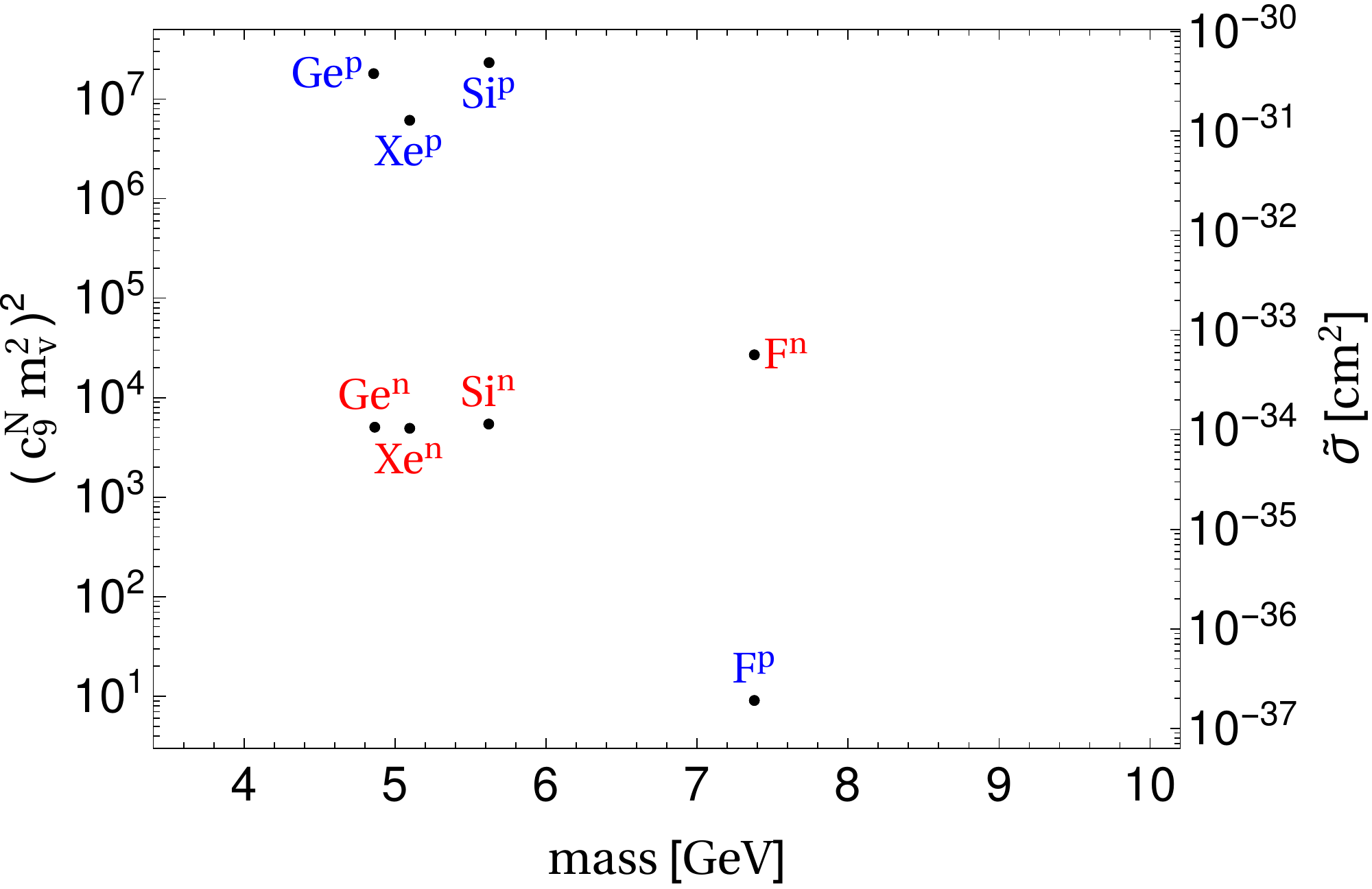} &
\includegraphics[height=4cm]{figures/8BmaxLike_O10.pdf} \\
\hspace{-1.5cm}\includegraphics[height=4cm]{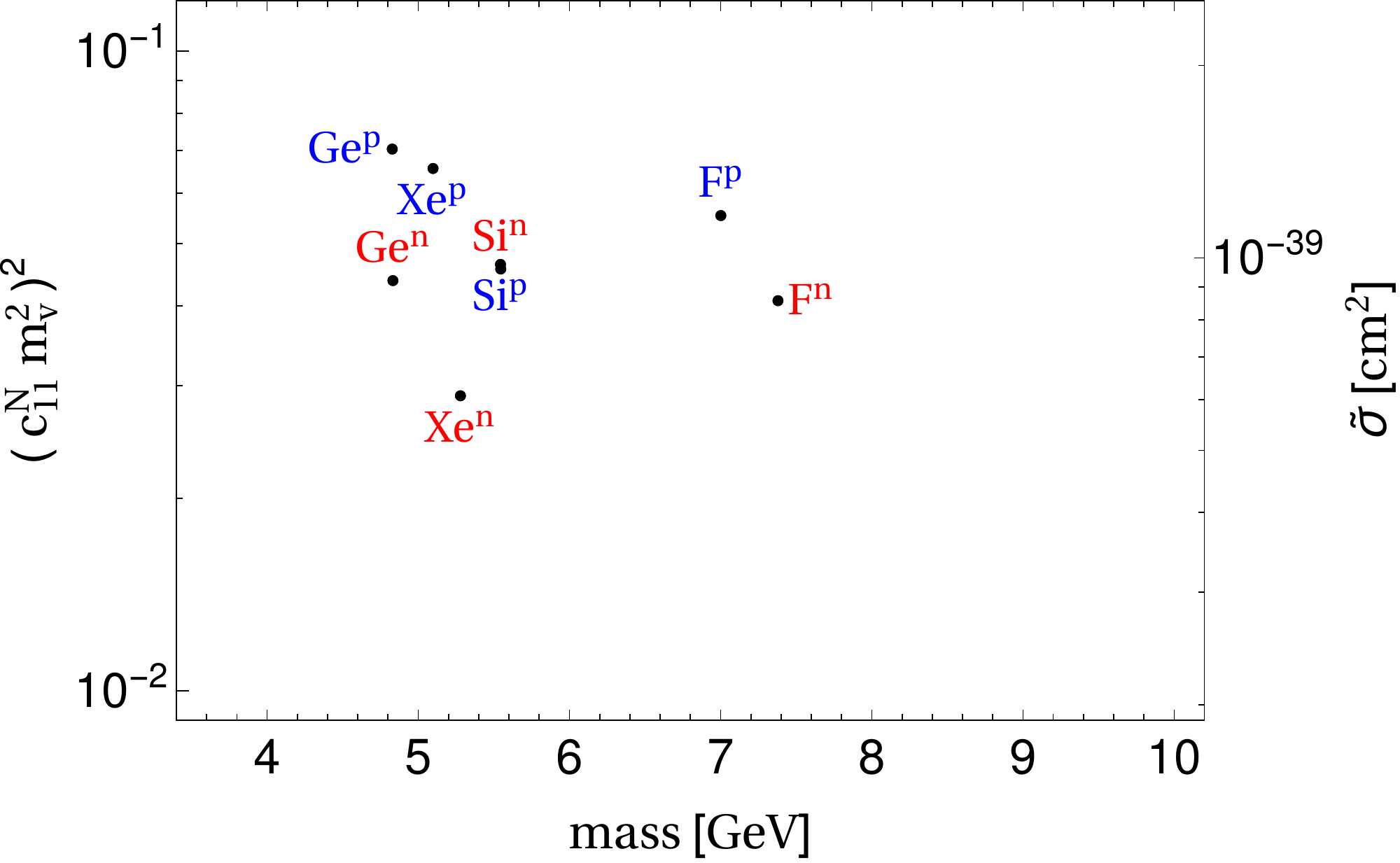} &
\includegraphics[height=4cm]{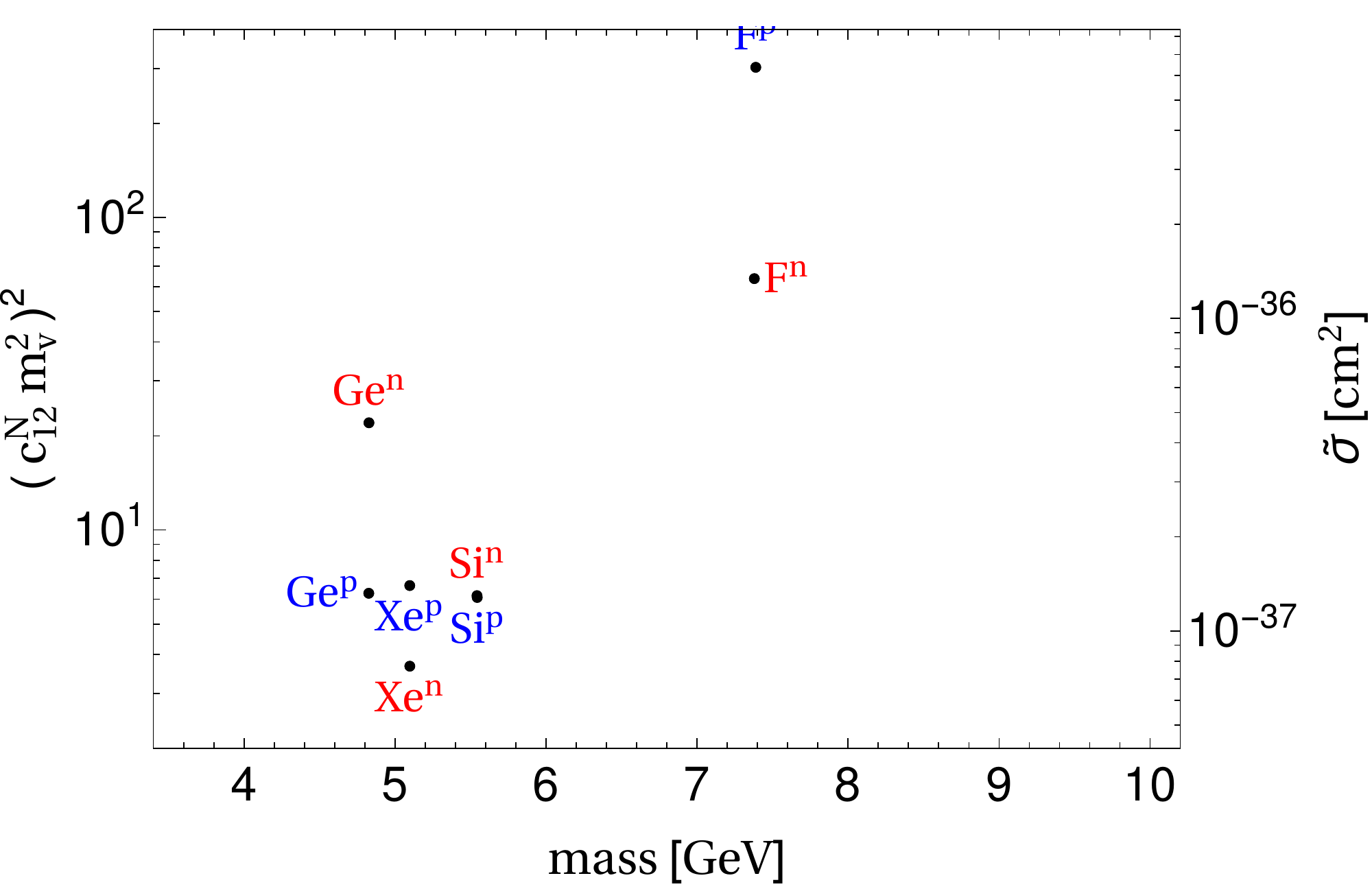} &
\includegraphics[height=4cm]{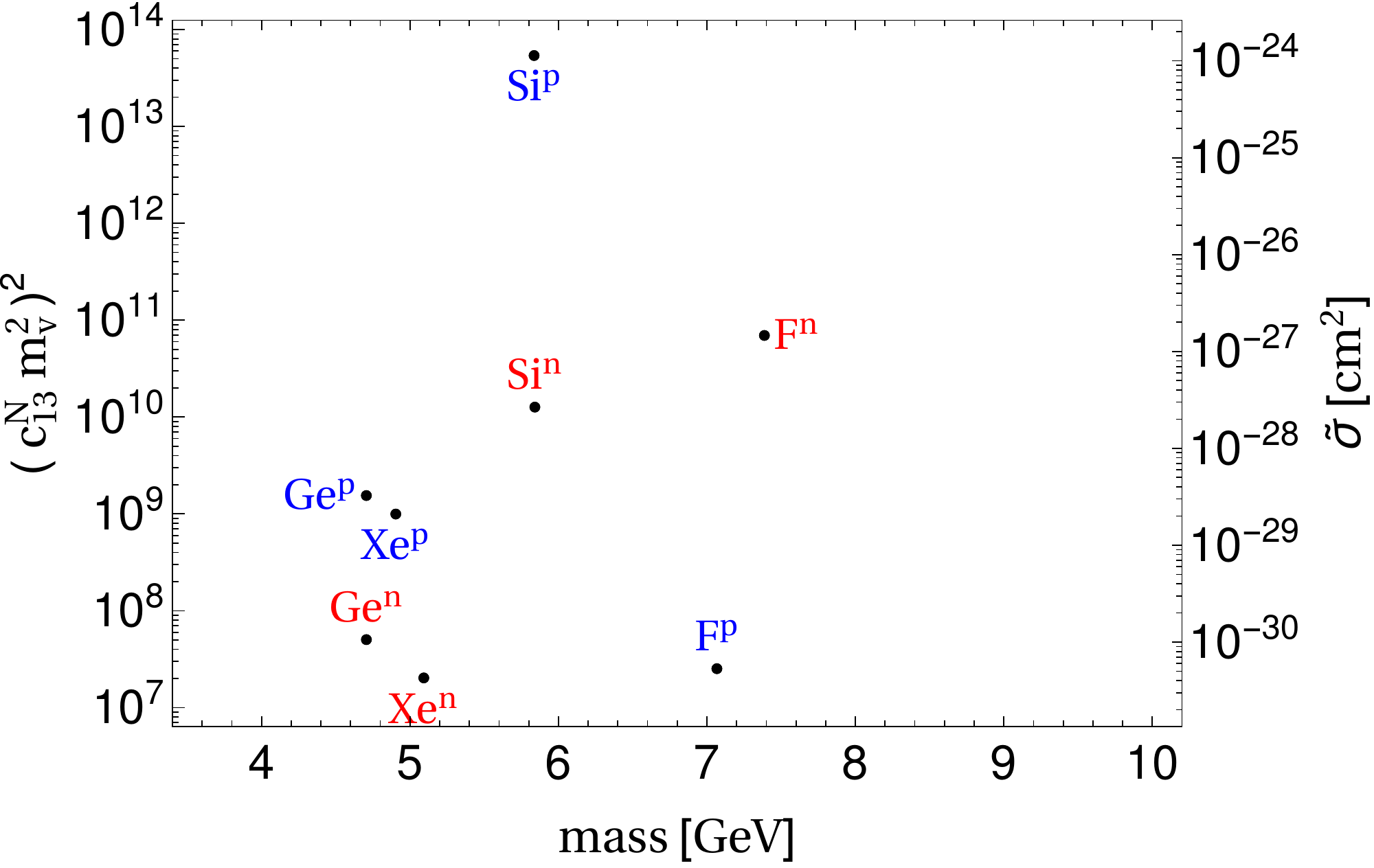} \\
\hspace{-1.5cm}\includegraphics[height=4cm]{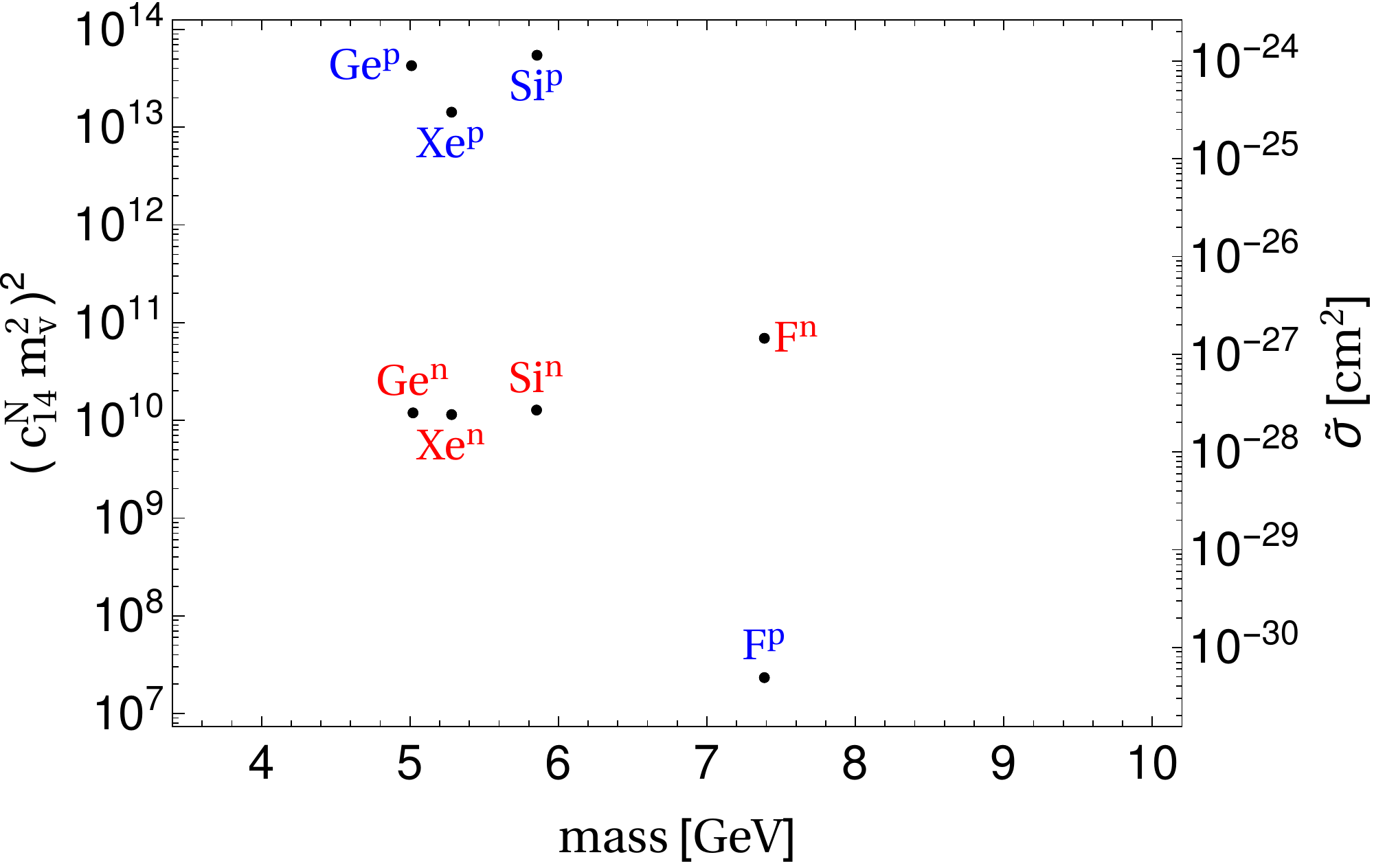} &
\includegraphics[height=4cm]{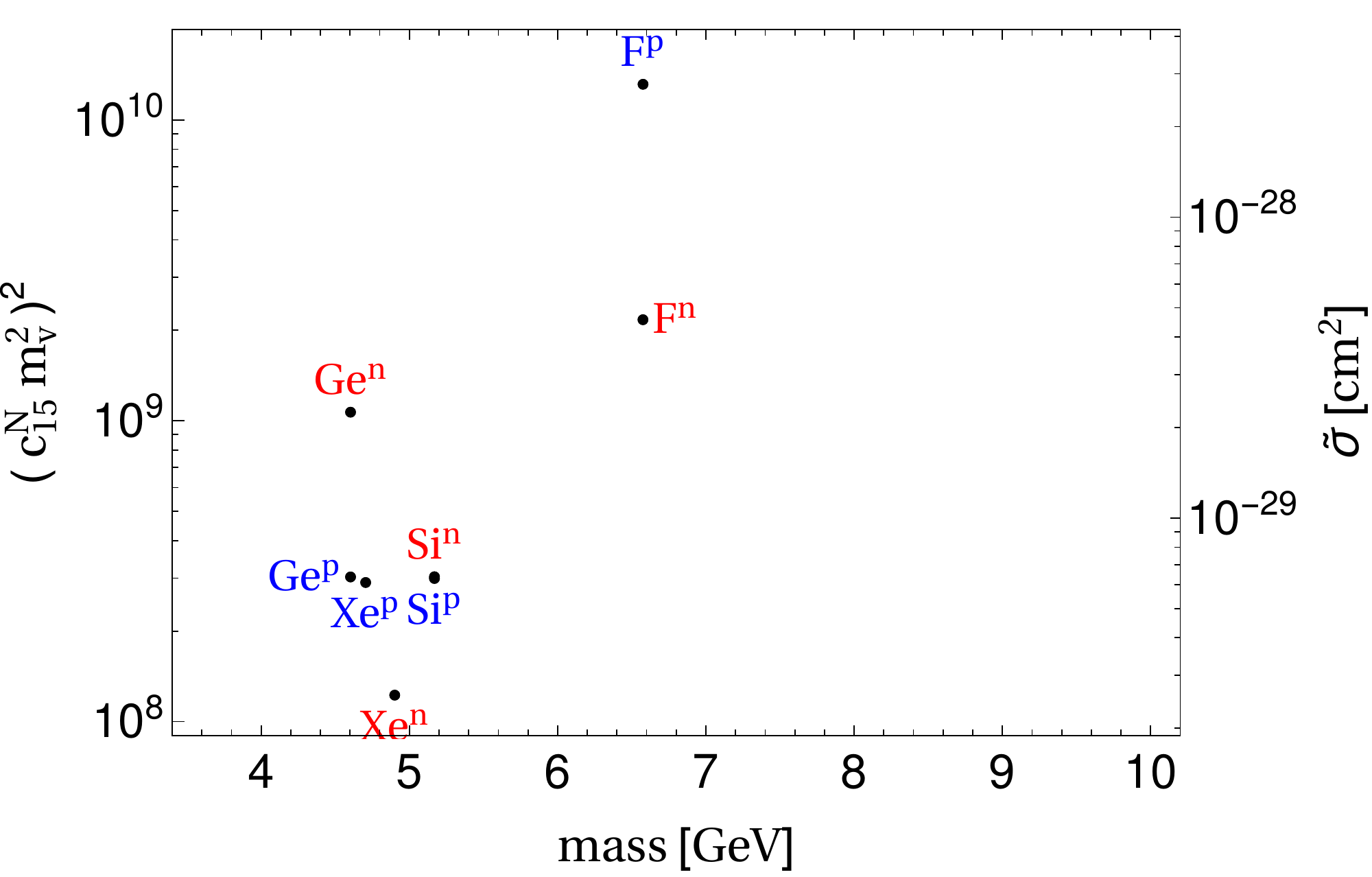} \\
\end{tabular}
\caption{Best fits of each operator to the $^8$B Solar neutrino rate for the four targets. }
\label{fig:8BbestFitMasses_all}
\end{figure}

\begin{figure}[ht]
\begin{tabular}{ccc}
\includegraphics[height=4cm]{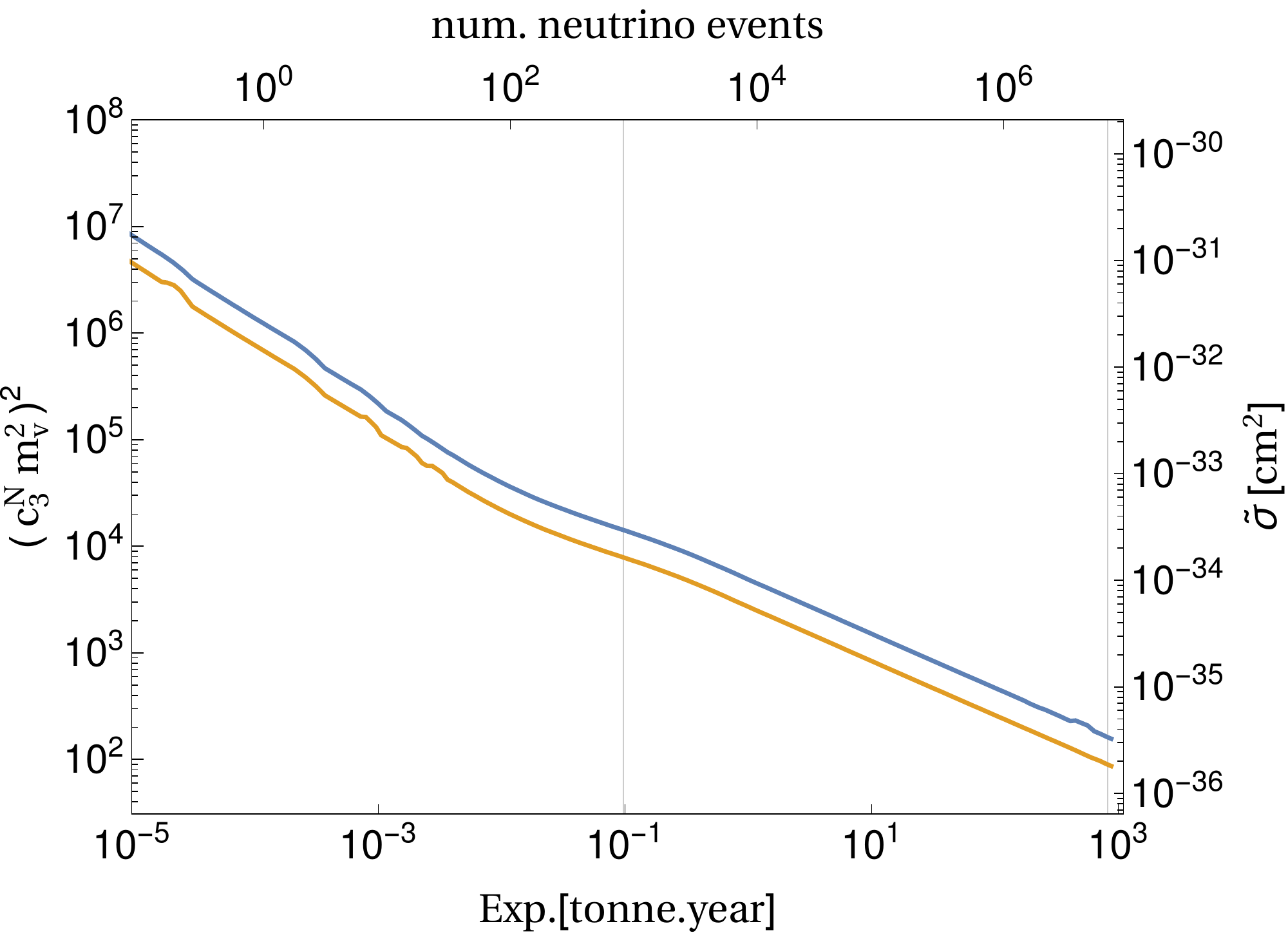}  &
\includegraphics[height=4cm]{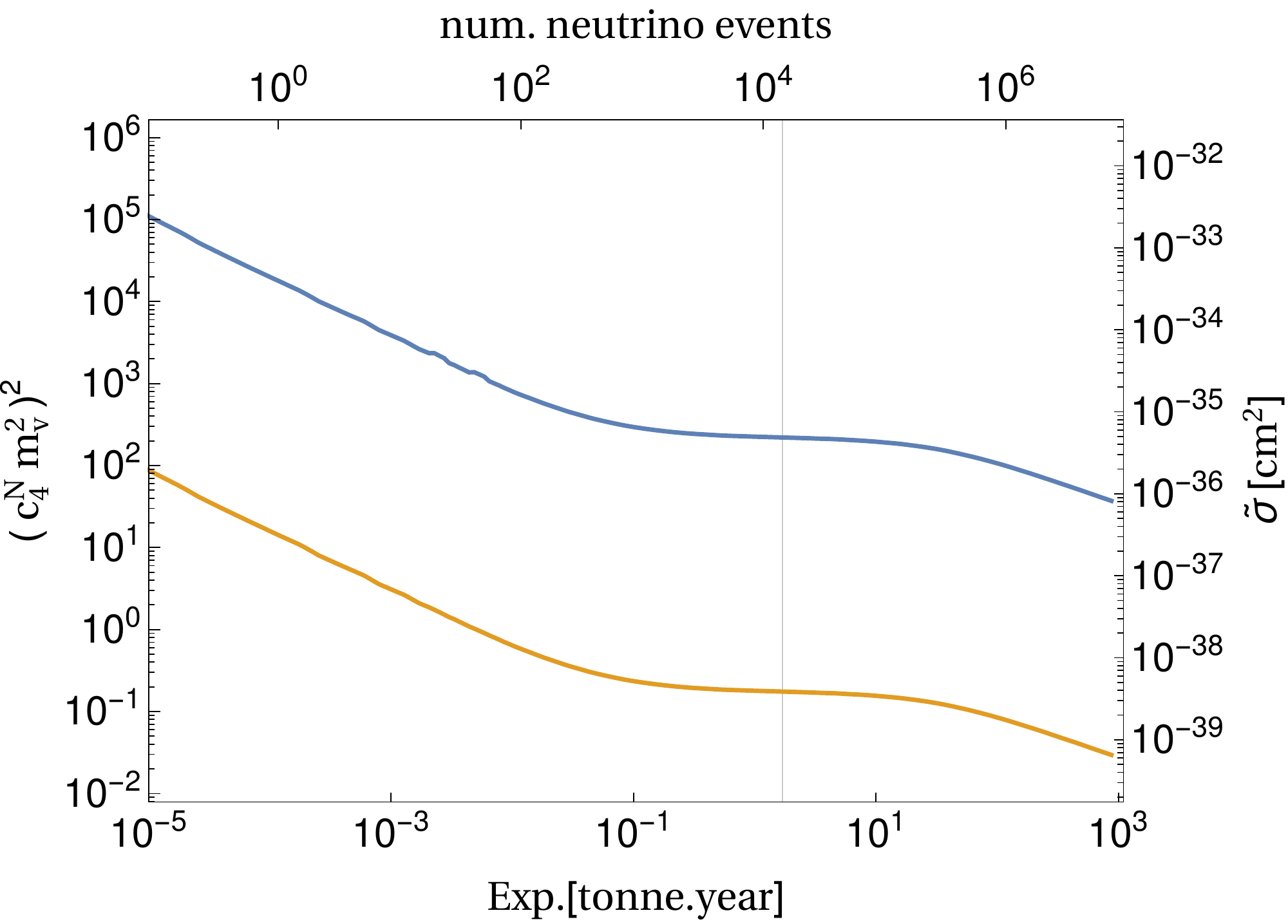}  &
\includegraphics[height=4cm]{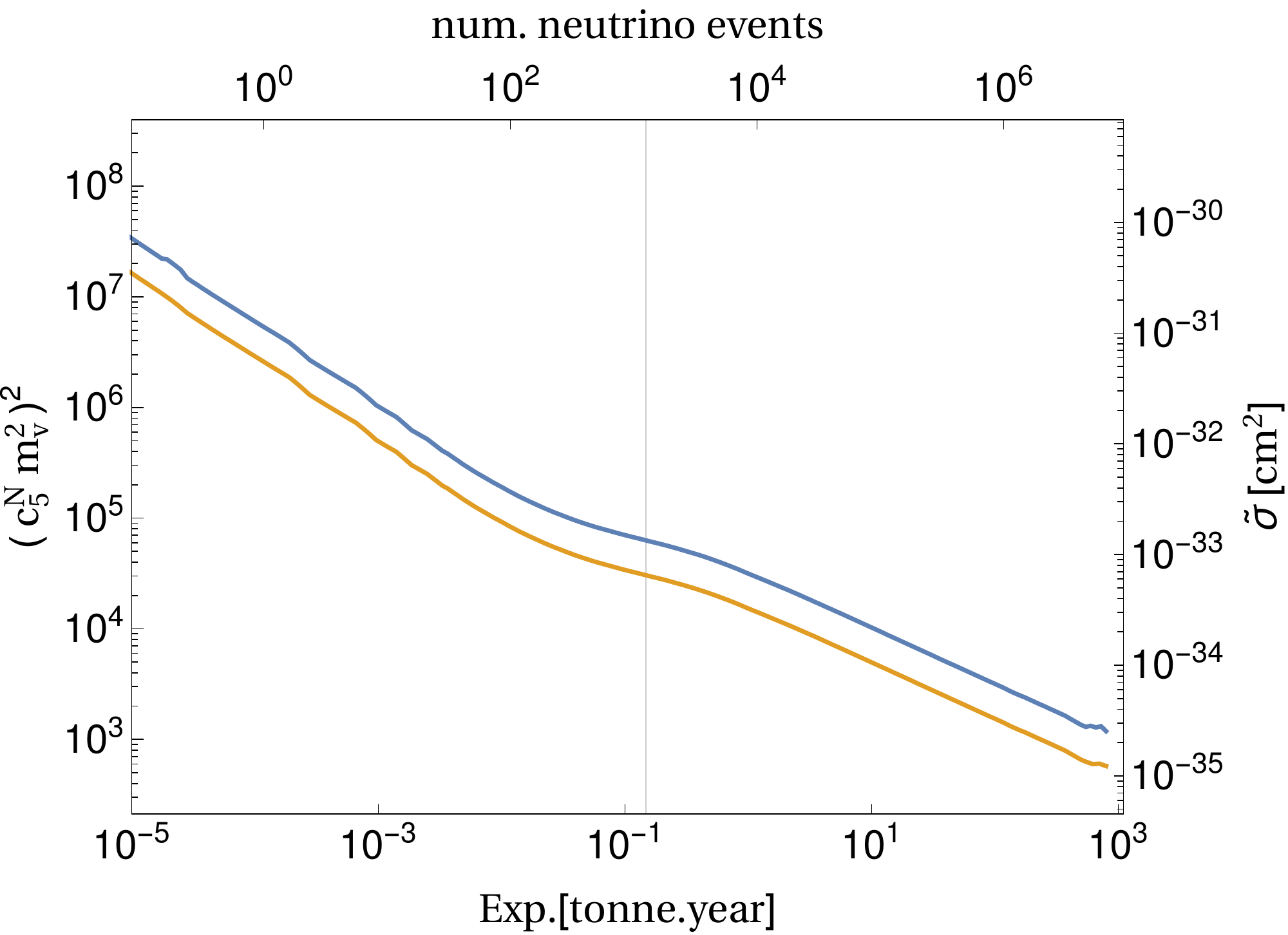} \\
\includegraphics[height=4cm]{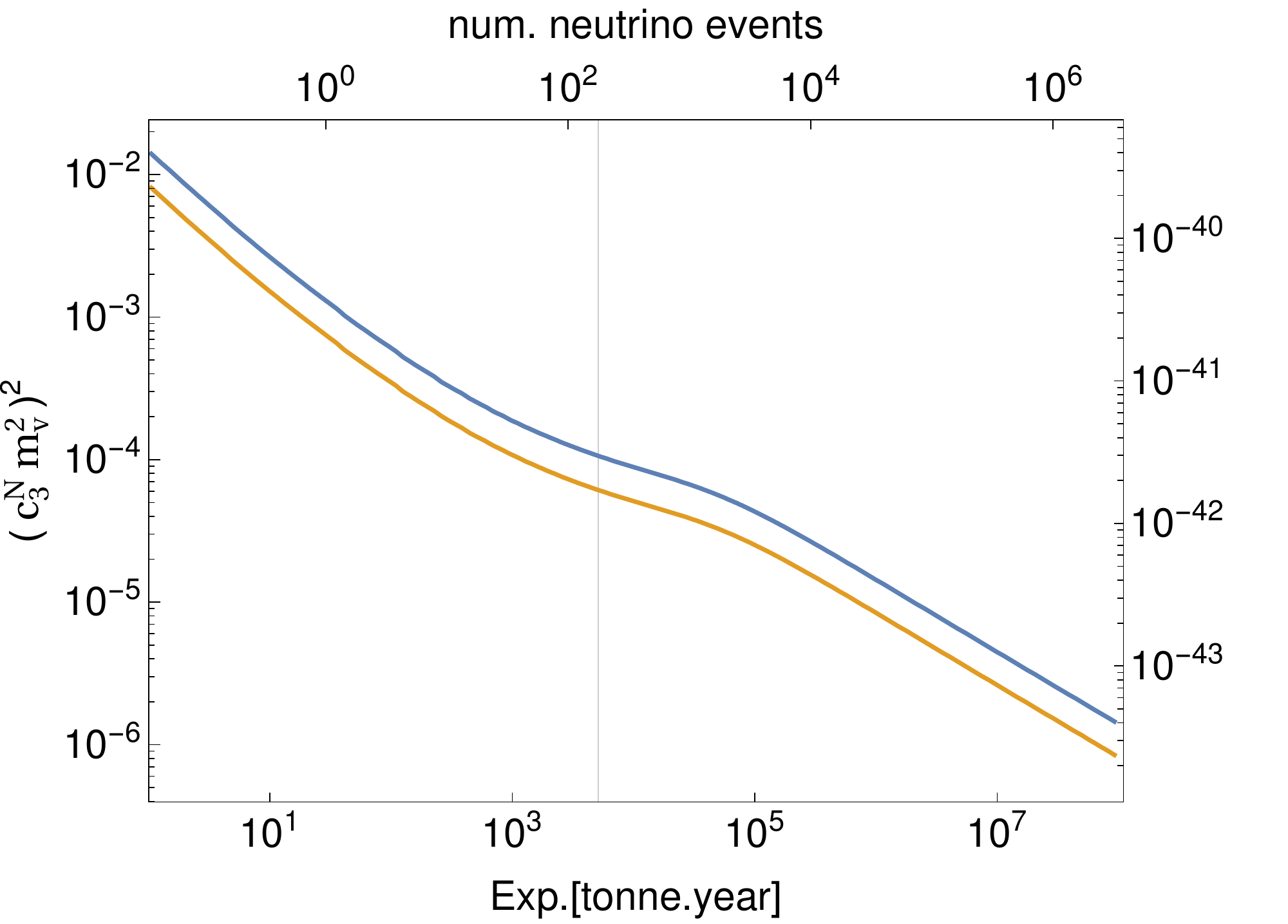} &
\includegraphics[height=4cm]{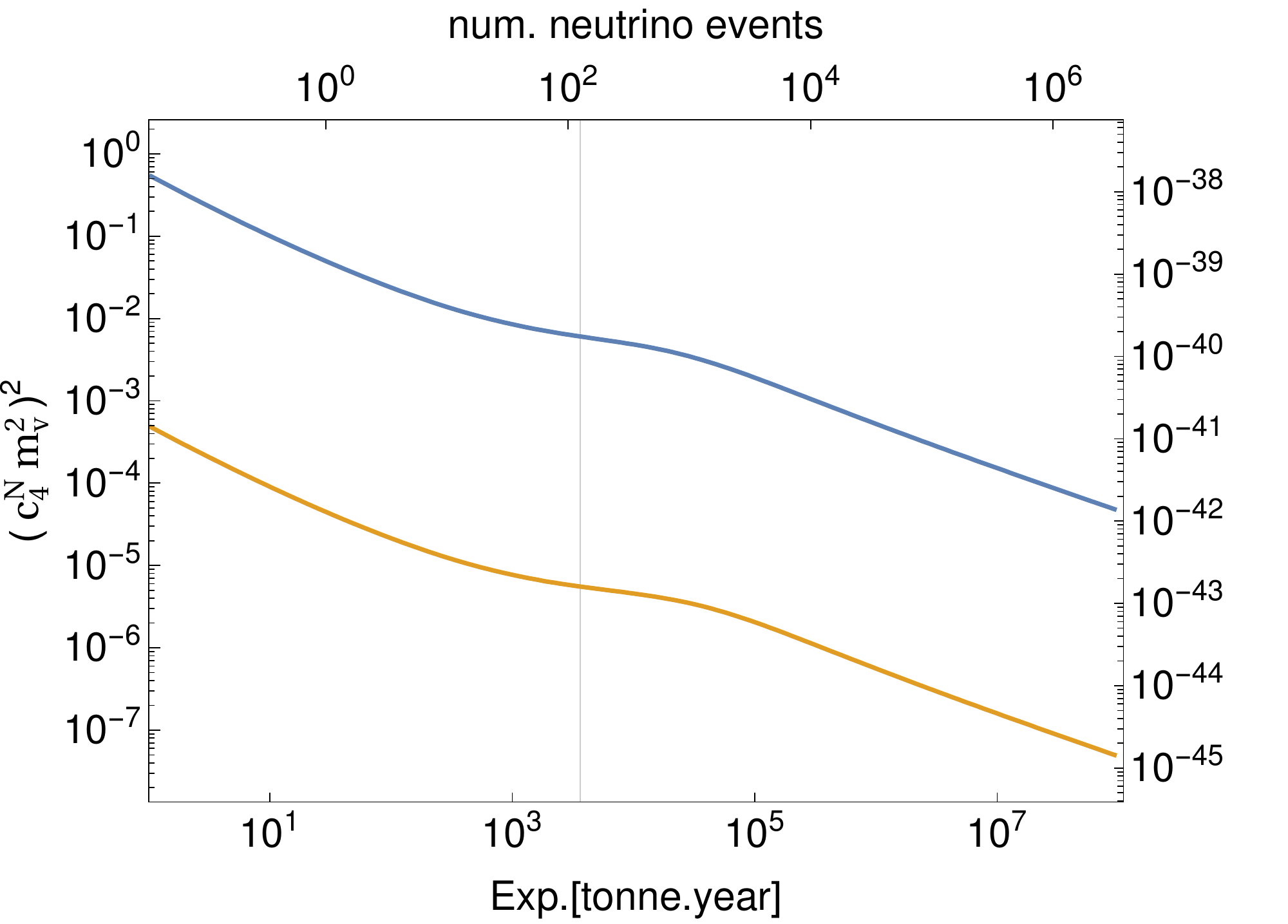} &
\includegraphics[height=4cm]{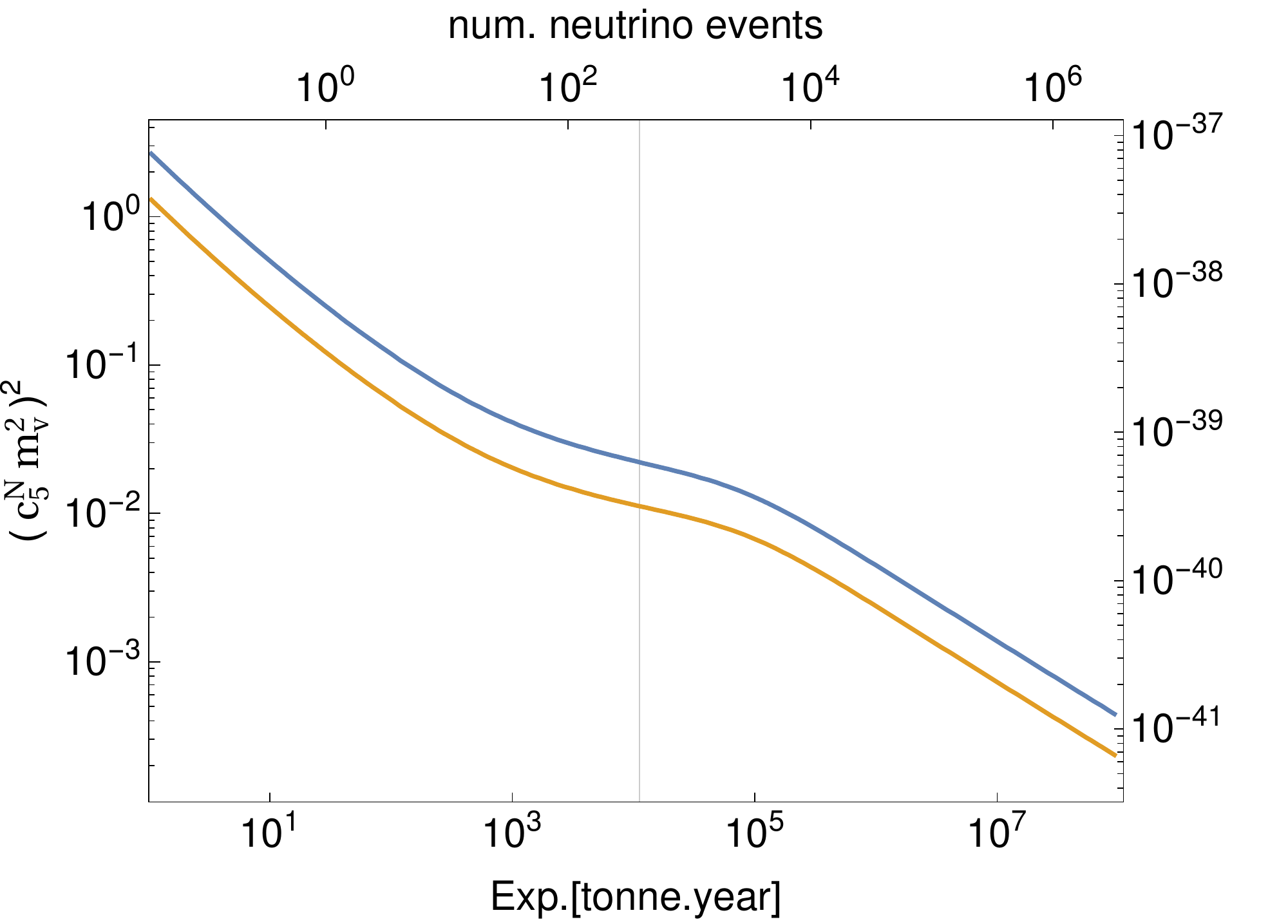} \\
\includegraphics[height=4cm]{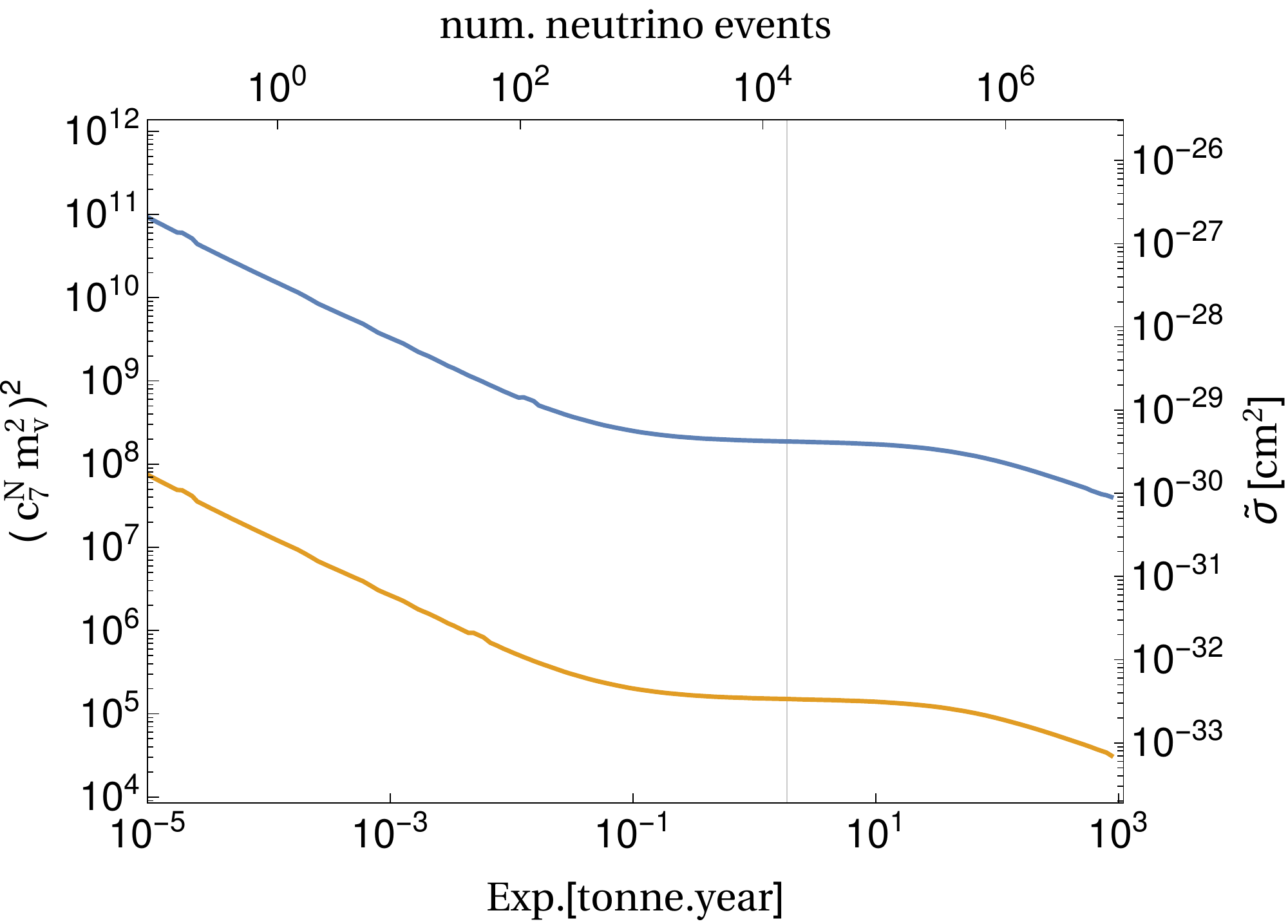}  &
\includegraphics[height=4cm]{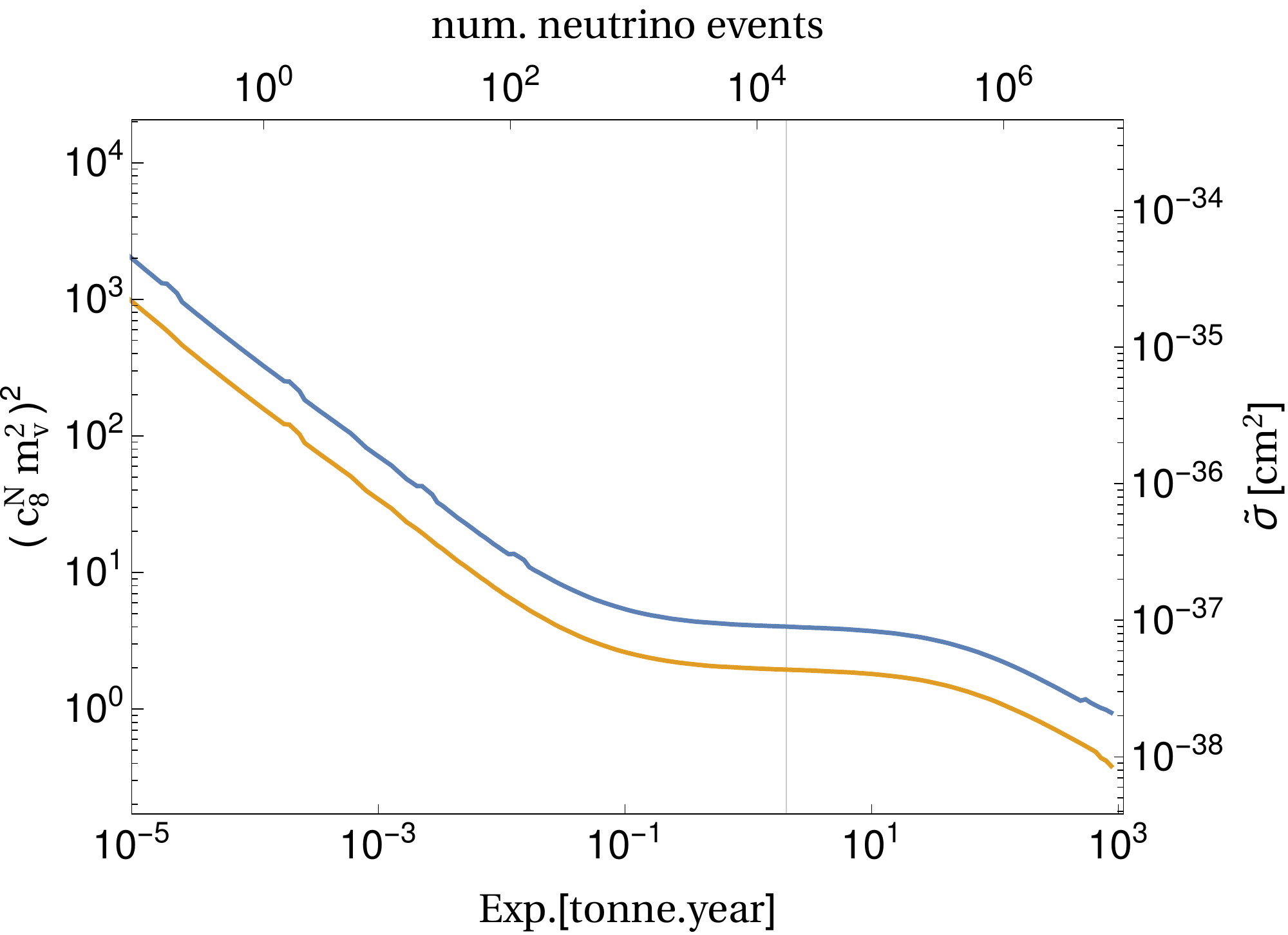}  &
\includegraphics[height=4cm]{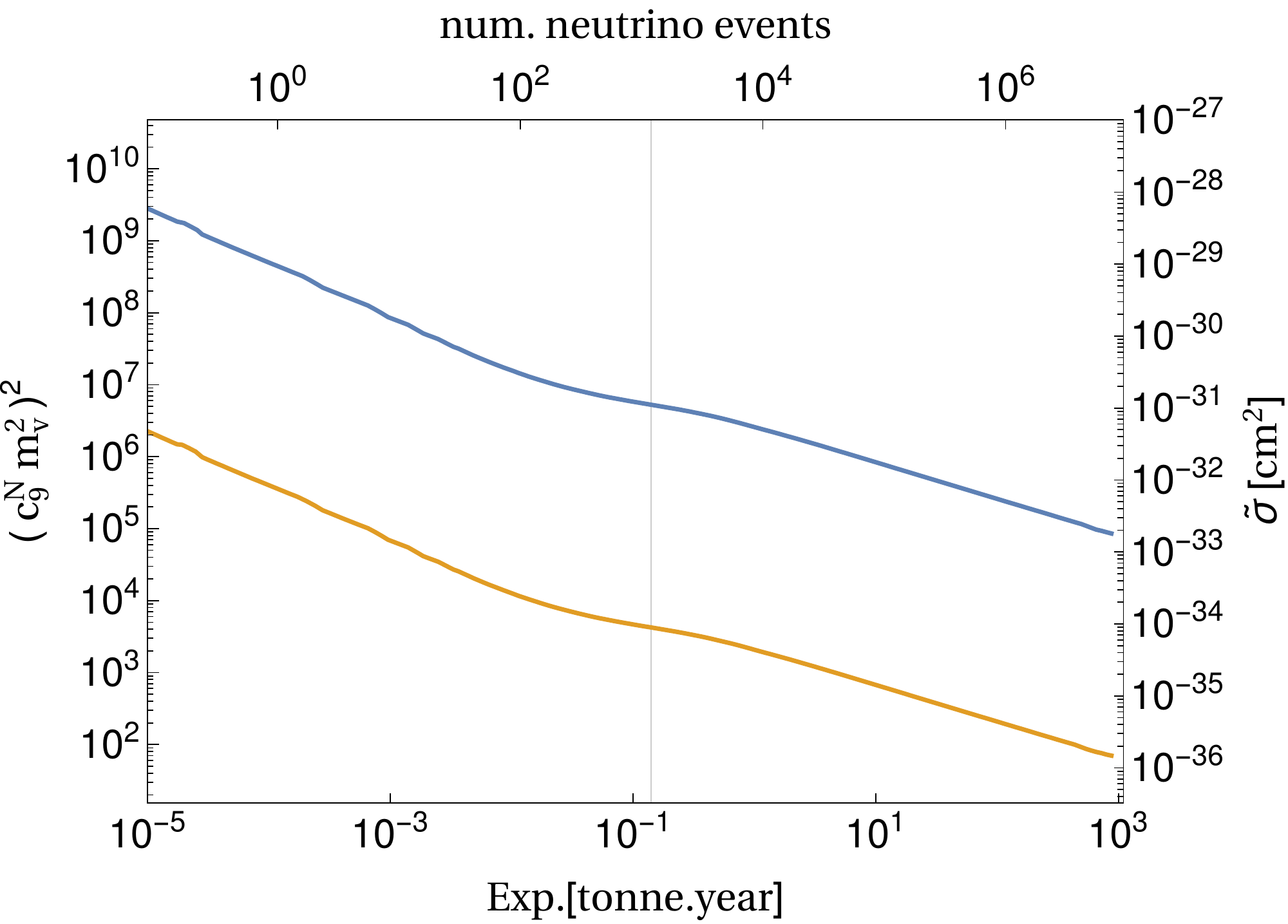} \\
\includegraphics[height=4cm]{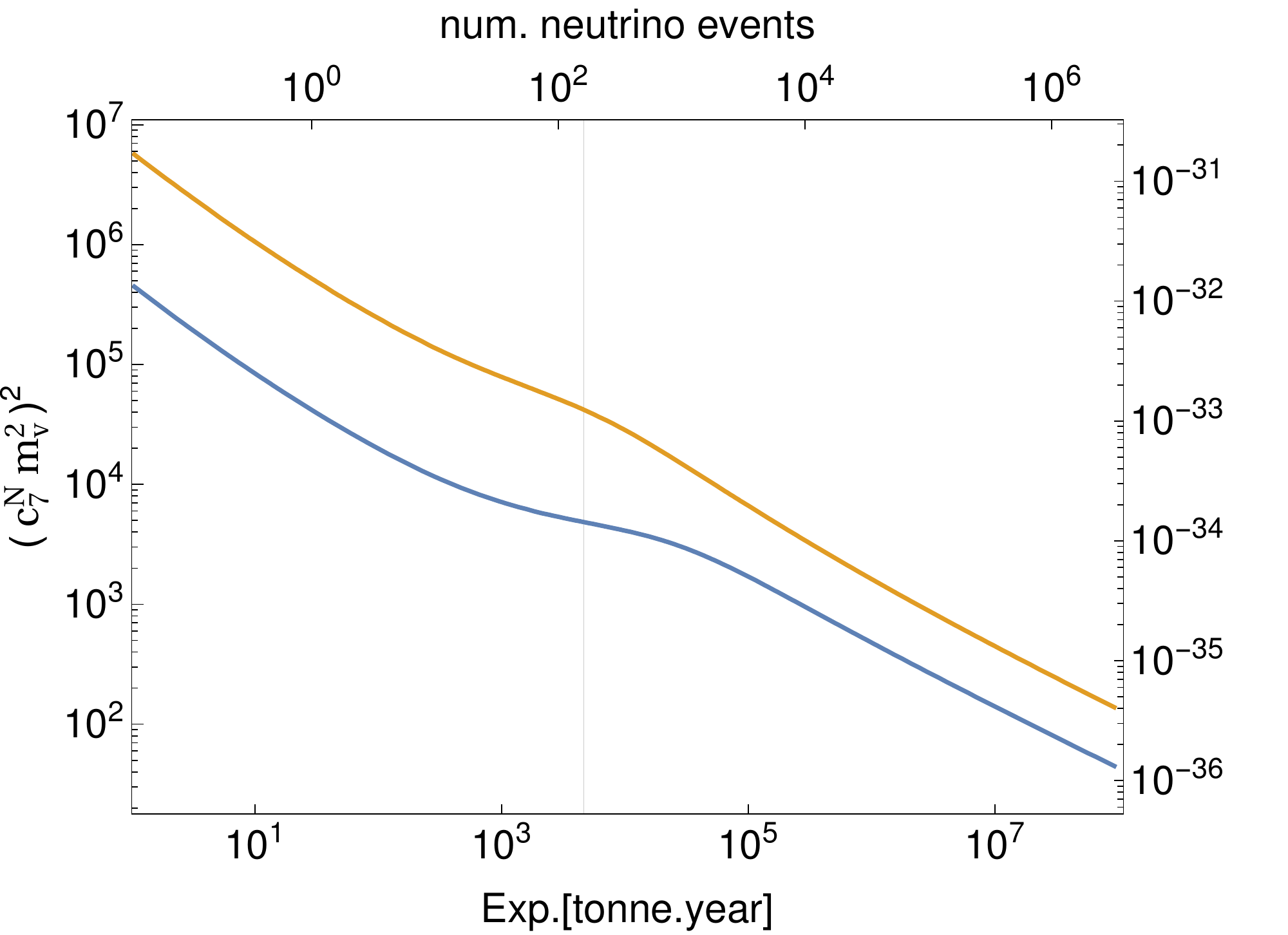} &
\includegraphics[height=4cm]{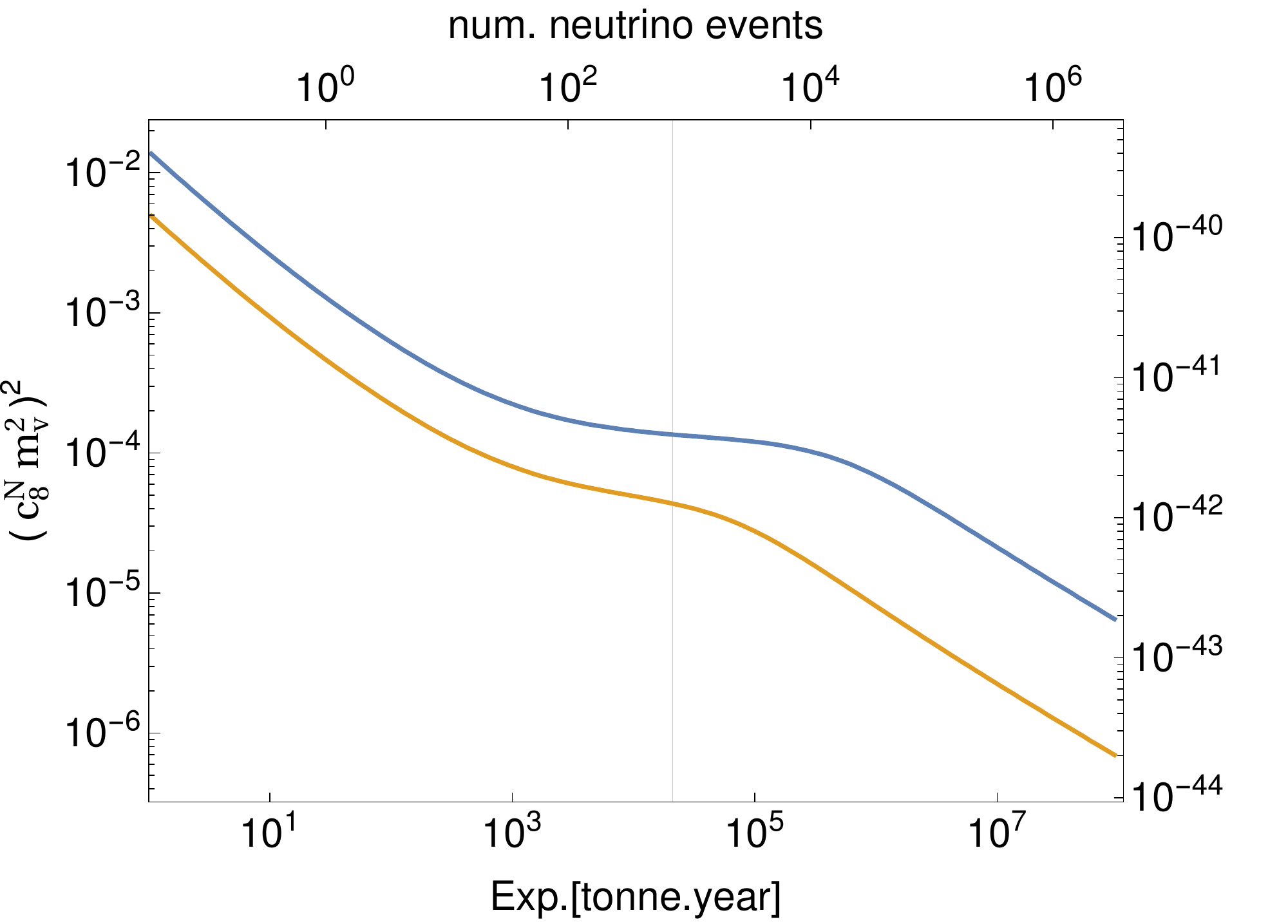} &
\includegraphics[height=4cm]{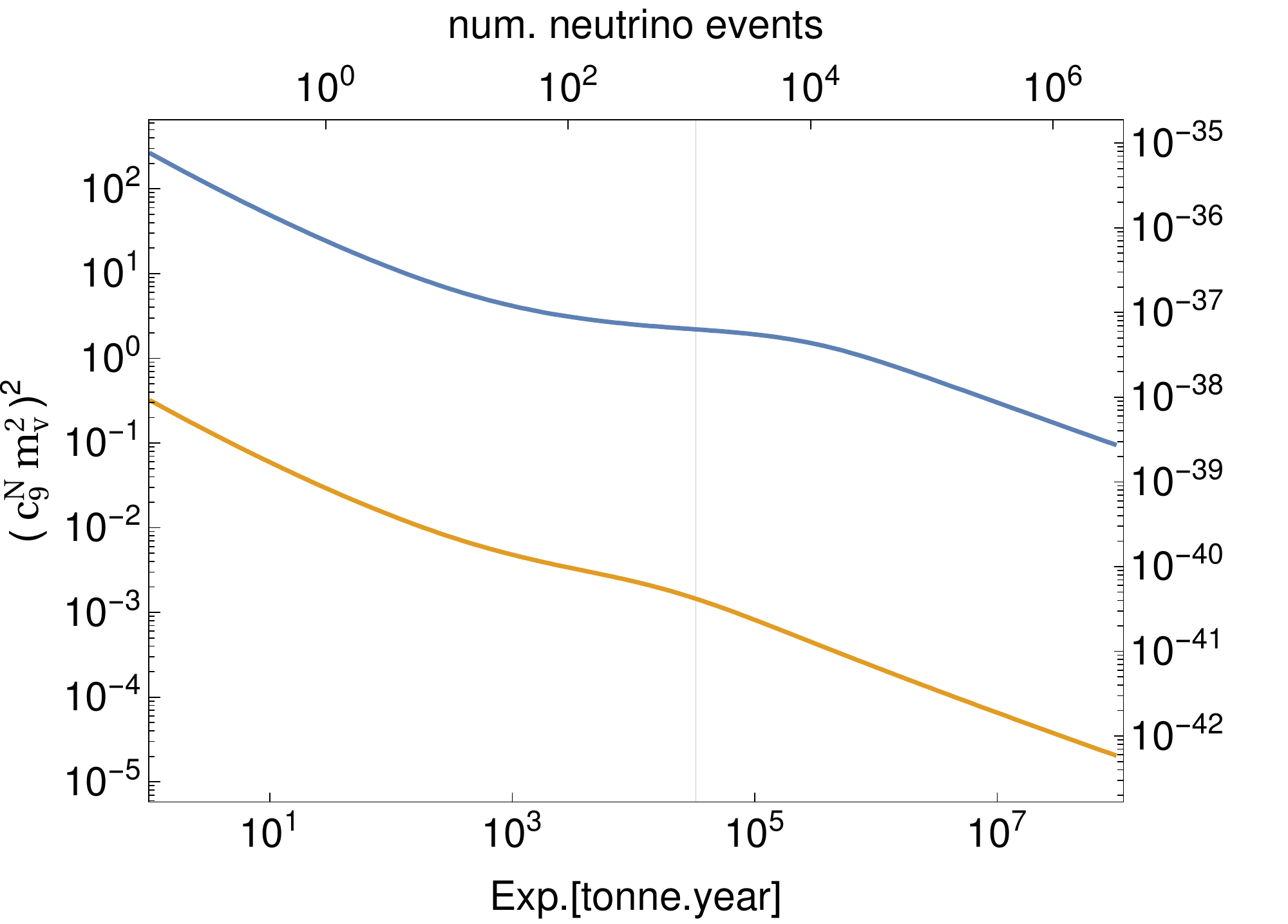} \\
\end{tabular}
\caption{Discovery evolution for the low mass region (first and third rows) and high mass region (second and fourth rows) for operators 3-9. The blue and yellow curves show the limits for proton and neutron scattering, respectively.}
\label{figDiscEvoFull3-9}
\end{figure}

\begin{figure}[ht]
\begin{tabular}{cc}
\includegraphics[height=5cm]{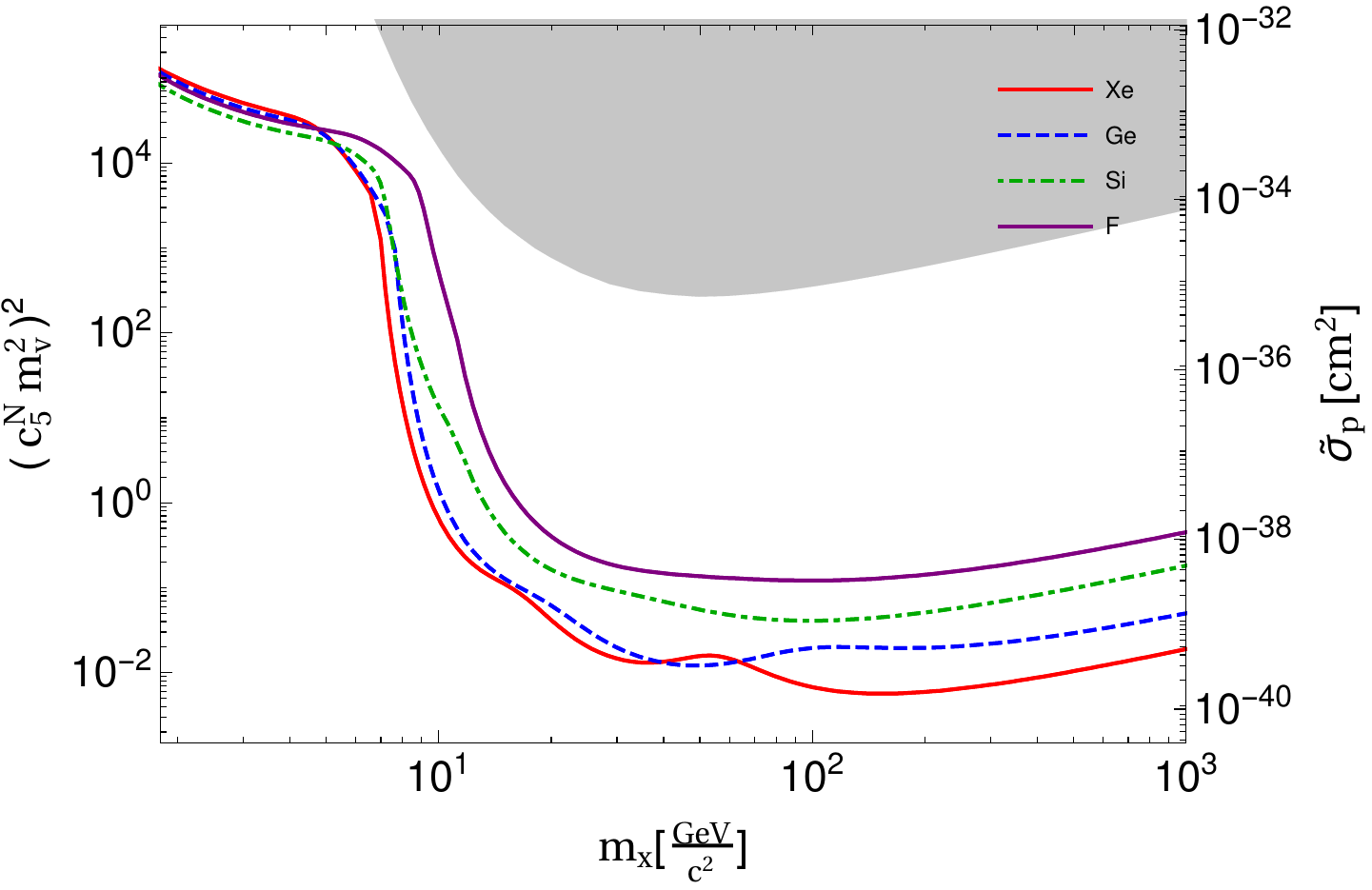} &
\includegraphics[height=5cm]{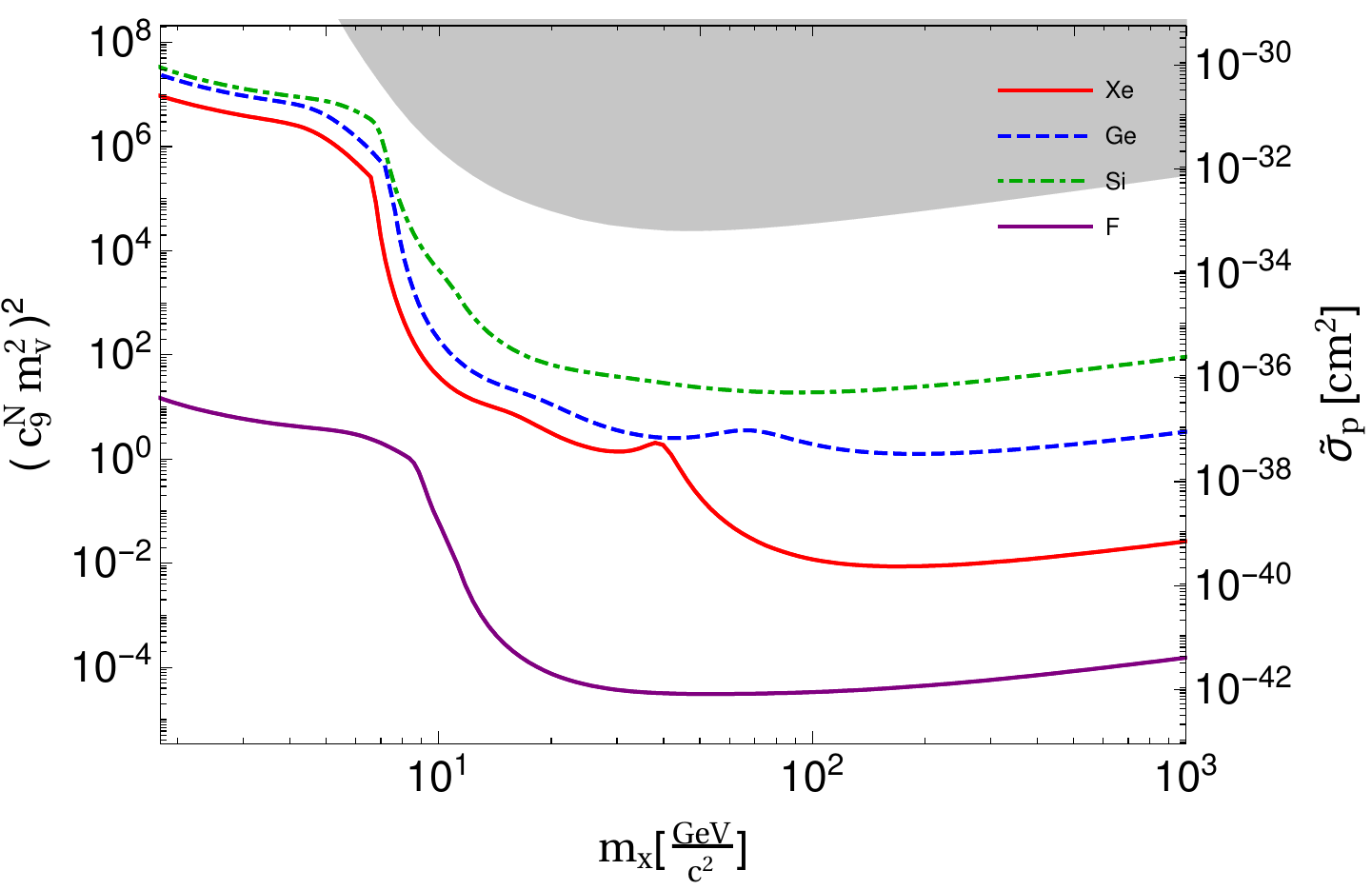} \\
\includegraphics[height=5cm]{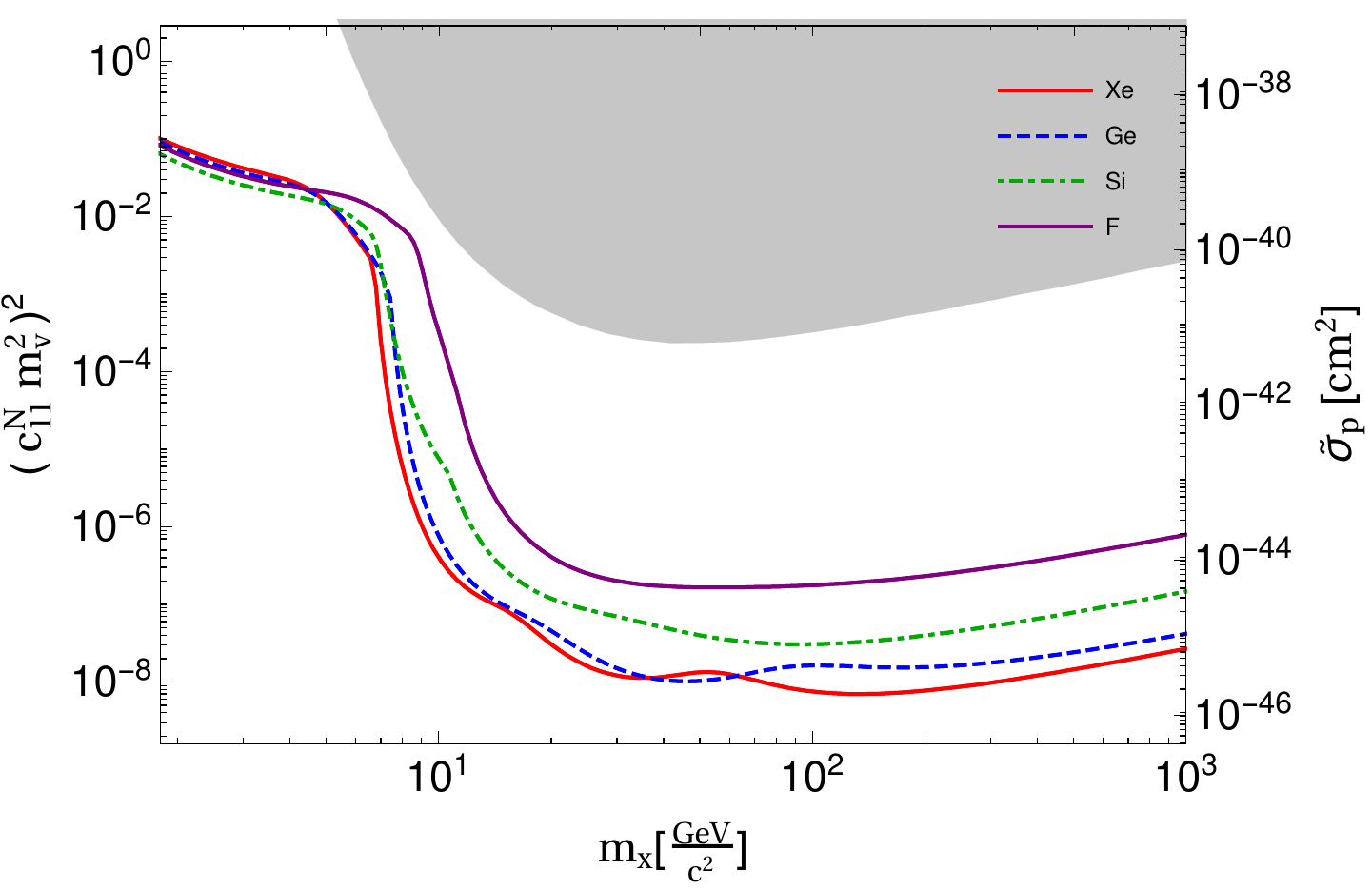} &
\includegraphics[height=5cm]{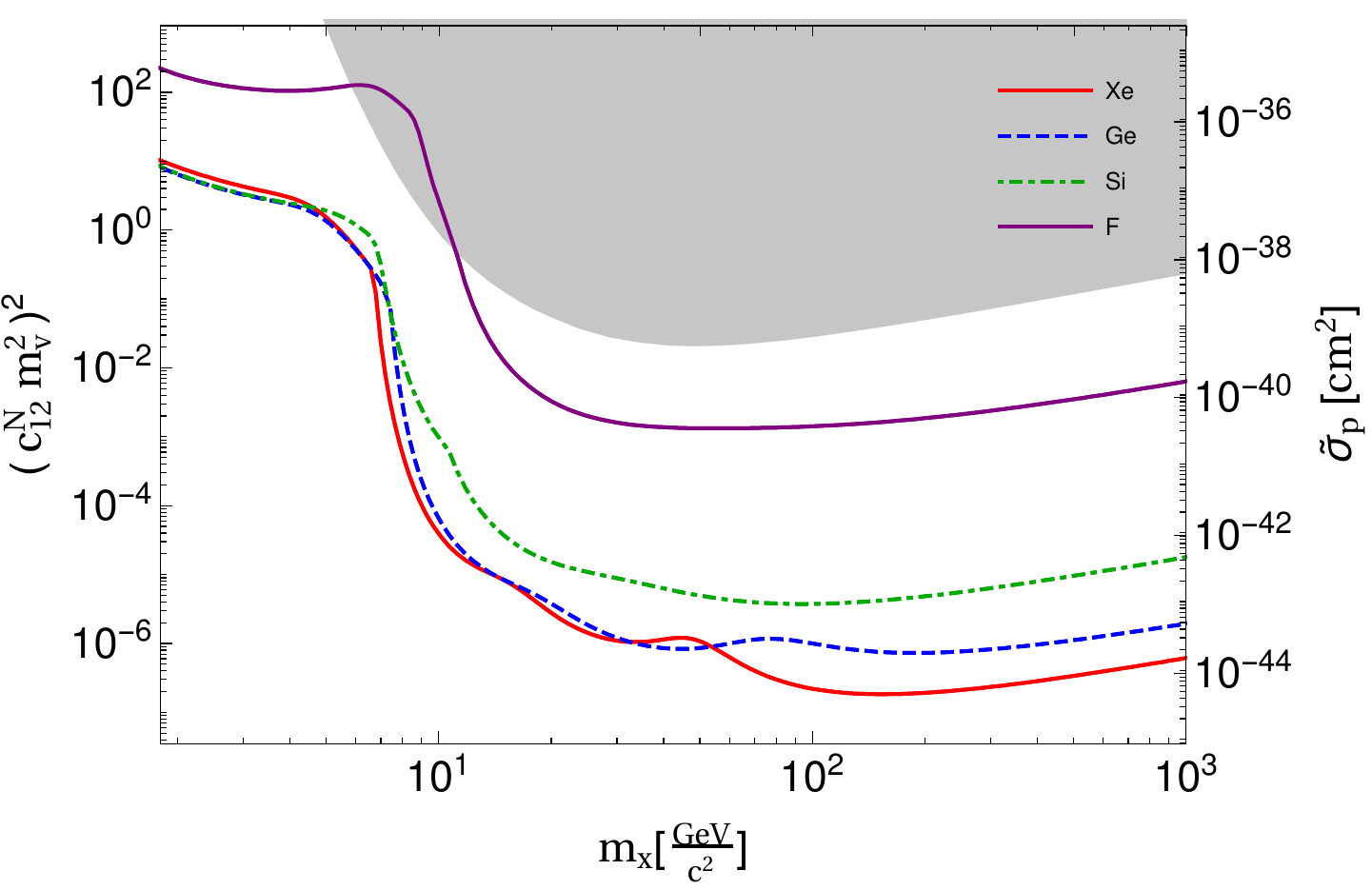} \\
\includegraphics[height=5cm]{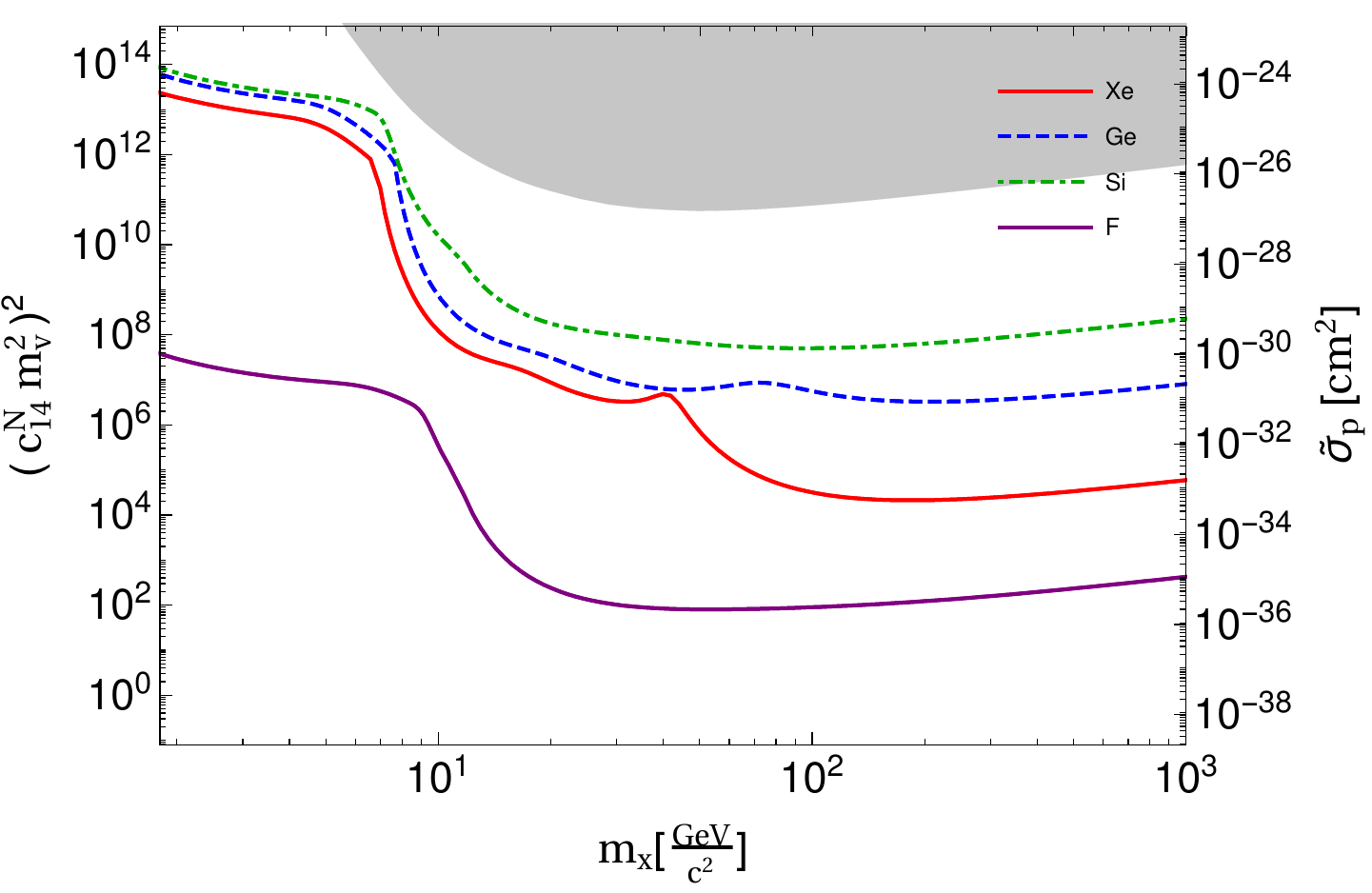} & \\
\end{tabular}
\caption{Discovery limits for group 2 operators interacting with protons.}
\label{figFloors2p}
\end{figure}

\begin{figure}[ht]
\begin{tabular}{cc}
\includegraphics[height=5cm]{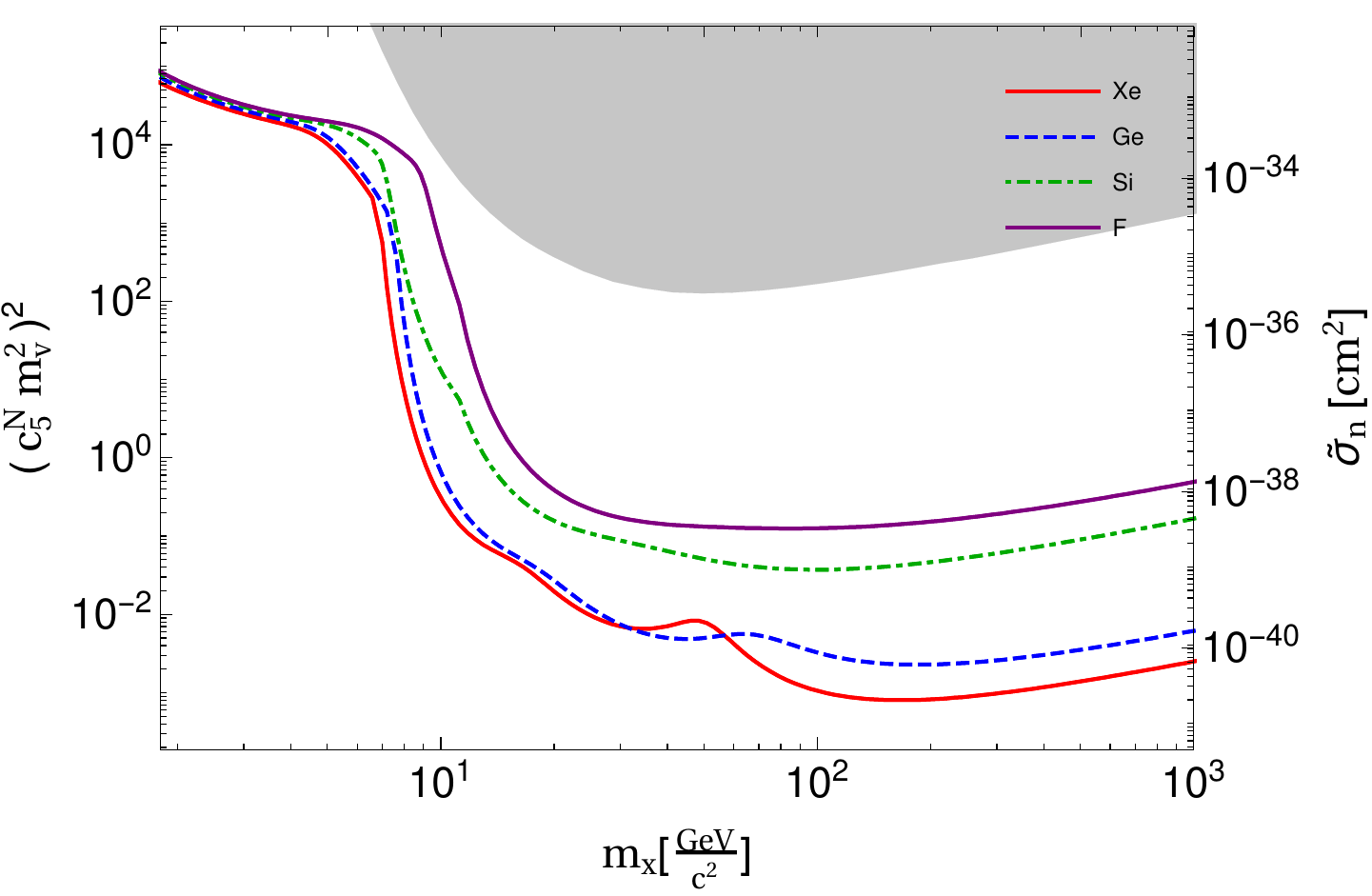} &
\includegraphics[height=5cm]{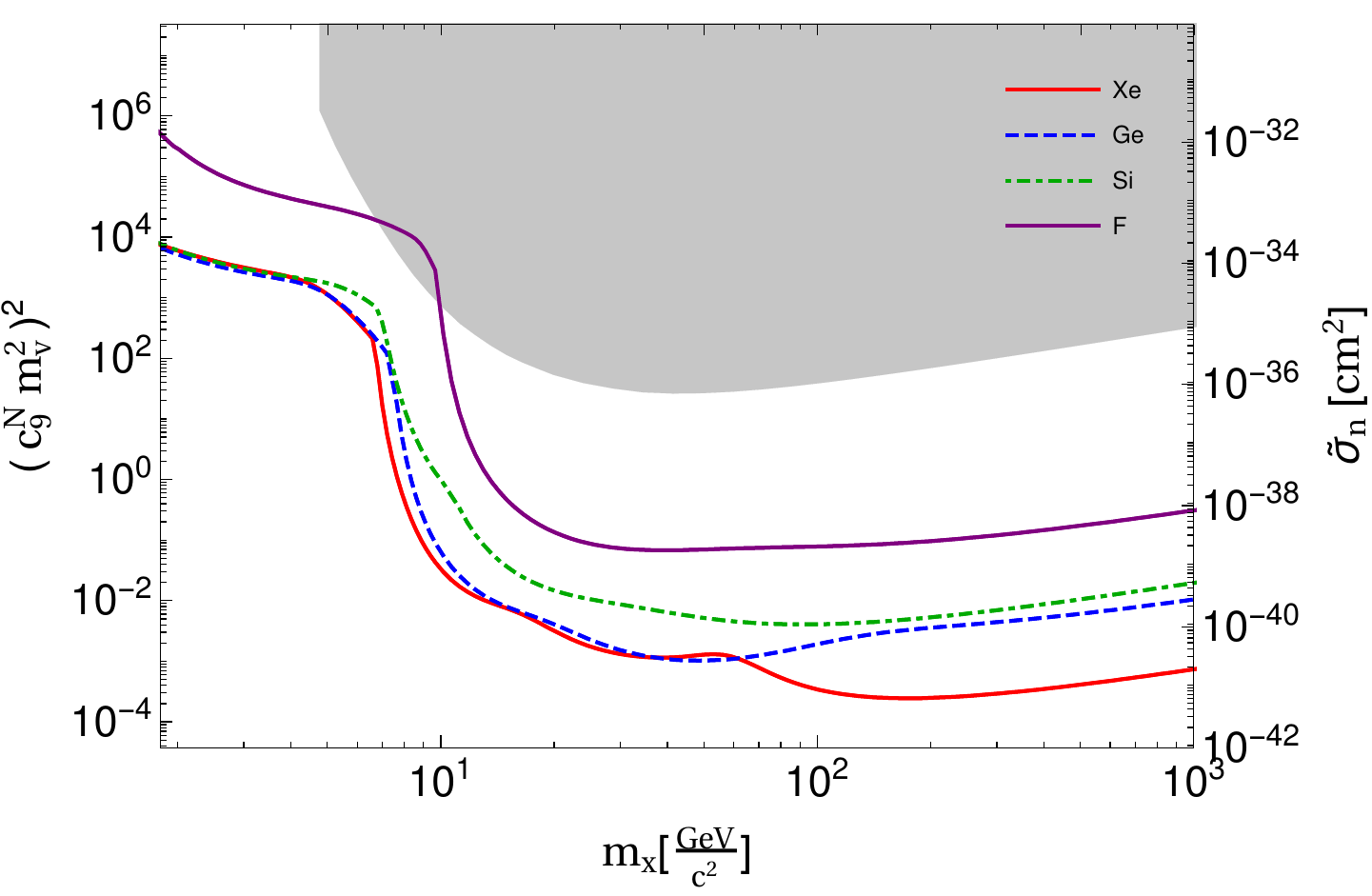} \\
\includegraphics[height=5cm]{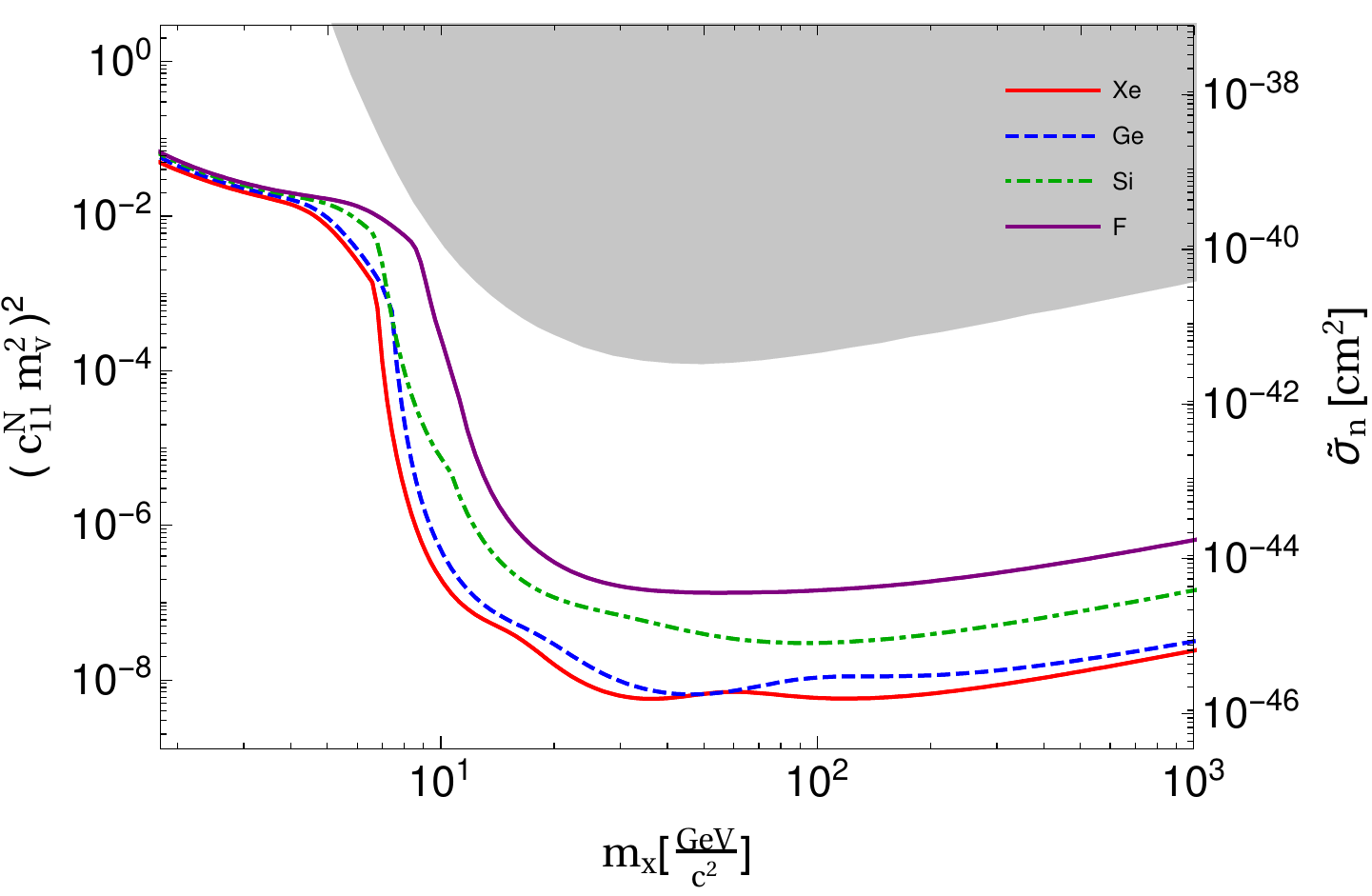} &
\includegraphics[height=5cm]{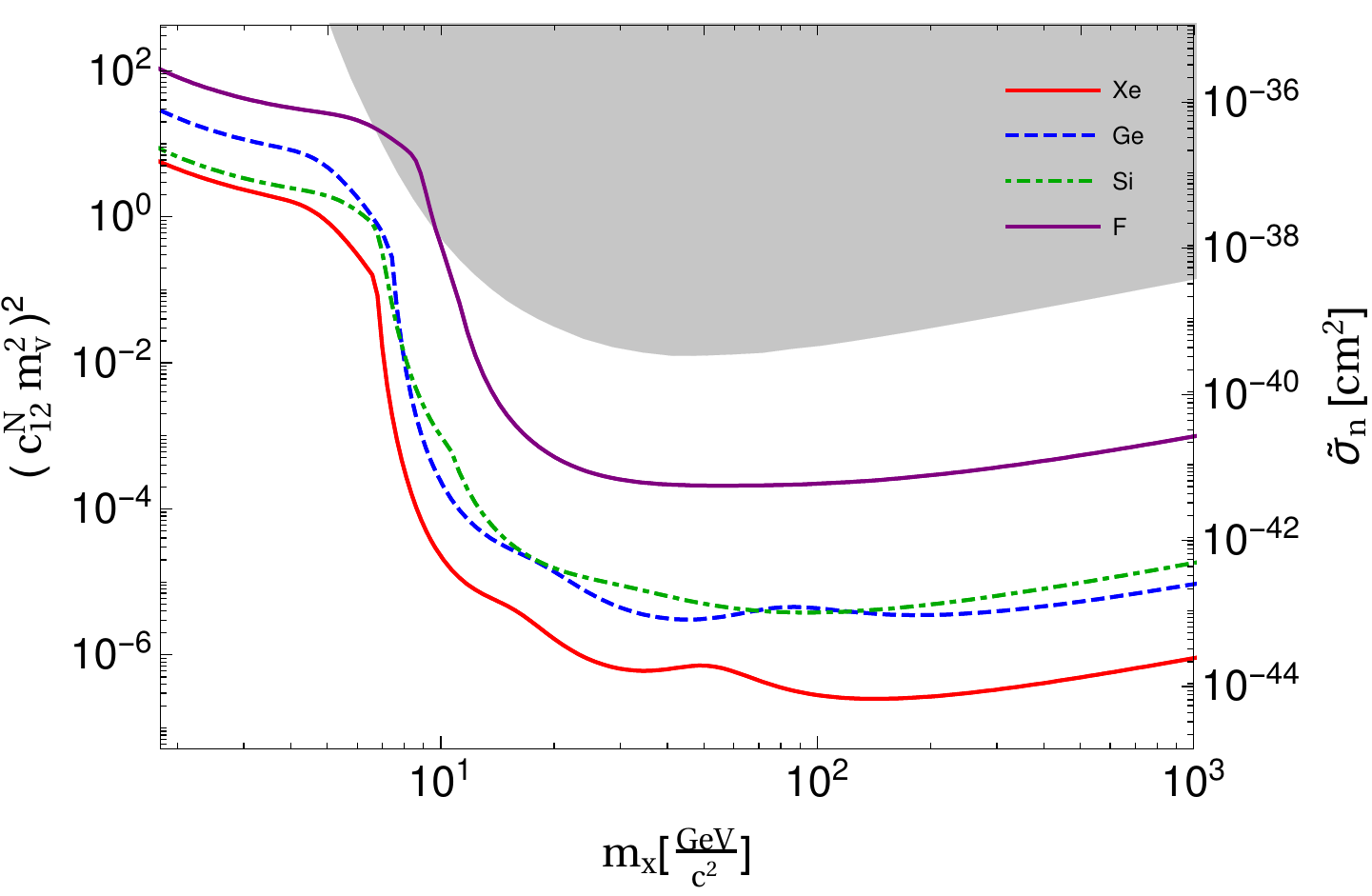} \\
\includegraphics[height=5cm]{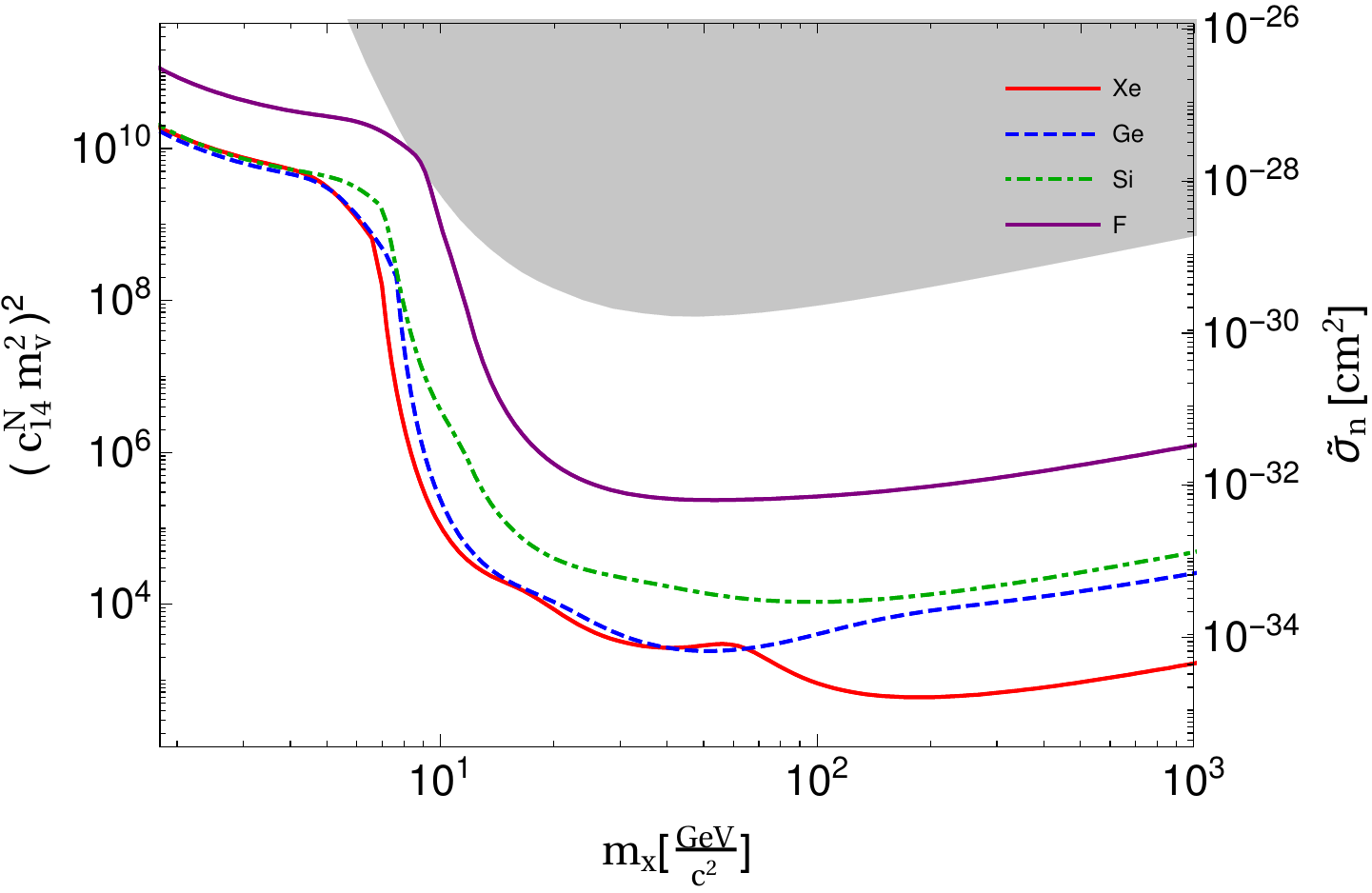} & \\
\end{tabular}
\caption{Discovery limits for group 2 operators interacting with neutrons.}
\label{figFloors2n}
\end{figure}

\begin{figure}[ht]
\begin{tabular}{cc}
\includegraphics[height=5cm]{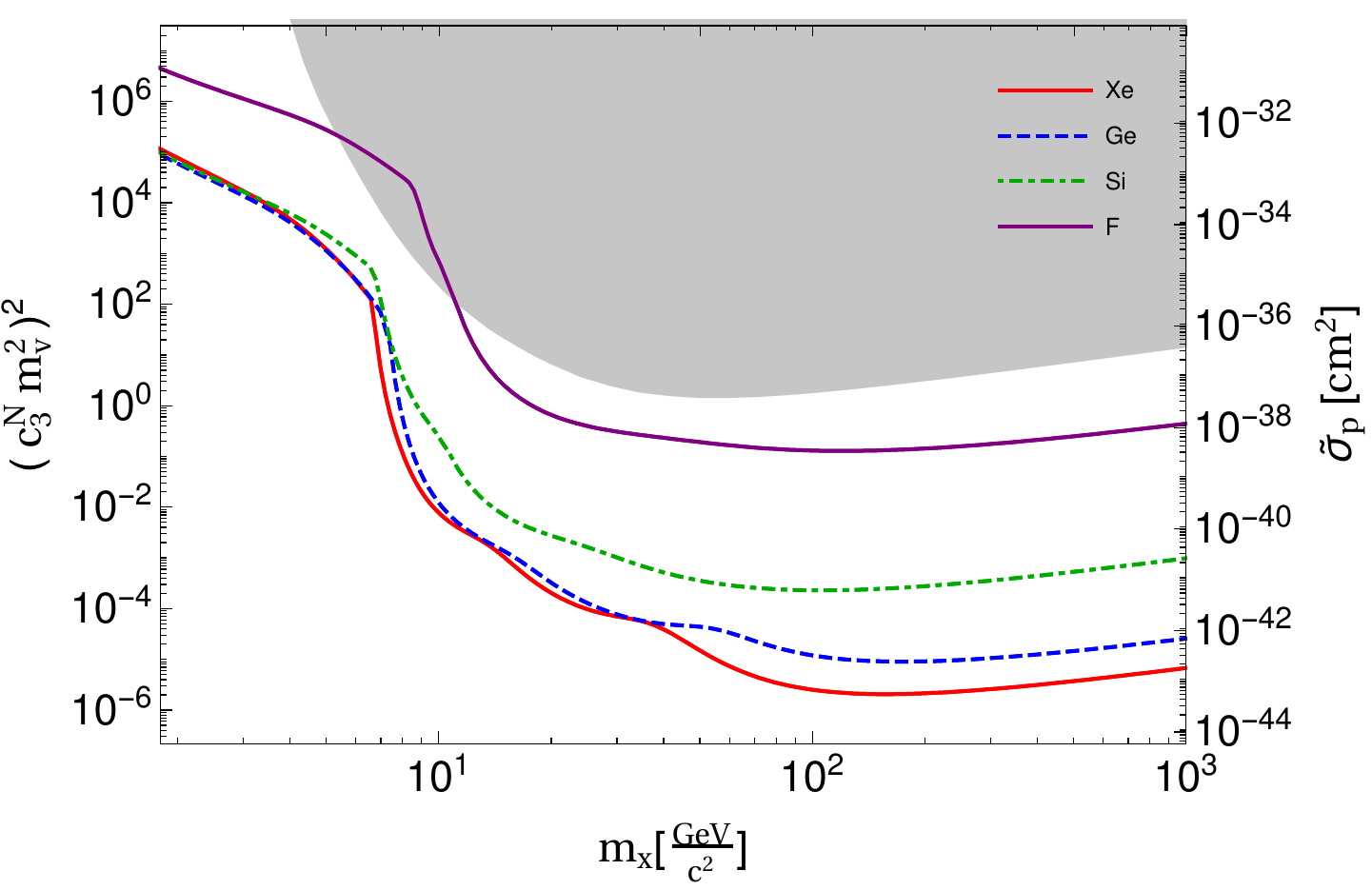} &
\includegraphics[height=5cm]{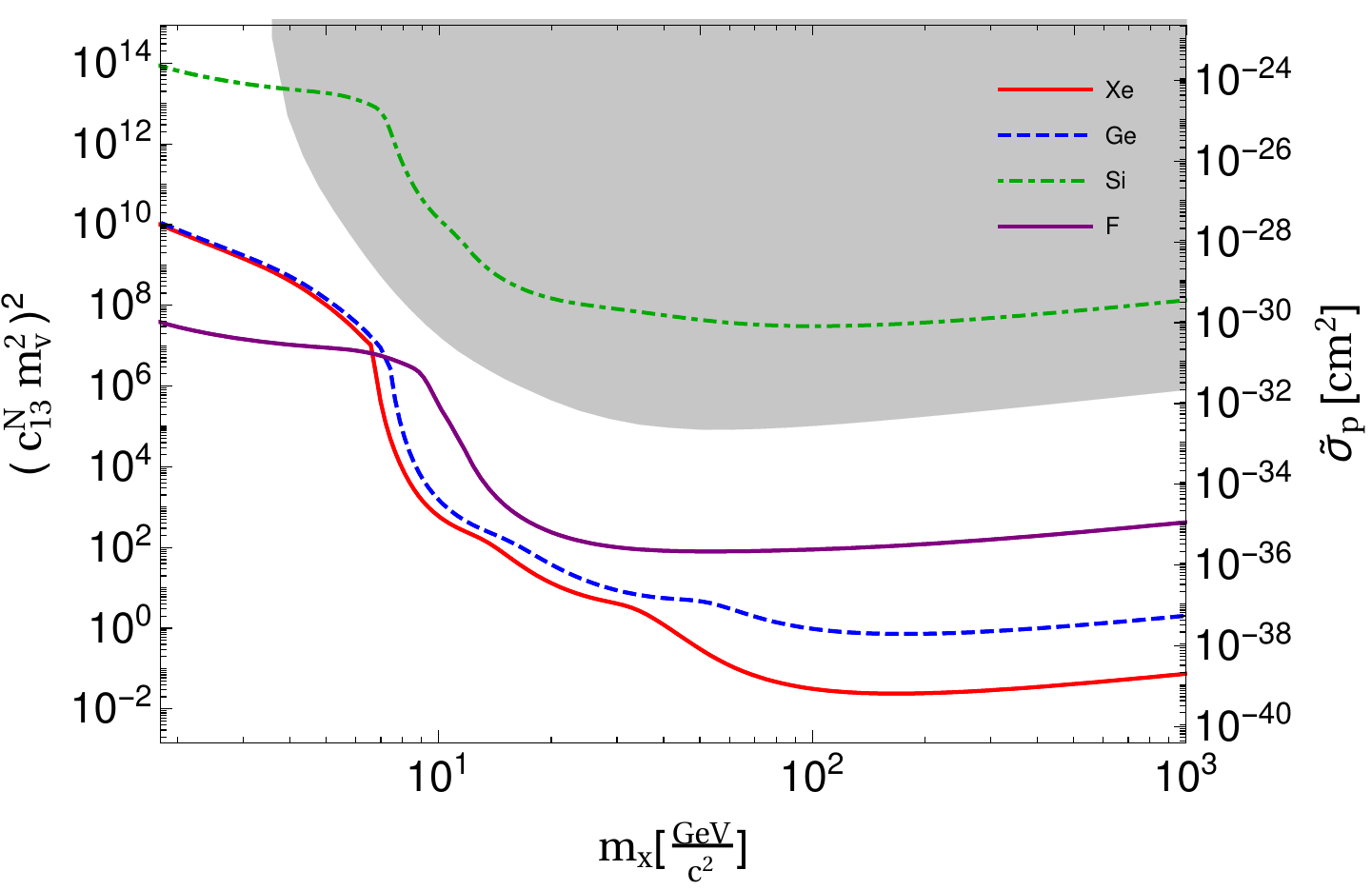} \\
\includegraphics[height=5cm]{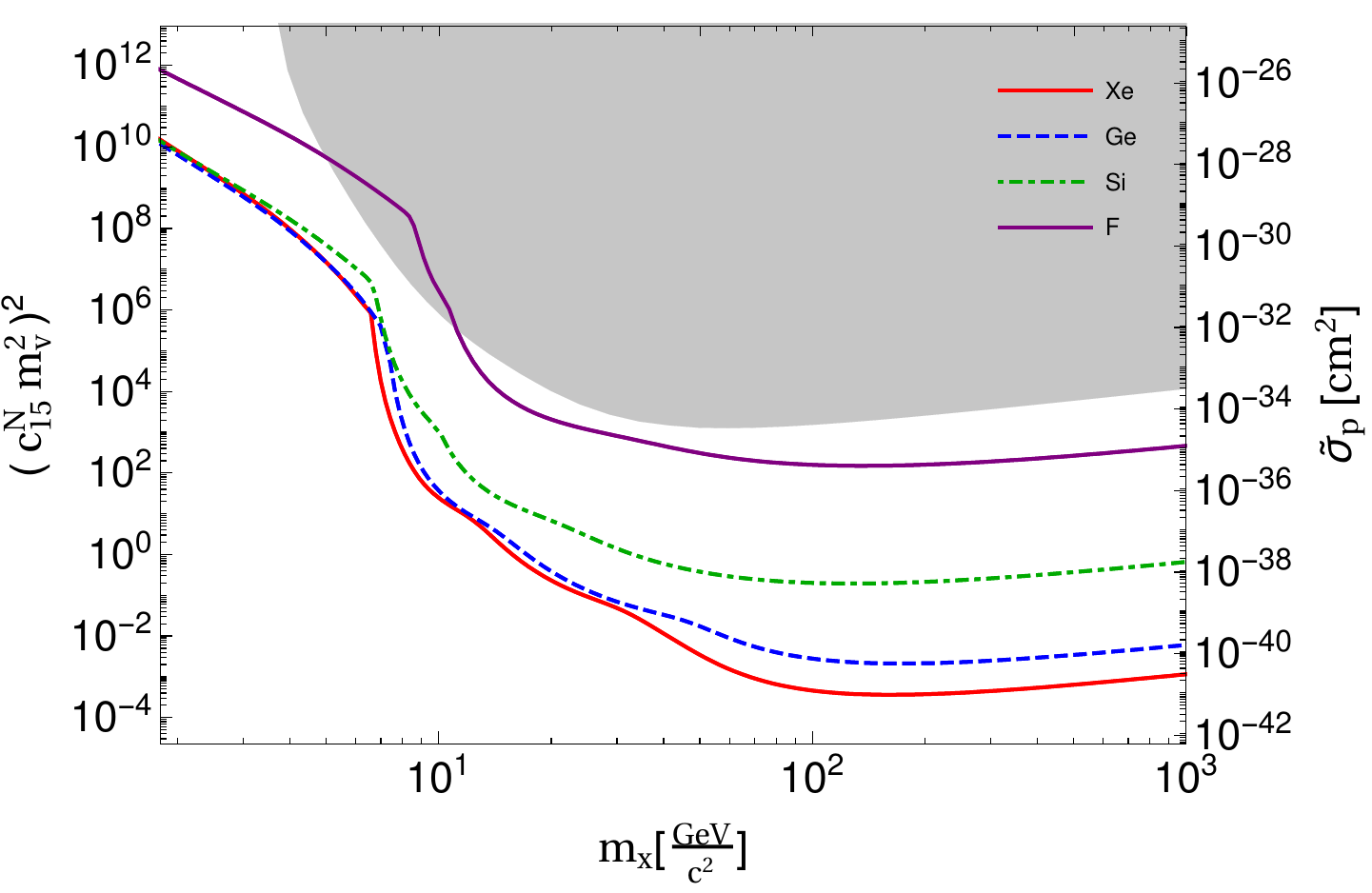} & \\
\end{tabular}
\caption{Discovery limits for group 3 operators interacting with protons.}
\label{figFloors3p}
\end{figure}

\begin{figure}[ht]
\begin{tabular}{cc}
\includegraphics[height=5cm]{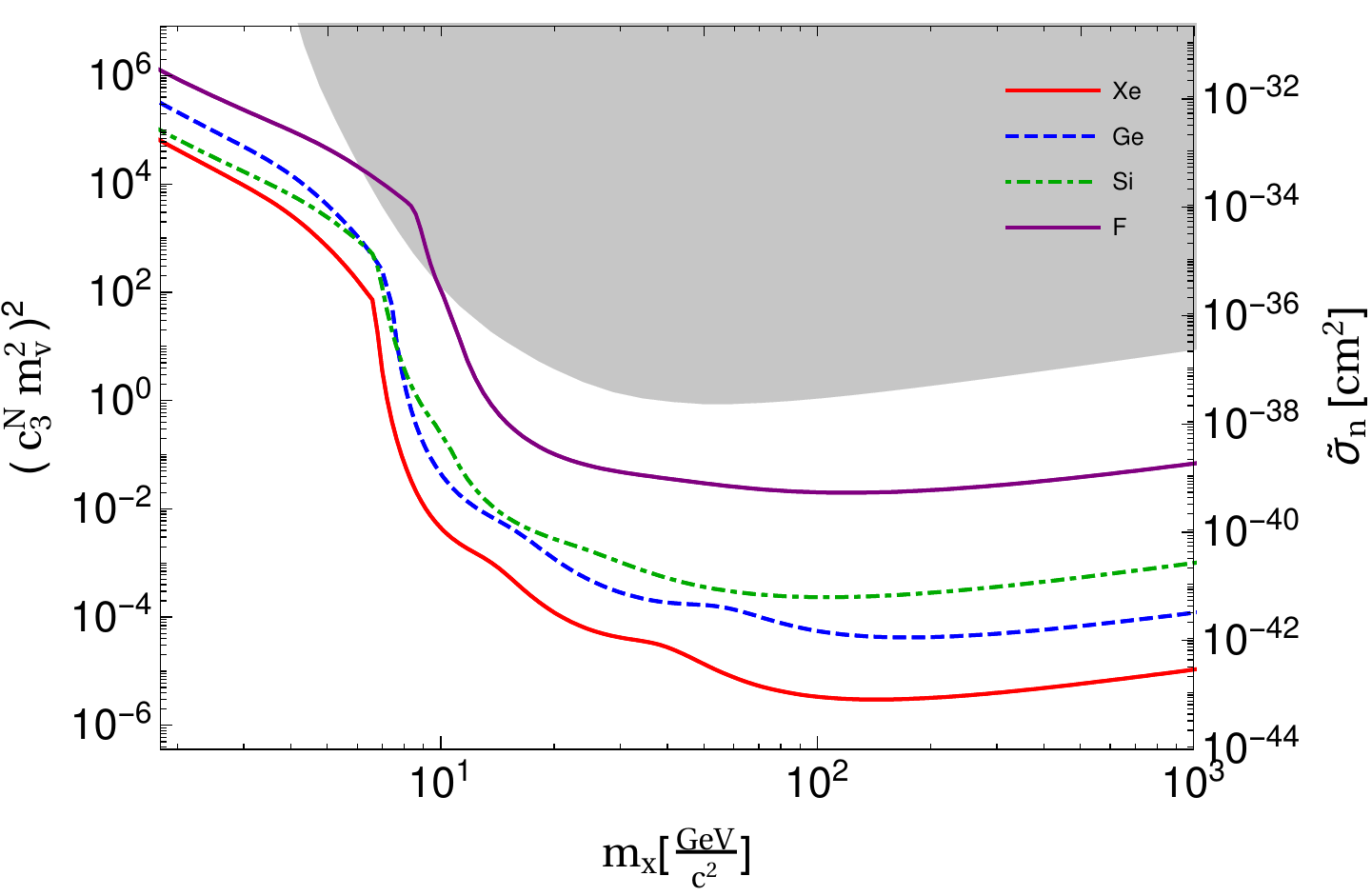} &
\includegraphics[height=5cm]{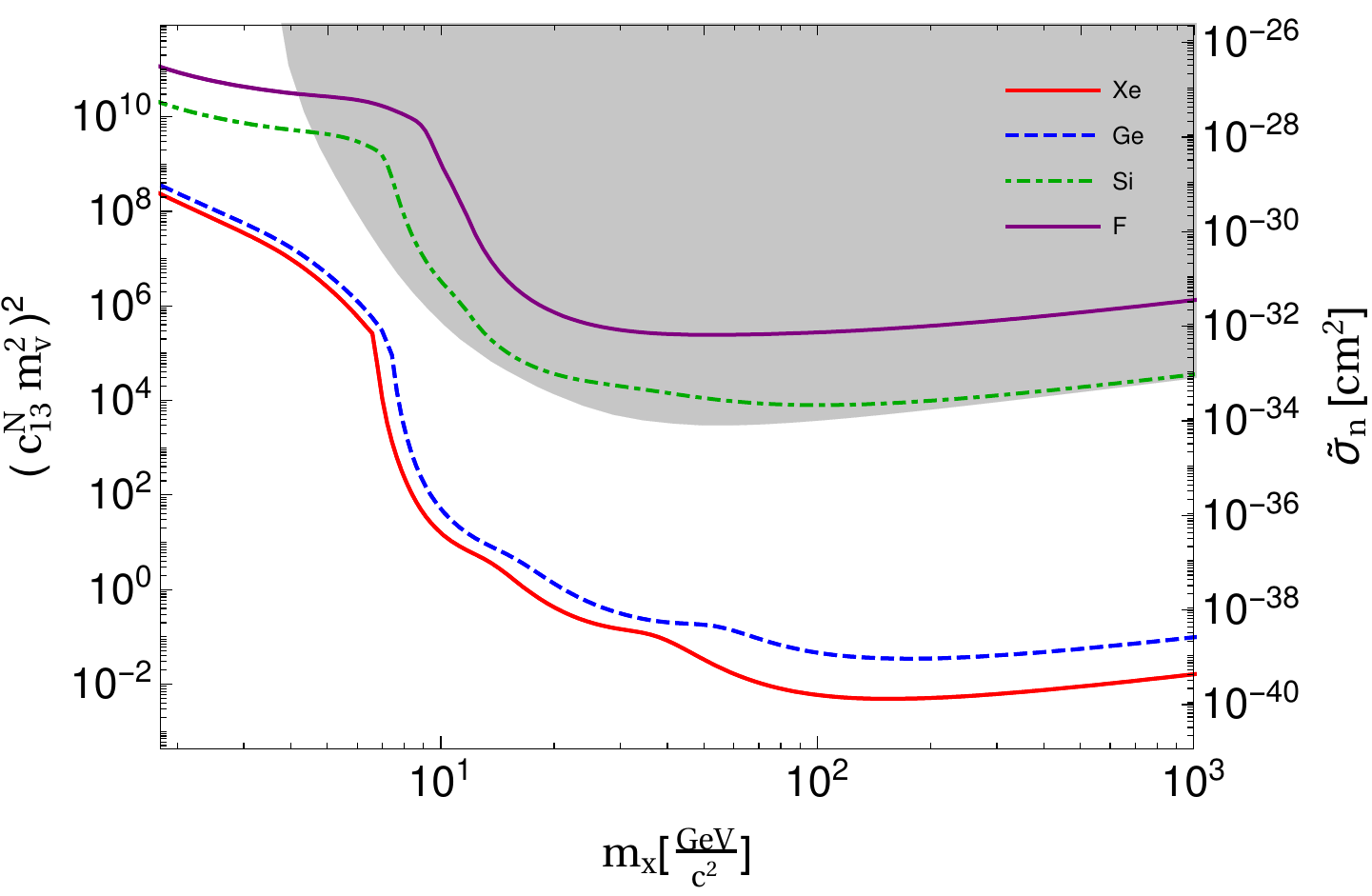} \\
\includegraphics[height=5cm]{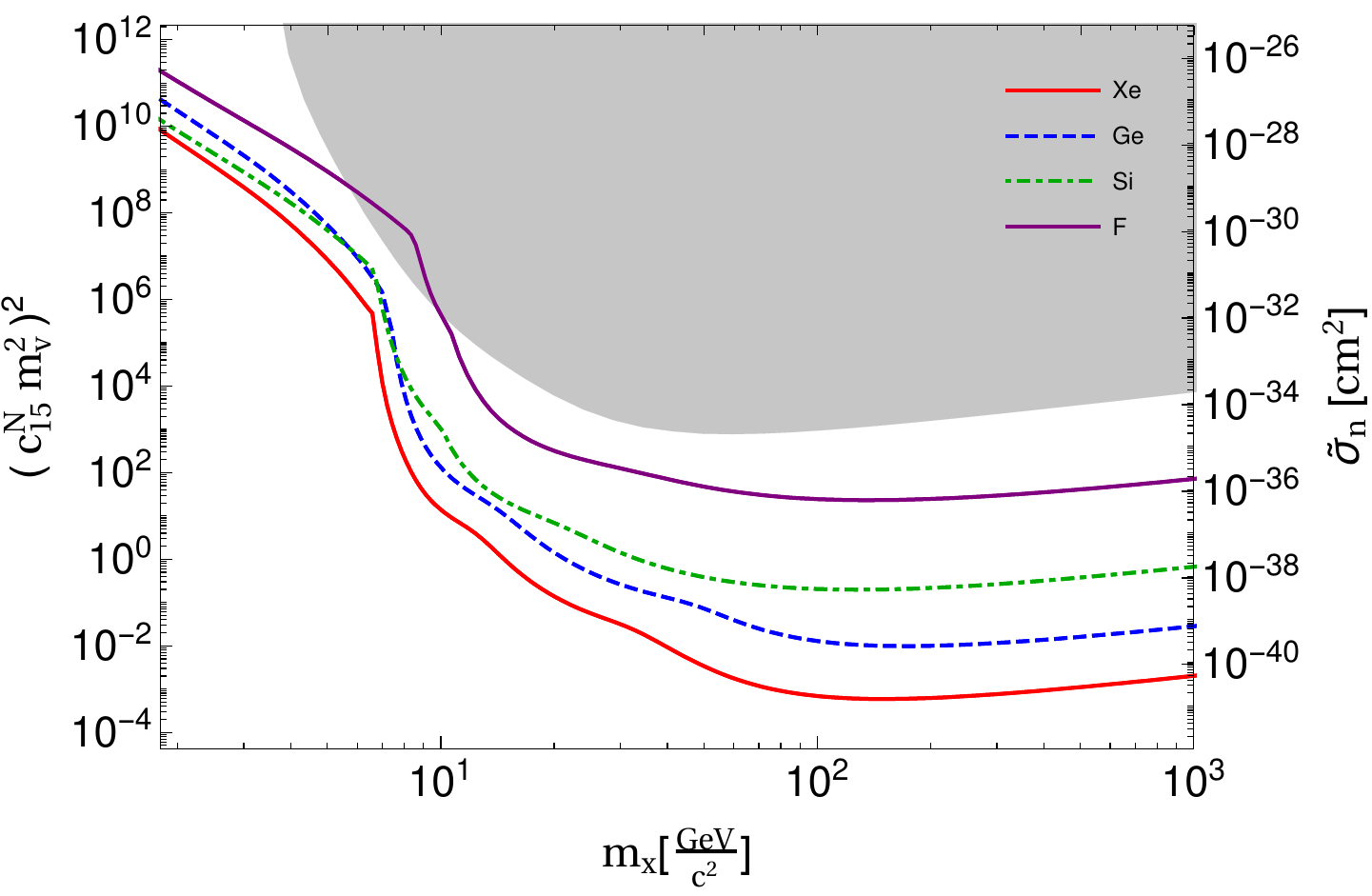} & \\
\end{tabular}
\caption{Discovery limits for group 3 operators interacting with neutrons.}
\label{figFloors3n}
\end{figure}

 In this appendix, we show best fitting masses and discovery limits for the operators that were not shown in the main text. These figures motivate the operator groupings that were presented above. Figure~\ref{fig:8BbestFitMasses_all} show the best fit masses to the $^8$B neutrino rate for the four targets. Figure~\ref{figDiscEvoFull3-9} shows the discovery evolution for the low mass and high mass WIMP region for operators ${\cal O}_3 - {\cal O}_9$, and Figure~\ref{figDiscEvoFull3-9} shows the discovery evolution for the low mass and high mass WIMP region for operators ${\cal O}_{11} - {\cal O}_{15}$. Figure~\ref{figFloors2p} shows the discovery limits for group 2 operators interacting with protons, and Figure~\ref{figFloors2n} shows the discovery limits for group 2 operators interacting with neutrons. Finally, Figure~\ref{figFloors3p} shows the discovery limits for group 2 operators interacting with protons, and Figure~\ref{figFloors3n} shows the discovery limits for group 2 operators interacting with neutrons.

\begin{figure}[ht]
\begin{tabular}{ccc}
\includegraphics[height=4cm]{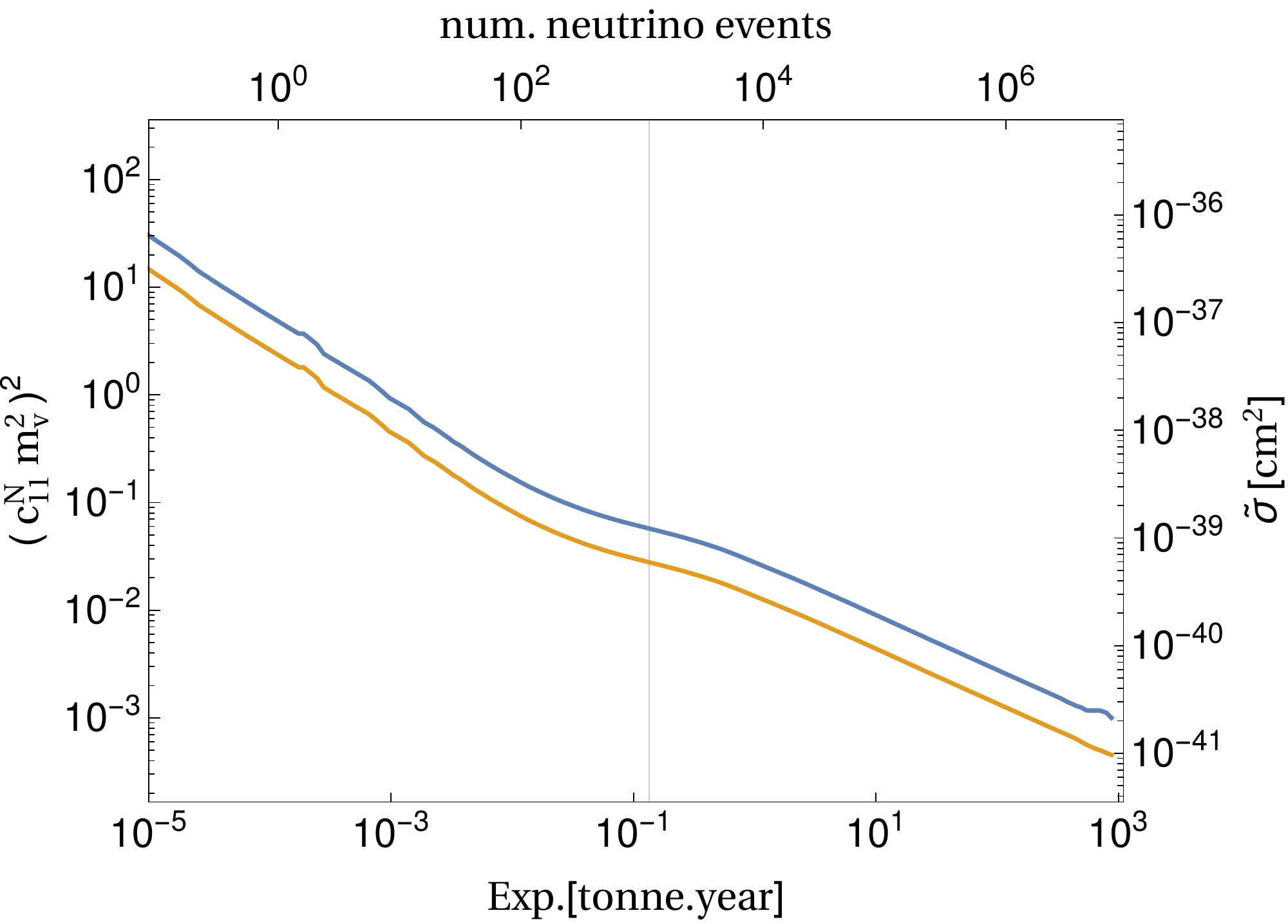}  &
\includegraphics[height=4cm]{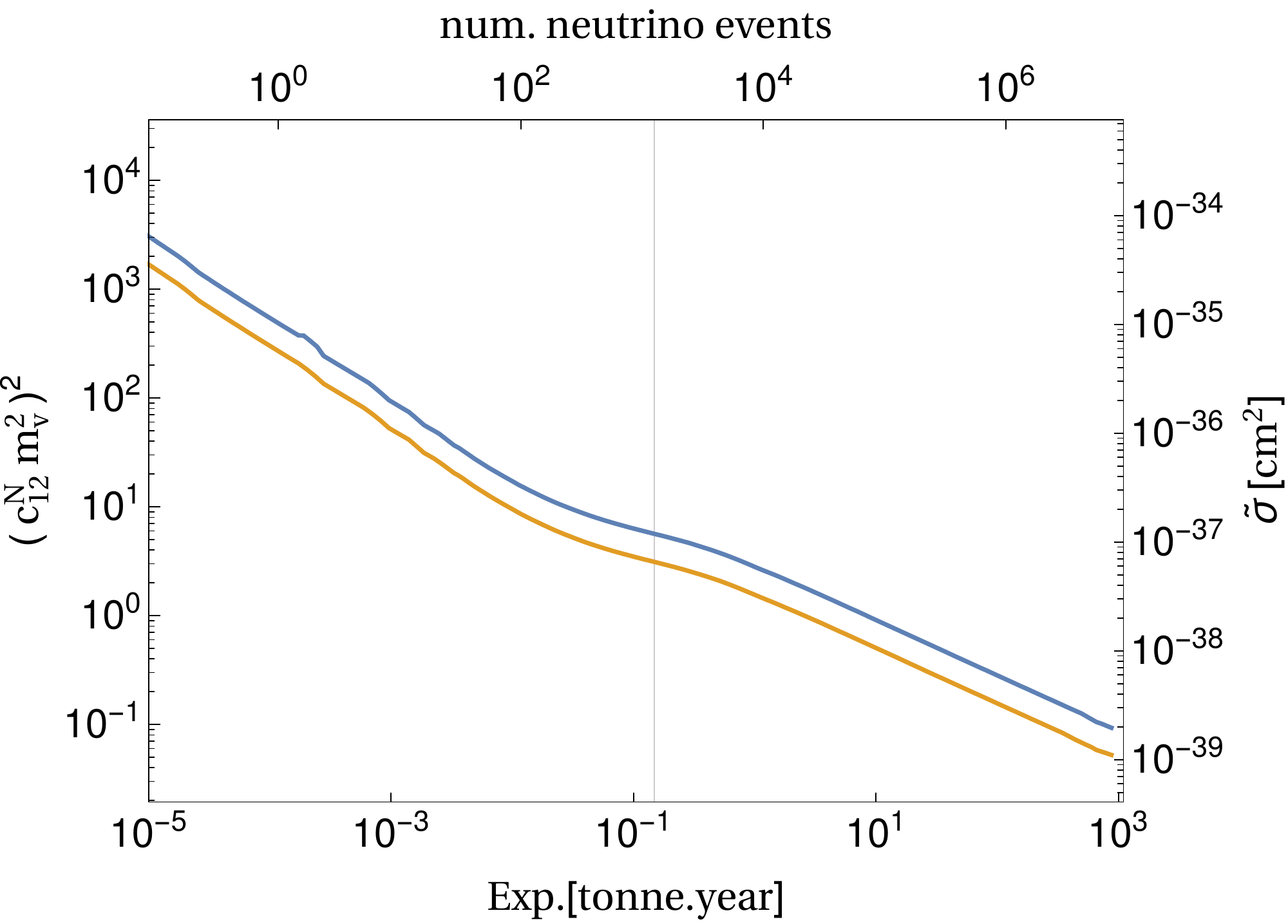}  &
\includegraphics[height=4cm]{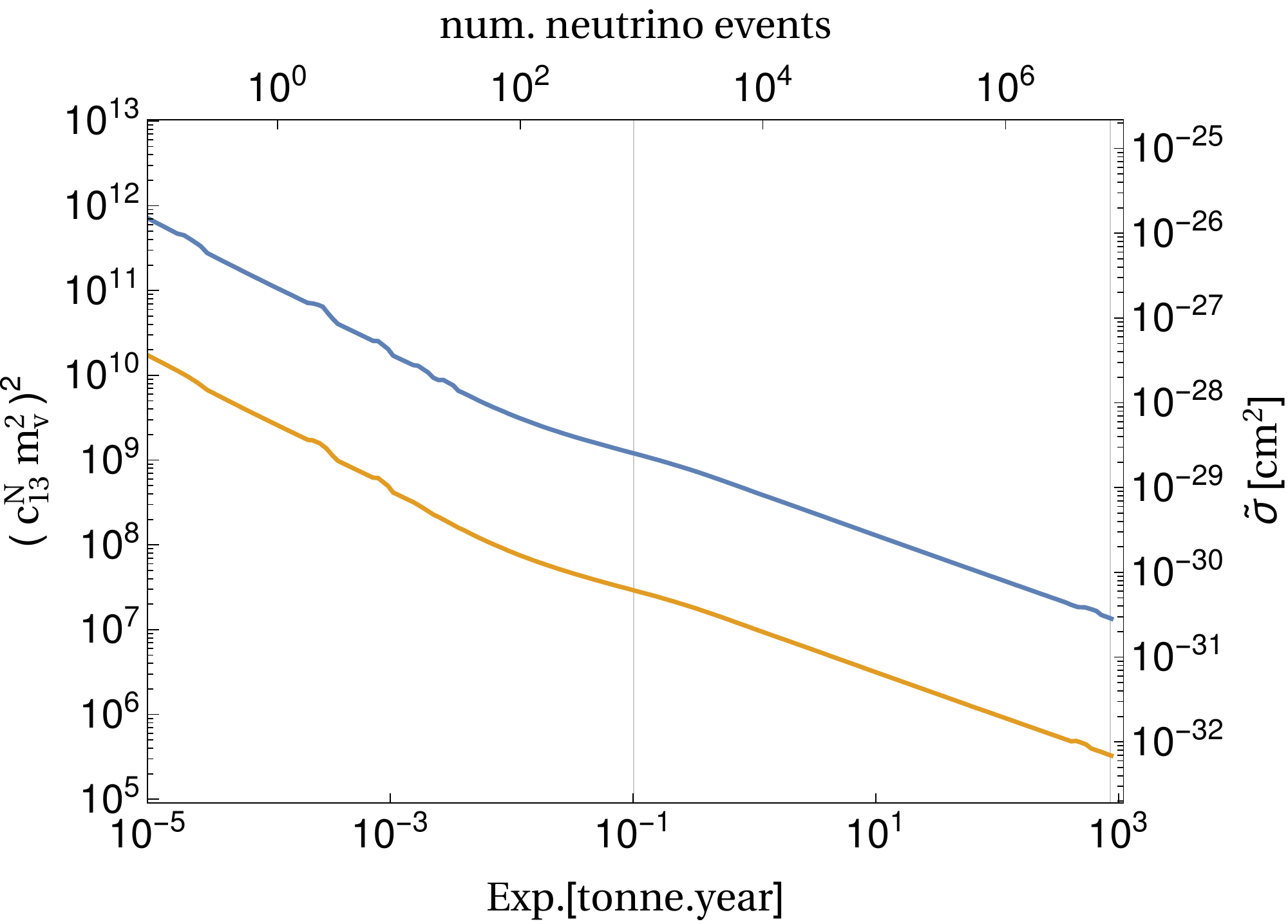} \\
\includegraphics[height=4cm]{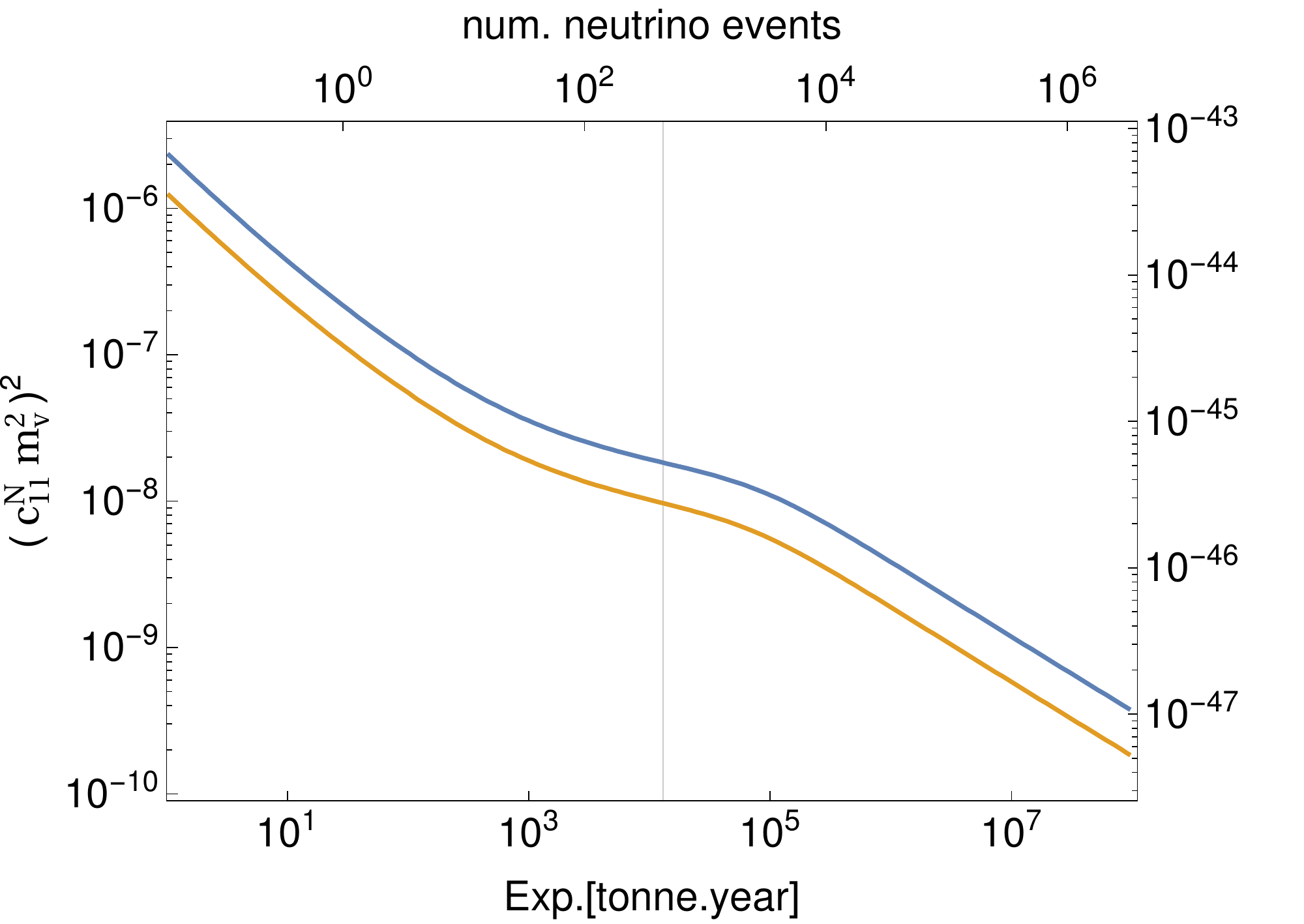} &
\includegraphics[height=4cm]{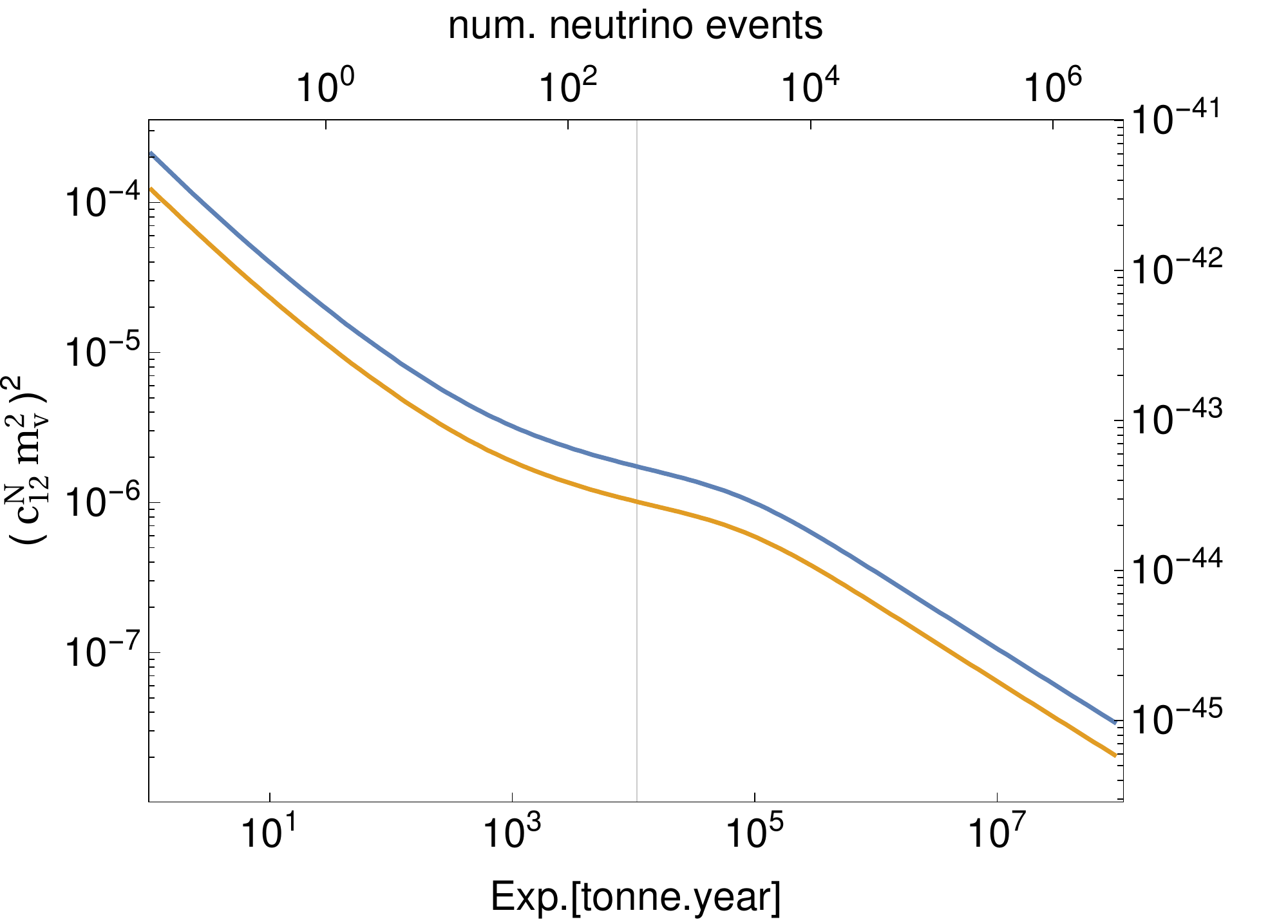} &
\includegraphics[height=4cm]{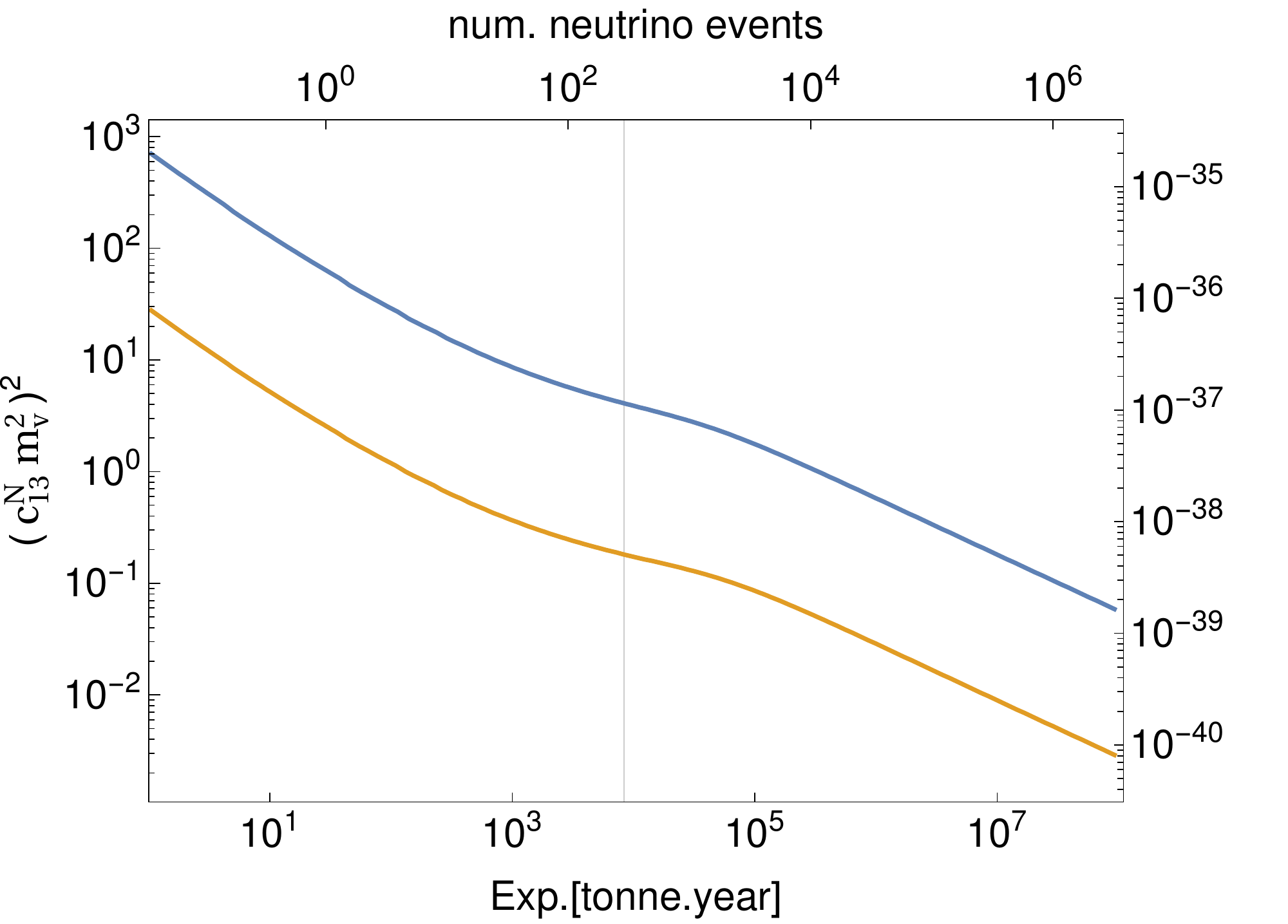} \\
\includegraphics[height=4cm]{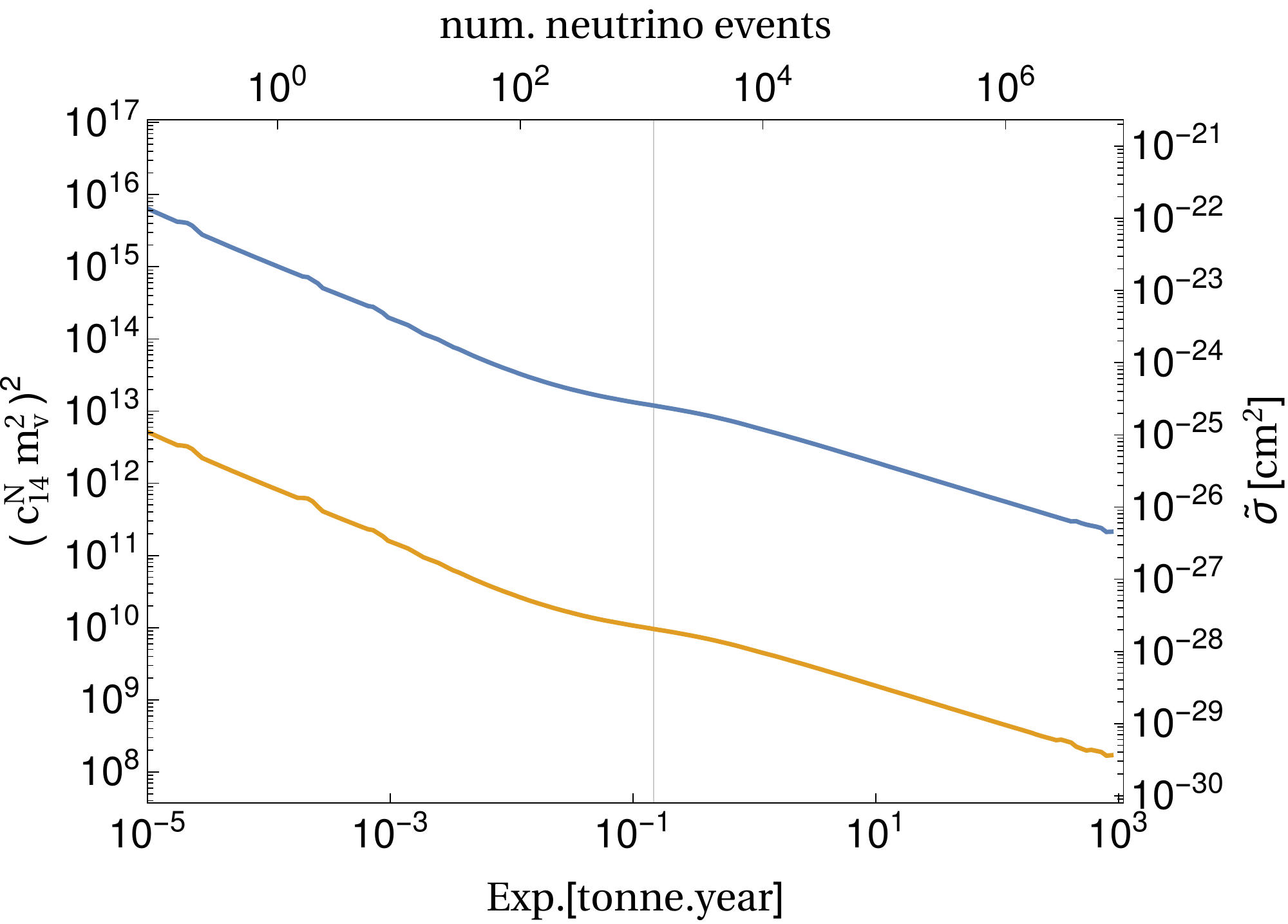}  &
\includegraphics[height=4cm]{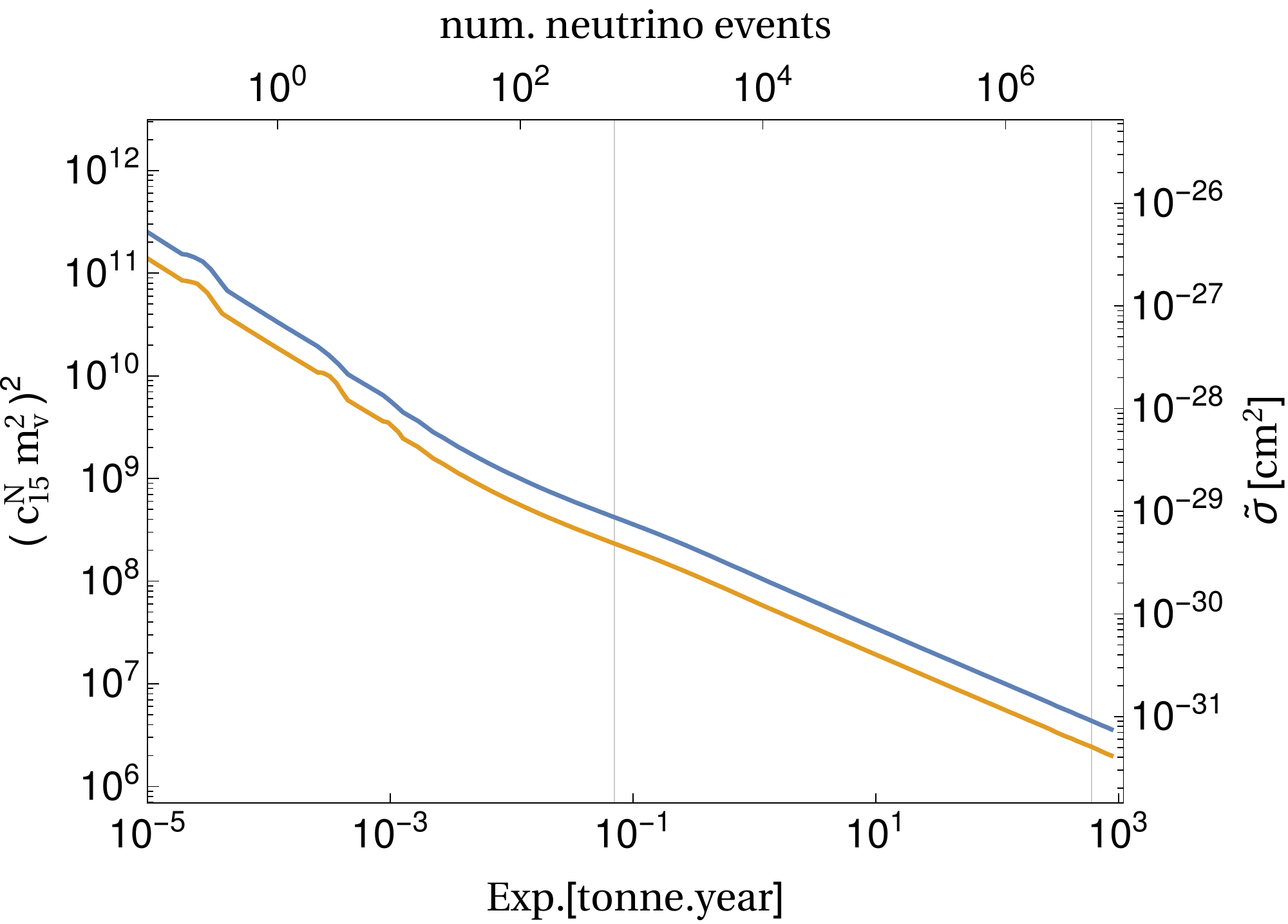}  \\ 
\includegraphics[height=4cm]{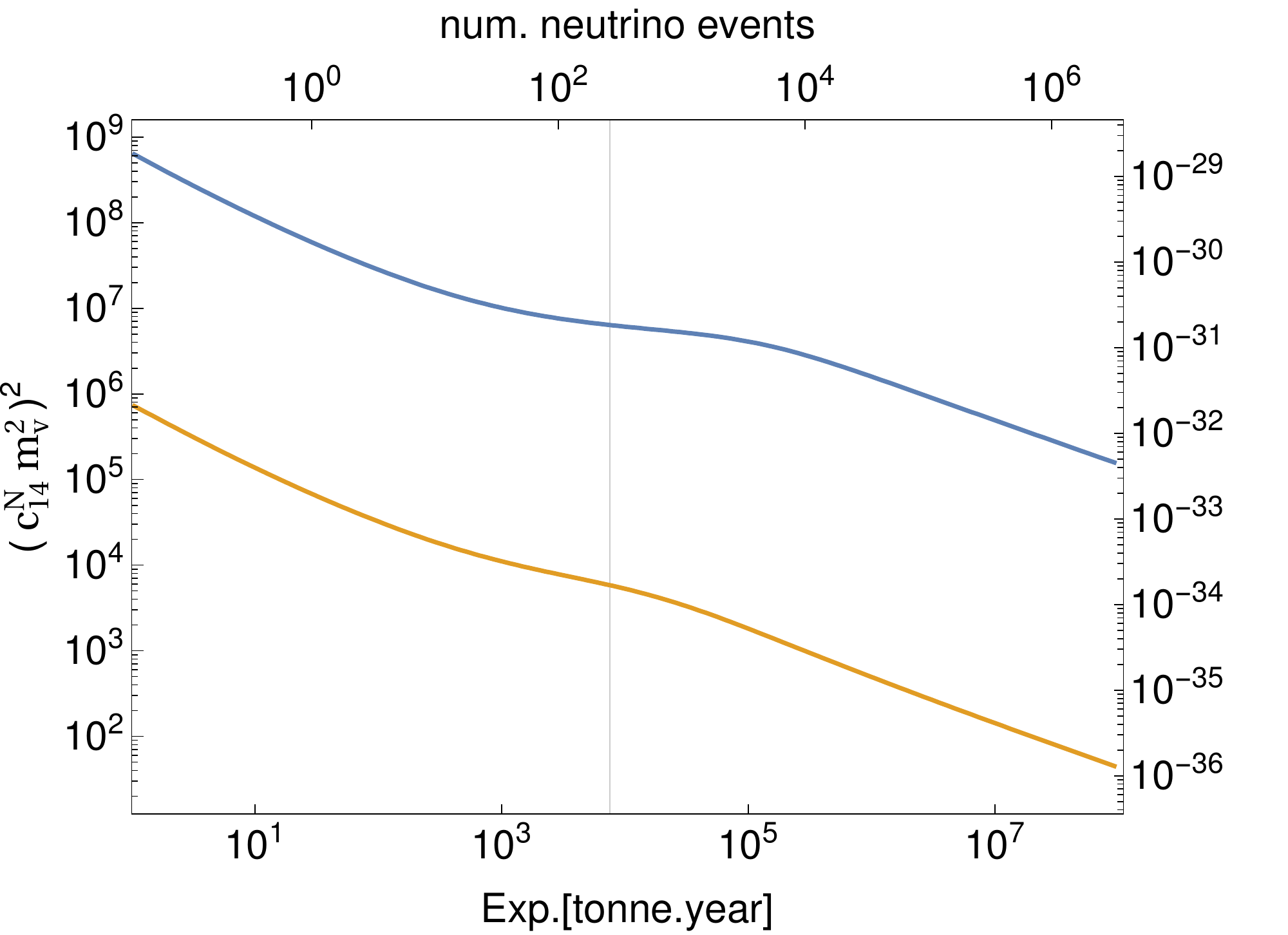} & 
\includegraphics[height=4cm]{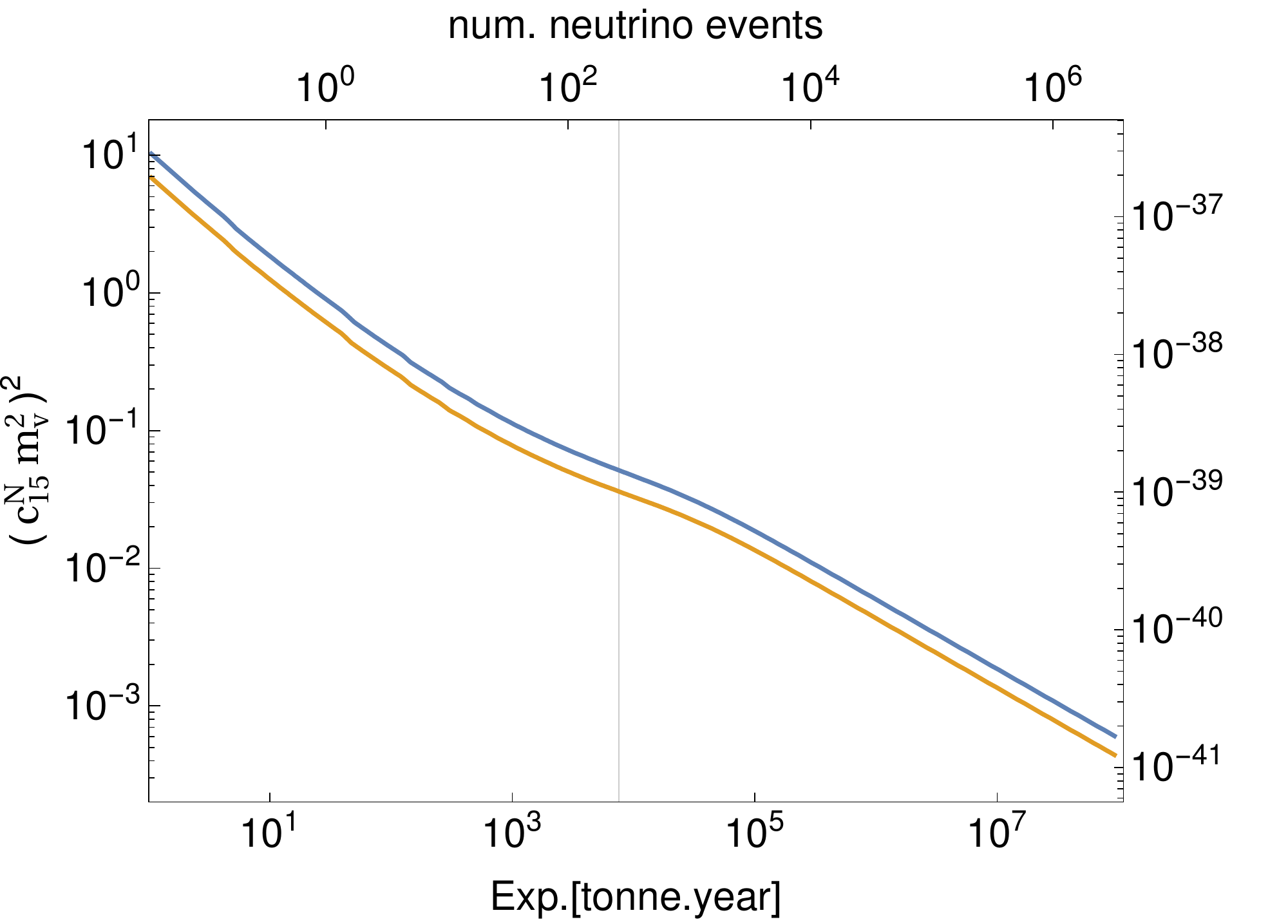} \\
\end{tabular}
\caption{Discovery evolution for the low mass region (first and third rows) and high mass region (second and fourth rows) for operators 11-15. The blue and yellow curves show the limits for proton and neutron scattering, respectively.}
\label{figDiscEvoFull11-15}
\end{figure}

\bibliography{PhysicsBibtex} 

\begin{thebibliography}{40}
\expandafter\ifx\csname natexlab\endcsname\relax\def\natexlab#1{#1}\fi
\expandafter\ifx\csname bibnamefont\endcsname\relax
  \def\bibnamefont#1{#1}\fi
\expandafter\ifx\csname bibfnamefont\endcsname\relax
  \def\bibfnamefont#1{#1}\fi
\expandafter\ifx\csname citenamefont\endcsname\relax
  \def\citenamefont#1{#1}\fi
\expandafter\ifx\csname url\endcsname\relax
  \def\url#1{\texttt{#1}}\fi
\expandafter\ifx\csname urlprefix\endcsname\relax\def\urlprefix{URL }\fi
\providecommand{\bibinfo}[2]{#2}
\providecommand{\eprint}[2][]{\url{#2}}

\bibitem[{\citenamefont{Akerib et~al.}(2015{\natexlab{a}})}]{Akerib:2015rjg}
\bibinfo{author}{\bibfnamefont{D.~S.} \bibnamefont{Akerib}}
  \bibnamefont{et~al.} (\bibinfo{collaboration}{LUX})
  (\bibinfo{year}{2015}{\natexlab{a}}), \eprint{1512.03506}.

\bibitem[{\citenamefont{Amole et~al.}(2015)}]{Amole:2015lsj}
\bibinfo{author}{\bibfnamefont{C.}~\bibnamefont{Amole}} \bibnamefont{et~al.}
  (\bibinfo{collaboration}{PICO}), \bibinfo{journal}{Phys. Rev. Lett.}
  \textbf{\bibinfo{volume}{114}}, \bibinfo{pages}{231302}
  (\bibinfo{year}{2015}), \eprint{1503.00008}.

\bibitem[{\citenamefont{Aprile et~al.}(2015)}]{Aprile:2015uzo}
\bibinfo{author}{\bibfnamefont{E.}~\bibnamefont{Aprile}} \bibnamefont{et~al.}
  (\bibinfo{collaboration}{XENON}), \bibinfo{journal}{Submitted to: JCAP}
  (\bibinfo{year}{2015}), \eprint{1512.07501}.

\bibitem[{\citenamefont{Akerib et~al.}(2015{\natexlab{b}})}]{Akerib:2015cja}
\bibinfo{author}{\bibfnamefont{D.~S.} \bibnamefont{Akerib}}
  \bibnamefont{et~al.} (\bibinfo{collaboration}{LZ})
  (\bibinfo{year}{2015}{\natexlab{b}}), \eprint{1509.02910}.

\bibitem[{\citenamefont{Goodman and Witten}(1985)}]{Goodman:1984dc}
\bibinfo{author}{\bibfnamefont{M.~W.} \bibnamefont{Goodman}} \bibnamefont{and}
  \bibinfo{author}{\bibfnamefont{E.}~\bibnamefont{Witten}},
  \bibinfo{journal}{Phys. Rev.} \textbf{\bibinfo{volume}{D31}},
  \bibinfo{pages}{3059} (\bibinfo{year}{1985}).

\bibitem[{\citenamefont{Lewin and Smith}(1996)}]{Lewin:1995rx}
\bibinfo{author}{\bibfnamefont{J.~D.} \bibnamefont{Lewin}} \bibnamefont{and}
  \bibinfo{author}{\bibfnamefont{P.~F.} \bibnamefont{Smith}},
  \bibinfo{journal}{Astropart. Phys.} \textbf{\bibinfo{volume}{6}},
  \bibinfo{pages}{87} (\bibinfo{year}{1996}).

\bibitem[{\citenamefont{Fitzpatrick et~al.}(2013)\citenamefont{Fitzpatrick,
  Haxton, Katz, Lubbers, and Xu}}]{Fitzpatrick:2012ix}
\bibinfo{author}{\bibfnamefont{A.~L.} \bibnamefont{Fitzpatrick}},
  \bibinfo{author}{\bibfnamefont{W.}~\bibnamefont{Haxton}},
  \bibinfo{author}{\bibfnamefont{E.}~\bibnamefont{Katz}},
  \bibinfo{author}{\bibfnamefont{N.}~\bibnamefont{Lubbers}}, \bibnamefont{and}
  \bibinfo{author}{\bibfnamefont{Y.}~\bibnamefont{Xu}}, \bibinfo{journal}{JCAP}
  \textbf{\bibinfo{volume}{1302}}, \bibinfo{pages}{004} (\bibinfo{year}{2013}),
  \eprint{1203.3542}.

\bibitem[{\citenamefont{Anand et~al.}(2014)\citenamefont{Anand, Fitzpatrick,
  and Haxton}}]{Anand:2013yka}
\bibinfo{author}{\bibfnamefont{N.}~\bibnamefont{Anand}},
  \bibinfo{author}{\bibfnamefont{A.~L.} \bibnamefont{Fitzpatrick}},
  \bibnamefont{and} \bibinfo{author}{\bibfnamefont{W.~C.}
  \bibnamefont{Haxton}}, \bibinfo{journal}{Phys. Rev.}
  \textbf{\bibinfo{volume}{C89}}, \bibinfo{pages}{065501}
  (\bibinfo{year}{2014}), \eprint{1308.6288}.

\bibitem[{\citenamefont{Gresham and Zurek}(2014)}]{Gresham:2014vja}
\bibinfo{author}{\bibfnamefont{M.~I.} \bibnamefont{Gresham}} \bibnamefont{and}
  \bibinfo{author}{\bibfnamefont{K.~M.} \bibnamefont{Zurek}},
  \bibinfo{journal}{Phys. Rev.} \textbf{\bibinfo{volume}{D89}},
  \bibinfo{pages}{123521} (\bibinfo{year}{2014}), \eprint{1401.3739}.

\bibitem[{\citenamefont{Catena}(2015)}]{Catena:2015vpa}
\bibinfo{author}{\bibfnamefont{R.}~\bibnamefont{Catena}},
  \bibinfo{journal}{JCAP} \textbf{\bibinfo{volume}{1507}}, \bibinfo{pages}{026}
  (\bibinfo{year}{2015}), \eprint{1505.06441}.

\bibitem[{\citenamefont{Kavanagh}(2015)}]{Kavanagh:2015jma}
\bibinfo{author}{\bibfnamefont{B.~J.} \bibnamefont{Kavanagh}},
  \bibinfo{journal}{Phys. Rev.} \textbf{\bibinfo{volume}{D92}},
  \bibinfo{pages}{023513} (\bibinfo{year}{2015}), \eprint{1505.07406}.

\bibitem[{\citenamefont{Monroe and Fisher}(2007)}]{Monroe:2007xp}
\bibinfo{author}{\bibfnamefont{J.}~\bibnamefont{Monroe}} \bibnamefont{and}
  \bibinfo{author}{\bibfnamefont{P.}~\bibnamefont{Fisher}},
  \bibinfo{journal}{Phys. Rev.} \textbf{\bibinfo{volume}{D76}},
  \bibinfo{pages}{033007} (\bibinfo{year}{2007}), \eprint{0706.3019}.

\bibitem[{\citenamefont{Strigari}(2009)}]{Strigari:2009bq}
\bibinfo{author}{\bibfnamefont{L.~E.} \bibnamefont{Strigari}},
  \bibinfo{journal}{New J. Phys.} \textbf{\bibinfo{volume}{11}},
  \bibinfo{pages}{105011} (\bibinfo{year}{2009}), \eprint{0903.3630}.

\bibitem[{\citenamefont{Billard et~al.}(2014)\citenamefont{Billard, Strigari,
  and Figueroa-Feliciano}}]{Billard:2013qya}
\bibinfo{author}{\bibfnamefont{J.}~\bibnamefont{Billard}},
  \bibinfo{author}{\bibfnamefont{L.}~\bibnamefont{Strigari}}, \bibnamefont{and}
  \bibinfo{author}{\bibfnamefont{E.}~\bibnamefont{Figueroa-Feliciano}},
  \bibinfo{journal}{Phys. Rev.} \textbf{\bibinfo{volume}{D89}},
  \bibinfo{pages}{023524} (\bibinfo{year}{2014}), \eprint{1307.5458}.

\bibitem[{\citenamefont{Ruppin et~al.}(2014)\citenamefont{Ruppin, Billard,
  Figueroa-Feliciano, and Strigari}}]{Ruppin:2014bra}
\bibinfo{author}{\bibfnamefont{F.}~\bibnamefont{Ruppin}},
  \bibinfo{author}{\bibfnamefont{J.}~\bibnamefont{Billard}},
  \bibinfo{author}{\bibfnamefont{E.}~\bibnamefont{Figueroa-Feliciano}},
  \bibnamefont{and} \bibinfo{author}{\bibfnamefont{L.}~\bibnamefont{Strigari}},
  \bibinfo{journal}{Phys. Rev.} \textbf{\bibinfo{volume}{D90}},
  \bibinfo{pages}{083510} (\bibinfo{year}{2014}), \eprint{1408.3581}.

\bibitem[{\citenamefont{Davis}(2015)}]{Davis:2014ama}
\bibinfo{author}{\bibfnamefont{J.~H.} \bibnamefont{Davis}},
  \bibinfo{journal}{JCAP} \textbf{\bibinfo{volume}{1503}}, \bibinfo{pages}{012}
  (\bibinfo{year}{2015}), \eprint{1412.1475}.

\bibitem[{\citenamefont{Freese et~al.}(2013)\citenamefont{Freese, Lisanti, and
  Savage}}]{Freese:2012xd}
\bibinfo{author}{\bibfnamefont{K.}~\bibnamefont{Freese}},
  \bibinfo{author}{\bibfnamefont{M.}~\bibnamefont{Lisanti}}, \bibnamefont{and}
  \bibinfo{author}{\bibfnamefont{C.}~\bibnamefont{Savage}},
  \bibinfo{journal}{Rev. Mod. Phys.} \textbf{\bibinfo{volume}{85}},
  \bibinfo{pages}{1561} (\bibinfo{year}{2013}), \eprint{1209.3339}.

\bibitem[{\citenamefont{Grothaus et~al.}(2014)\citenamefont{Grothaus,
  Fairbairn, and Monroe}}]{Grothaus:2014hja}
\bibinfo{author}{\bibfnamefont{P.}~\bibnamefont{Grothaus}},
  \bibinfo{author}{\bibfnamefont{M.}~\bibnamefont{Fairbairn}},
  \bibnamefont{and} \bibinfo{author}{\bibfnamefont{J.}~\bibnamefont{Monroe}},
  \bibinfo{journal}{Phys. Rev.} \textbf{\bibinfo{volume}{D90}},
  \bibinfo{pages}{055018} (\bibinfo{year}{2014}), \eprint{1406.5047}.

\bibitem[{\citenamefont{O'Hare et~al.}(2015)\citenamefont{O'Hare, Green,
  Billard, Figueroa-Feliciano, and Strigari}}]{O'Hare:2015mda}
\bibinfo{author}{\bibfnamefont{C.~A.~J.} \bibnamefont{O'Hare}},
  \bibinfo{author}{\bibfnamefont{A.~M.} \bibnamefont{Green}},
  \bibinfo{author}{\bibfnamefont{J.}~\bibnamefont{Billard}},
  \bibinfo{author}{\bibfnamefont{E.}~\bibnamefont{Figueroa-Feliciano}},
  \bibnamefont{and} \bibinfo{author}{\bibfnamefont{L.~E.}
  \bibnamefont{Strigari}}, \bibinfo{journal}{Phys. Rev.}
  \textbf{\bibinfo{volume}{D92}}, \bibinfo{pages}{063518}
  (\bibinfo{year}{2015}), \eprint{1505.08061}.

\bibitem[{\citenamefont{Read}(2014)}]{Read:2014qva}
\bibinfo{author}{\bibfnamefont{J.~I.} \bibnamefont{Read}}, \bibinfo{journal}{J.
  Phys.} \textbf{\bibinfo{volume}{G41}}, \bibinfo{pages}{063101}
  (\bibinfo{year}{2014}), \eprint{1404.1938}.

\bibitem[{\citenamefont{Strigari}(2013)}]{Strigari:2013iaa}
\bibinfo{author}{\bibfnamefont{L.~E.} \bibnamefont{Strigari}},
  \bibinfo{journal}{Phys. Rept.} \textbf{\bibinfo{volume}{531}},
  \bibinfo{pages}{1} (\bibinfo{year}{2013}), \eprint{1211.7090}.

\bibitem[{\citenamefont{Mao et~al.}(2013)\citenamefont{Mao, Strigari, Wechsler,
  Wu, and Hahn}}]{Mao:2012hf}
\bibinfo{author}{\bibfnamefont{Y.-Y.} \bibnamefont{Mao}},
  \bibinfo{author}{\bibfnamefont{L.~E.} \bibnamefont{Strigari}},
  \bibinfo{author}{\bibfnamefont{R.~H.} \bibnamefont{Wechsler}},
  \bibinfo{author}{\bibfnamefont{H.-Y.} \bibnamefont{Wu}}, \bibnamefont{and}
  \bibinfo{author}{\bibfnamefont{O.}~\bibnamefont{Hahn}},
  \bibinfo{journal}{Astrophys. J.} \textbf{\bibinfo{volume}{764}},
  \bibinfo{pages}{35} (\bibinfo{year}{2013}), \eprint{1210.2721}.

\bibitem[{\citenamefont{Mao et~al.}(2014)\citenamefont{Mao, Strigari, and
  Wechsler}}]{Mao:2013nda}
\bibinfo{author}{\bibfnamefont{Y.-Y.} \bibnamefont{Mao}},
  \bibinfo{author}{\bibfnamefont{L.~E.} \bibnamefont{Strigari}},
  \bibnamefont{and} \bibinfo{author}{\bibfnamefont{R.~H.}
  \bibnamefont{Wechsler}}, \bibinfo{journal}{Phys. Rev.}
  \textbf{\bibinfo{volume}{D89}}, \bibinfo{pages}{063513}
  (\bibinfo{year}{2014}), \eprint{1304.6401}.

\bibitem[{\citenamefont{Piffl et~al.}(2014)}]{Piffl:2013mla}
\bibinfo{author}{\bibfnamefont{T.}~\bibnamefont{Piffl}} \bibnamefont{et~al.},
  \bibinfo{journal}{Astron. Astrophys.} \textbf{\bibinfo{volume}{562}},
  \bibinfo{pages}{A91} (\bibinfo{year}{2014}), \eprint{1309.4293}.

\bibitem[{\citenamefont{Catena}(2014)}]{Catena:2014hla}
\bibinfo{author}{\bibfnamefont{R.}~\bibnamefont{Catena}},
  \bibinfo{journal}{JCAP} \textbf{\bibinfo{volume}{1409}}, \bibinfo{pages}{049}
  (\bibinfo{year}{2014}), \eprint{1407.0127}.

\bibitem[{\citenamefont{Catena and Gondolo}(2015)}]{Catena:2015uua}
\bibinfo{author}{\bibfnamefont{R.}~\bibnamefont{Catena}} \bibnamefont{and}
  \bibinfo{author}{\bibfnamefont{P.}~\bibnamefont{Gondolo}},
  \bibinfo{journal}{JCAP} \textbf{\bibinfo{volume}{1508}}, \bibinfo{pages}{022}
  (\bibinfo{year}{2015}), \eprint{1504.06554}.

\bibitem[{\citenamefont{Schneck et~al.}(2015)}]{Schneck:2015eqa}
\bibinfo{author}{\bibfnamefont{K.}~\bibnamefont{Schneck}} \bibnamefont{et~al.}
  (\bibinfo{collaboration}{SuperCDMS}), \bibinfo{journal}{Phys. Rev.}
  \textbf{\bibinfo{volume}{D91}}, \bibinfo{pages}{092004}
  (\bibinfo{year}{2015}), \eprint{1503.03379}.

\bibitem[{\citenamefont{Dent et~al.}(2015)\citenamefont{Dent, Krauss, Newstead,
  and Sabharwal}}]{Dent:2015zpa}
\bibinfo{author}{\bibfnamefont{J.~B.} \bibnamefont{Dent}},
  \bibinfo{author}{\bibfnamefont{L.~M.} \bibnamefont{Krauss}},
  \bibinfo{author}{\bibfnamefont{J.~L.} \bibnamefont{Newstead}},
  \bibnamefont{and}
  \bibinfo{author}{\bibfnamefont{S.}~\bibnamefont{Sabharwal}},
  \bibinfo{journal}{Phys. Rev.} \textbf{\bibinfo{volume}{D92}},
  \bibinfo{pages}{063515} (\bibinfo{year}{2015}), \eprint{1505.03117}.

\bibitem[{\citenamefont{Agrawal et~al.}(2010)\citenamefont{Agrawal, Chacko,
  Kilic, and Mishra}}]{Agrawal:2010fh}
\bibinfo{author}{\bibfnamefont{P.}~\bibnamefont{Agrawal}},
  \bibinfo{author}{\bibfnamefont{Z.}~\bibnamefont{Chacko}},
  \bibinfo{author}{\bibfnamefont{C.}~\bibnamefont{Kilic}}, \bibnamefont{and}
  \bibinfo{author}{\bibfnamefont{R.~K.} \bibnamefont{Mishra}}
  (\bibinfo{year}{2010}), \eprint{1003.1912}.

\bibitem[{\citenamefont{Dienes et~al.}(2014)\citenamefont{Dienes, Kumar,
  Thomas, and Yaylali}}]{Dienes:2013xya}
\bibinfo{author}{\bibfnamefont{K.~R.} \bibnamefont{Dienes}},
  \bibinfo{author}{\bibfnamefont{J.}~\bibnamefont{Kumar}},
  \bibinfo{author}{\bibfnamefont{B.}~\bibnamefont{Thomas}}, \bibnamefont{and}
  \bibinfo{author}{\bibfnamefont{D.}~\bibnamefont{Yaylali}},
  \bibinfo{journal}{Phys. Rev.} \textbf{\bibinfo{volume}{D90}},
  \bibinfo{pages}{015012} (\bibinfo{year}{2014}), \eprint{1312.7772}.

\bibitem[{\citenamefont{Hill and Solon}(2015)}]{Hill:2014yxa}
\bibinfo{author}{\bibfnamefont{R.~J.} \bibnamefont{Hill}} \bibnamefont{and}
  \bibinfo{author}{\bibfnamefont{M.~P.} \bibnamefont{Solon}},
  \bibinfo{journal}{Phys. Rev.} \textbf{\bibinfo{volume}{D91}},
  \bibinfo{pages}{043505} (\bibinfo{year}{2015}), \eprint{1409.8290}.

\bibitem[{\citenamefont{Hoferichter et~al.}(2015)\citenamefont{Hoferichter,
  Klos, and Schwenk}}]{Hoferichter:2015ipa}
\bibinfo{author}{\bibfnamefont{M.}~\bibnamefont{Hoferichter}},
  \bibinfo{author}{\bibfnamefont{P.}~\bibnamefont{Klos}}, \bibnamefont{and}
  \bibinfo{author}{\bibfnamefont{A.}~\bibnamefont{Schwenk}},
  \bibinfo{journal}{Phys. Lett.} \textbf{\bibinfo{volume}{B746}},
  \bibinfo{pages}{410} (\bibinfo{year}{2015}), \eprint{1503.04811}.

\bibitem[{\citenamefont{Newstead et~al.}(2013)\citenamefont{Newstead, Jacques,
  Krauss, Dent, and Ferrer}}]{Newstead:2013pea}
\bibinfo{author}{\bibfnamefont{J.~L.} \bibnamefont{Newstead}},
  \bibinfo{author}{\bibfnamefont{T.~D.} \bibnamefont{Jacques}},
  \bibinfo{author}{\bibfnamefont{L.~M.} \bibnamefont{Krauss}},
  \bibinfo{author}{\bibfnamefont{J.~B.} \bibnamefont{Dent}}, \bibnamefont{and}
  \bibinfo{author}{\bibfnamefont{F.}~\bibnamefont{Ferrer}},
  \bibinfo{journal}{Phys. Rev.} \textbf{\bibinfo{volume}{D88}},
  \bibinfo{pages}{076011} (\bibinfo{year}{2013}), \eprint{1306.3244}.

\bibitem[{\citenamefont{Freedman}(1974)}]{Freedman:1973yd}
\bibinfo{author}{\bibfnamefont{D.~Z.} \bibnamefont{Freedman}},
  \bibinfo{journal}{Phys. Rev.} \textbf{\bibinfo{volume}{D9}},
  \bibinfo{pages}{1389} (\bibinfo{year}{1974}).

\bibitem[{\citenamefont{Cabrera et~al.}(1985)\citenamefont{Cabrera, Krauss, and
  Wilczek}}]{Cabrera:1984rr}
\bibinfo{author}{\bibfnamefont{B.}~\bibnamefont{Cabrera}},
  \bibinfo{author}{\bibfnamefont{L.~M.} \bibnamefont{Krauss}},
  \bibnamefont{and} \bibinfo{author}{\bibfnamefont{F.}~\bibnamefont{Wilczek}},
  \bibinfo{journal}{Phys. Rev. Lett.} \textbf{\bibinfo{volume}{55}},
  \bibinfo{pages}{25} (\bibinfo{year}{1985}).

\bibitem[{\citenamefont{Dutta et~al.}(2015)\citenamefont{Dutta, Mahapatra,
  Strigari, and Walker}}]{Dutta:2015vwa}
\bibinfo{author}{\bibfnamefont{B.}~\bibnamefont{Dutta}},
  \bibinfo{author}{\bibfnamefont{R.}~\bibnamefont{Mahapatra}},
  \bibinfo{author}{\bibfnamefont{L.~E.} \bibnamefont{Strigari}},
  \bibnamefont{and} \bibinfo{author}{\bibfnamefont{J.~W.} \bibnamefont{Walker}}
  (\bibinfo{year}{2015}), \eprint{1508.07981}.

\bibitem[{\citenamefont{Cowan et~al.}(2011)\citenamefont{Cowan, Cranmer, Gross,
  and Vitells}}]{Cowan:2010js}
\bibinfo{author}{\bibfnamefont{G.}~\bibnamefont{Cowan}},
  \bibinfo{author}{\bibfnamefont{K.}~\bibnamefont{Cranmer}},
  \bibinfo{author}{\bibfnamefont{E.}~\bibnamefont{Gross}}, \bibnamefont{and}
  \bibinfo{author}{\bibfnamefont{O.}~\bibnamefont{Vitells}},
  \bibinfo{journal}{Eur. Phys. J.} \textbf{\bibinfo{volume}{C71}},
  \bibinfo{pages}{1554} (\bibinfo{year}{2011}), \bibinfo{note}{[Erratum: Eur.
  Phys. J.C73,2501(2013)]}, \eprint{1007.1727}.

\bibitem[{\citenamefont{Haxton et~al.}(2013)\citenamefont{Haxton,
  Hamish~Robertson, and Serenelli}}]{Robertson:2012ib}
\bibinfo{author}{\bibfnamefont{W.~C.} \bibnamefont{Haxton}},
  \bibinfo{author}{\bibfnamefont{R.~G.} \bibnamefont{Hamish~Robertson}},
  \bibnamefont{and} \bibinfo{author}{\bibfnamefont{A.~M.}
  \bibnamefont{Serenelli}}, \bibinfo{journal}{Ann. Rev. Astron. Astrophys.}
  \textbf{\bibinfo{volume}{51}}, \bibinfo{pages}{21} (\bibinfo{year}{2013}),
  \eprint{1208.5723}.

\bibitem[{\citenamefont{Savage et~al.}(2015)\citenamefont{Savage, Scaffidi,
  White, and Williams}}]{Savage:2015xta}
\bibinfo{author}{\bibfnamefont{C.}~\bibnamefont{Savage}},
  \bibinfo{author}{\bibfnamefont{A.}~\bibnamefont{Scaffidi}},
  \bibinfo{author}{\bibfnamefont{M.}~\bibnamefont{White}}, \bibnamefont{and}
  \bibinfo{author}{\bibfnamefont{A.~G.} \bibnamefont{Williams}},
  \bibinfo{journal}{Phys. Rev.} \textbf{\bibinfo{volume}{D92}},
  \bibinfo{pages}{103519} (\bibinfo{year}{2015}), \eprint{1502.02667}.

\bibitem[{\citenamefont{Gluscevic et~al.}(2015)\citenamefont{Gluscevic,
  Gresham, McDermott, Peter, and Zurek}}]{Gluscevic:2015sqa}
\bibinfo{author}{\bibfnamefont{V.}~\bibnamefont{Gluscevic}},
  \bibinfo{author}{\bibfnamefont{M.~I.} \bibnamefont{Gresham}},
  \bibinfo{author}{\bibfnamefont{S.~D.} \bibnamefont{McDermott}},
  \bibinfo{author}{\bibfnamefont{A.~H.~G.} \bibnamefont{Peter}},
  \bibnamefont{and} \bibinfo{author}{\bibfnamefont{K.~M.} \bibnamefont{Zurek}},
  \bibinfo{journal}{JCAP} \textbf{\bibinfo{volume}{1512}}, \bibinfo{pages}{057}
  (\bibinfo{year}{2015}), \eprint{1506.04454}.

\end{thebibliography}
\end{document}